\documentclass[reprint,aps,prx,superscriptaddress]{revtex4-2}
\usepackage{graphicx}
\usepackage{dcolumn}
\usepackage{bm}
\usepackage{mathrsfs}
\usepackage{tikz}
\usepackage{amssymb}
\usepackage{amsmath}
\usepackage{amsthm}
\usepackage{amsopn}
\usepackage[normalem]{ulem}
\usepackage{cancel}
\usepackage{color}
\usepackage{comment}
\usepackage{physics}
\usepackage{orcidlink}
\usepackage[margin=1.7cm]{geometry}
\usepackage{array}
\usepackage{tabularx}
\usepackage{multirow}
\usepackage{hhline}

\newtheorem{lemma}{Lemma}
\newtheorem{claim}{Claim}

\newenvironment{heuristicderivation}
  {\begin{proof}[Heuristic derivation]}
  {\end{proof}}

\usepackage{hyperref}
\hypersetup{
    colorlinks=true,    
    linkcolor=blue,
    citecolor=blue,
    filecolor=magenta,
    urlcolor=blue
}

\begin{document}

\title{The arrival position problem in quantum mechanics}
\author{Ali Ayatollah Rafsanjani\,\orcidlink{0000-0002-3717-718X}}
	\email{aliayat@physics.sharif.edu}
	\affiliation{Department of Physics and Energy Engineering, Amirkabir University of Technology, Tehran, Iran}
	\affiliation{Department of Physics, Sharif University of Technology, Tehran, Iran}
\author{Will Cavendish\,\orcidlink{0009-0005-7578-756X}}
    \email{willcavendish@johnbellinstitute.org}
    \affiliation{John Bell Institute for the Foundations of Physics, New York, NY 10003, USA}
\date{\today}

\begin{abstract} The problem of making unambiguous probabilistic predictions about experiments involving waiting ``always on'' detectors remains a challenge for quantum theory. While most research on this problem studies arrival \emph{time}, i.e., predicting the distribution of \emph{when} detection events occur, this paper studies the arrival \emph{position} problem, which is the complementary challenge of predicting the distribution of \emph{where} detection events occur. Despite the widespread recognition of the arrival time problem, the inability of standard quantum theory to unambiguously address the arrival position problem remains a pervasive theoretical blind spot. In this paper, we compare quantitative arrival position predictions derived from prominent proposed solutions to the screen problem. As we show, these models yield distinguishable predictions even in relatively simple experiments achievable with current technology. Notably, many of these discrepancies persist even in the far-field limit, where standard semiclassical approximations are typically assumed to be valid.
\end{abstract}

\maketitle

\section{Introduction}

Suppose that a particle, initially confined to a trap, is released in the presence of a waiting detector screen. If it is eventually detected by the screen, its ``arrival position'' is recorded. Depending on the experimental protocol, this position can be determined either directly, by the localized detectors comprising the screen, or indirectly, e.g., by triangulating the light scintillated at the screen's surface. How can we predict the probability distribution of these observed positions? Though many experiments fitting this description have been carried out \cite{lawall1995three,carcy2019momentum,bucker2009single,manning2014single}, standard quantum mechanics provides no unambiguous prescription. The difficulty is that the Dirac-von Neumann axioms are intended to describe measurements of observables executed at an \textit{instant} of time, not experiments occurring \textit{across} time (see \cite{sudbery1984observation, sudbery1986continuous, mielnik1994screen, home1997conceptual, sudbery2002verdammte} for discussions of this issue). Mielnik called the inadequacy of the orthodox quantum formalism to address the question of when and where detection events occur the ``screen problem'' \cite{mielnik1994screen}. It remains unsolved.

Much of the work on the screen problem to date has focused on the arrival time problem, i.e., the challenge of predicting the probability distribution of \textit{when} a detection event occurs \cite{muga2000arrival, book:Muga, vona2013does,Das2019Arrival}. Here we take up the complementary theoretical challenge of predicting the probability distribution of \textit{where} that detection event occurs, which we call the \textit{arrival position problem}. This distribution is formally a marginal of the joint distribution asked for in the screen problem. In many familiar settings, e.g., the double-slit experiment, the empirical data of interest are precisely this marginal: the positions of dots on the screen, recorded without any reference to time. However, even this aspect of the screen problem lies beyond the reach of quantum measurement theory as it is typically formulated. In particular, a waiting screen does not ``perform a position measurement'' in the textbook sense. The textbook formalism of instantaneous position measurement addresses the question: ``Given a state preparation procedure, what is the probability distribution of outcomes for position measurements performed at a specific \emph{instant} of time?'' In the arrival position problem, on the other hand, there is no ``instant of measurement'' specified by the procedure or chosen by the experimenter. The experimenter simply initiates each experimental trial and then waits for a dot to appear.

The gap presented by the screen problem is more than a curiosity waiting for a satisfactory theoretical resolution. It is an acute empirical problem that only data can settle. The leading attempts to solve this problem make flatly incompatible quantitative predictions about what experiments will show, a theme that has been well-explored in the arrival time literature \cite{ayatollah2023can, naidon2024inequivalence, Das2025DSERevisited}. In this paper, we show that strong contrasts between different proposals are present in the case of the arrival position problem as well. We do this by quantitatively comparing several proposed solutions to this problem, including the semiclassical proposal \cite{NewtonScattering,gliserin2016high,wolf2000ion,kurtsiefer1995time,copley1993neutron,kothe2013time}, the quantum flux proposal \cite{halliwell2009quantum, boonchui2013arrival, leavens1993arrival, damborenea2002measurement,muga1995time,Hannstein2005}, the ``standard'' (or Kijowski) proposal \cite{egusquiza2008standard,Kijowski1974,baute2000time, leon2000time}, the complex absorbing potential (or continuous measurement-based) proposal \cite{allcock1969time, halliwell1999arrival, muga2004complex, halliwell2009arrival, mackrory2010reflection}, the absorbing boundary rule \cite{werner1987arrival,Dubey2021Quantum,Tumulka2022Detection, Tumulka2022Distribution, tumulka2023absorbing, tahvildar2024detection}, and the path-integral with absorbing boundary proposal \cite{Marchewka2001Survival,Marchewka1998Feynman,Marchewka2000Path,Marchewka2002}. We demonstrate that these models yield empirically distinguishable arrival position marginals in surprisingly simple situations, allowing us to evaluate resolutions of the screen problem while sidestepping theory-laden debates regarding the ``observability'' of time.

The experiments we propose are grounded in conventional atom trapping and detection schemes. For state preparation, the protocols we discuss are based on single and double-well trapping techniques that are well-established for both single atoms \cite{kierig2008single,endres2016atom,florshaim2024spatial} and ensembles \cite{andrews1997observation,Shin2004Atom,schumm2005matter,Jo2007long,dussarrat2017two,Colciaghi2023Einstein,dai2016generation,folling2007direct}. For particle detection, our setup allows for different ``screen'' implementations, and the contrasting predictions we calculate are pronounced enough to be accessible using a variety of single-atom detection techniques involving ``waiting'' detectors \cite{bucker2009single,berrada2013integrated,erne2018universal,bergschneider2018spin,perrin2012hanbury,pigneur2018relaxation,borselli2021two,fuhrmanek2010measurement}. To ensure each framework yields an unambiguous prediction, we have chosen experimental setups that circumvent several well-known theoretical issues. Most importantly, the examples we consider exhibit current positivity along the screen locus (see Appendix~\ref{app:CPC}), which allows us to apply the quantum flux proposal without truncation \cite{vona2013does}.

Our formulation of this experimental scheme leaves the means of detection open by design. Any detection scheme with the following two features falls within the scope of the present framework. First, it must employ an ``always on'' screen, detector array, or light sheet, rather than instantaneous measurements executed at a time controlled by the experimenter (see, e.g., \cite{lloyd2026quantum} for an example of a protocol that lacks this feature but is intended to address closely related questions). Second, each trial must end either in a single localized detection event or in a ``non-detection'' outcome (see Section~\ref{sec:proposedExperiment} for details). Experiments that record a sequence of detection events per trial, such as recent Bose-Einstein condensate proposals \cite{naidon2026arrival}, fall outside the present framework, which assigns to each initial wave function a probability distribution over outcomes with at most one detection event per trial.

This agnosticism about detector details should not be read as an implicit claim that those details are unimportant. On the contrary, each of the proposals surveyed may work well for some screen implementations and not others. As we show below, these proposals make mutually incompatible predictions, even in the far-field regime, where standard scattering theory has been broadly successful. Since the predictions of standard scattering theory are detector agnostic, the existence of any detection scheme exhibiting the novel far-field behavior predicted by several of these proposals would demonstrate the limits of standard predictive techniques. Many relatively simple experiments could thus take us beyond our current understanding of how particles behave in the presence of waiting detectors.

 \begin{figure*}[ht!]
    \centering
    \begin{tikzpicture}
        \node[anchor=south west, inner sep=0] (img) at (0,0)
            {\includegraphics[width=1.0\linewidth]{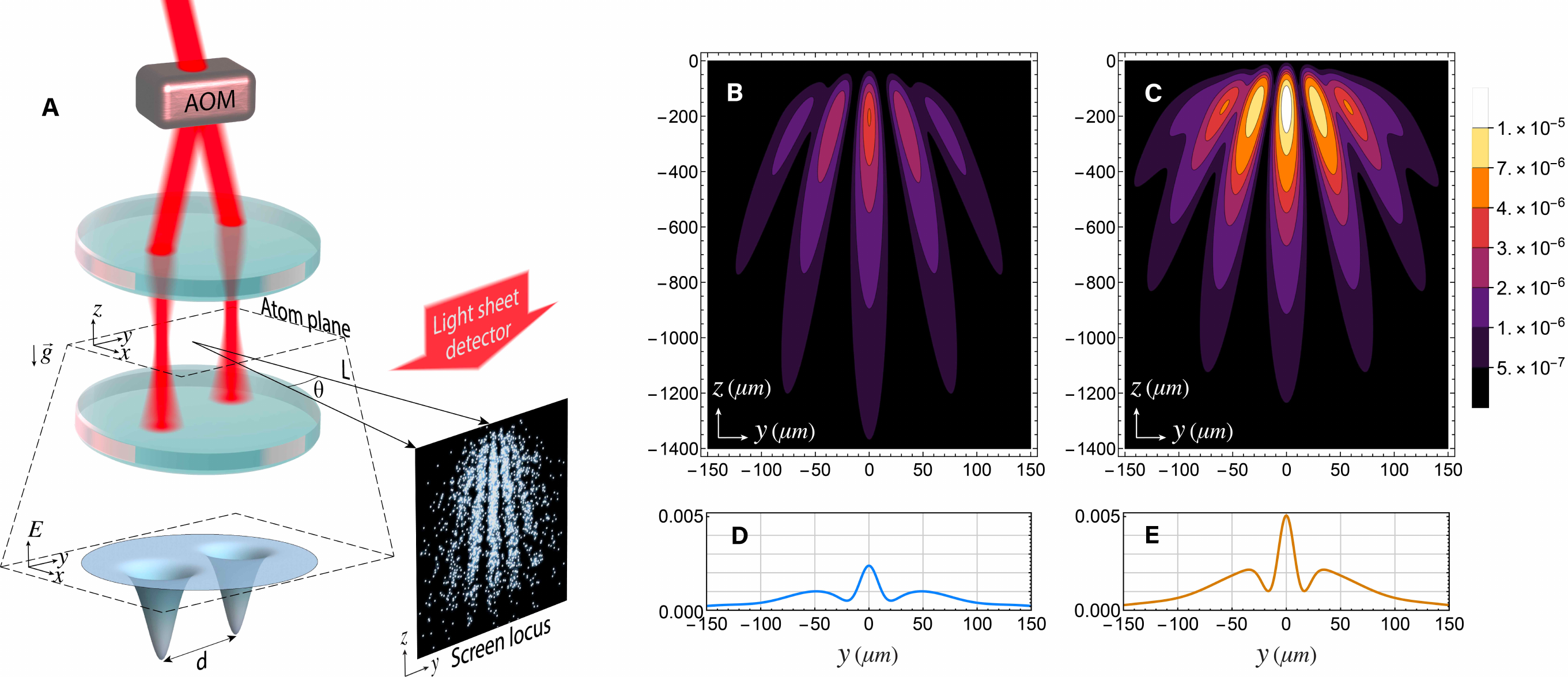}};
    \end{tikzpicture}
    \caption{A realization of the proposed experimental scheme and arrival position distributions for two proposals. Panel A shows a schematic of the setup. An atom is prepared in the delocalized ground state of an optical double-well potential \cite{kaufman2014two}, which is formed by two laser beams generated by an acousto-optic modulator (AOM) driven by two radio-frequency (rf) tones. Lenses following the AOM control the beam geometry. The separation between the two wells is precisely tuned by the frequency difference of the rf drives. The atom is released from the trap in the presence of a screen in the plane $x\!=\!L$, where the origin of the coordinate system is taken to be the center point between the two wells. In the implementation shown, the detector screen is realized by a light sheet \cite{bucker2009single, berrada2013integrated,erne2018universal,borselli2021two}. Panels B and C show the theoretical arrival position distributions for ${}^4\mathrm{He}$ given by the quantum flux (QF) proposal and Marchewka and Schuss's (MS) proposal for $\lambda \approx 2.73\,\sigma$ (see Section~\ref{sec:singleDouble}), respectively, with $\sigma=1\,\mu\mathrm{m}$ (the standard deviation of the Gaussian wells), $d=20\sigma$ (the well separation), $L=100\sigma$ (the distance to the screen), $T=100\,\mathrm{ms}$ (the cutoff time), and $g=9.81\,\mathrm{m/s}^2$. Panels D and E show the associated $y$-position marginals.}
    \label{SchematicSetup}
\end{figure*}


\section{Proposed experimental scheme}\label{sec:proposedExperiment}

In this section, we put forward a flexible experimental scheme for testing proposed solutions to the screen problem. The main components of the experiment are a particle trap used to confine the particle, and a ``screen'' designed to detect the particle along a planar surface $\mathbb{S}_L=\{(L,y,z)~|~y,z\in \mathbb{R}\}$ after it is released (see Fig.~\ref{SchematicSetup}). As we pointed out in the introduction, the proposals make different predictions for this experimental scheme across a range of trap and screen geometries.

The experimental procedure begins with the preparation of the target particle. The aim of the preparation phase is to produce a predictable initial state of the system, e.g., by cooling the trapped particle to its three-dimensional motional ground state using established laser techniques \cite{kaufman2012cooling,thompson2013coherence,norcia2018microscopic,spence2022preparation}. In the case of a single-well trap, this cooling phase is sufficient to complete the state preparation. To arrive at a controlled state of the double well, the potential can be adiabatically transformed from a single well to a double well with a delocalized ground state \cite{kierig2008single,florshaim2024spatial}. Trapping single atoms or ensembles of atoms to generate spatial superpositions is now a routine procedure that has been reported in numerous experiments \cite{parazzoli2012observation,folling2007direct,sebby2006lattice,kierig2008single,kaufman2014two,dai2016generation,endres2016atom,albiez2005direct,florshaim2024spatial}. These two trap geometries, the single and double well, serve as the basis of our quantitative analyses in Section~\ref{comp}.

The next step in the procedure is to turn off the trap at a time $t_i$. This allows the particle to disperse in the $x$ and $y$ directions and to fall under the force of gravity. The screen is then monitored for detection events up until a time $t_f$, where $t_f-t_i$ is a predetermined experimental duration $T$ that is consistent across trials. Each trial ends either with a detection event at time $t_d$ and position $\vb{s}$, recorded as $(t_d-t_i,\vb{s})$, or with no detection by time $t_f$, recorded as $\mathrm{ND}$. Repeating this protocol yields an empirical estimate of the total (non)detection probability and the joint distribution of detection times and positions.

In general, this distribution may depend on the full details of how the experimental procedure is implemented. In the description above, the detection times and positions are defined in terms of the \textit{data} gathered directly by the equipment in the lab. This sidesteps theoretically thorny questions about what can or cannot be ``measured'' or what is or is not ``observable.'' Any theory claiming to be empirically adequate must be able to account for the data generated by this procedure, at least once sufficient details of its implementation in the lab are supplied. The data themselves can be collected into a histogram, and a predicted probability distribution function describing that histogram (perhaps together with guidance about when it applies) is what empirical adequacy requires.

In a given implementation of this experimental scheme with a planar screen locus $\mathbb{S}_L\cong \mathbb{R}^2$, the data yield an observed distribution defined on the space of possible outcomes $\Omega\equiv\{\mathrm{ND}\}\cup\left(\mathbb{R}^2\times[0,T]\right)$. The problem of describing how this joint distribution depends on the experimental procedure is an instance of the screen problem in the special case of a planar screen. If we ignore (or simply do not record) the time component of the data points, the outcome space simplifies to $\{\mathrm{ND}\}\cup \mathbb{R}^2$, so the marginal probabilities of interest are the total (non)detection probability and the spatial distribution of detection events. We will be interested mainly in the distribution along the $y$-axis, so we will typically integrate out the $z$-dependence and focus on the distribution of the arrival position $y$-coordinate. Using the angular coordinate $\theta\in (-\pi/2,\pi/2)$, where $y=L\tan(\theta)$, the outcome space becomes $\Omega_\theta=\{\mathrm{ND}\}\cup(-\pi/2,\pi/2)$. Distributions on this outcome space will be compared in Section~\ref{comp}.

In an ideal implementation of the experiment, the free-falling particle is isolated from everything other than the screen up until the cutoff time $T$. The characteristic time scale of the dispersion of the wave function scales linearly with the particle mass $m$. It follows that the distance the particle falls to reach a given amount of dispersion scales as $m^2$. Consequently, given the height constraints of a fixed apparatus, a larger portion of the angular distribution can be recorded with lighter particles, e.g., single-atom helium \cite{manning2014single,manning2015wheeler}. Given that our primary interest is in the $y$-coordinate marginal, using potentials confining the particle in the $z$-direction as in \cite{gaunt2013bose,ville2017loading} would provide an alternative implementation without the need to consider vertical displacement due to gravity.

The proposals we survey make distinct predictions even about the total probability of detection, so we will not condition on detection in what follows \footnote{Note that the conditional distribution \textit{given} that a detection has occurred is all that is available in some closely related experimental protocols, e.g., the standard double-slit experiment where particles are not prepared one at a time, but are produced at a variable rate by an oven or electrode or with variable count in a condensate or atom array. For the ``trap and release'' protocols discussed here, however, the total number of trials can be known exactly, and, as we will see, provides useful information for distinguishing theoretical proposals.}. We will also typically consider the distributions these proposals predict in the ``long accumulation-time limit,'' i.e., in the limit where $T$ goes to infinity. While this is done for convenience, it is worth bearing in mind that these distributions do, in principle, depend on $T$, and that in practice shorter $T$ cutoffs may be needed because of the empirical challenges described above (see Appendix~\ref{ExperimentalConsiderations}).

As emphasized in the introduction, it is essential that the ``screen'' is \textit{always on} in this procedure. To be relevant to the screen problem, the procedure must involve \textit{waiting} for a detection event to occur, rather than imaging the particle with a pulse over a short interval of time, as is common in many absorption imaging experiments \cite{roati2004atom,gunther2007atom,streed2012absorption,muntinga2013interferometry,bergschneider2018spin,brown2023time}. This distinction separates the \textit{arrival position problem} from the textbook theory of instantaneous position measurements. Although the arrival position marginal is our ultimate target, the relevant experiments are those in which time-resolved data (however coarse-grained) could be gathered in principle, even if they are not gathered in practice.


\section{Predictive models}\label{predictiveModels}

Several distinct theoretical approaches provide predictive models for the screen problem. These include extensions of the standard formalism of quantum observables \cite{Aharonov-Bohm,Kijowski1974,Werner_Screen,grot1996time,Hegerfeldt_2010,hegerfeldt2010manufacturing, baute2000time, leon2000time, baute2002quantum, egusquiza2008standard}, models based on particle trajectories \cite{leavens1993arrival,Leavens_BohmianTOA, ruggenthaler2005times, nitta2008time,Das2019Arrival,kazemi2023detection,Ayatollah2024nonlocal, Das2025DSERevisited, naidon2024inequivalence}, approaches based on non-unitary linear dynamical models of quantum subsystems \cite{werner1987arrival, halliwell2008path, ruschhaupt2004exact, damborenea2002measurement, Dubey2021Quantum,Tumulka2022Detection,Tumulka2022Distribution,tumulka2023absorbing}, and models derived from path integrals \cite{Marchewka1998Feynman,Marchewka2000Path,Marchewka2001Survival,Marchewka2002}. Each of these proposals has the standard form of a quantum-mechanical prediction for a ``prepare-and-measure'' experiment, in that each provides a mapping from initial wave functions $\Psi$ to experimental outcome probabilities.

While these frameworks are typically applied to the arrival time problem, each class contains proposals that address the full screen problem by providing unambiguous joint distributions of both arrival time and position. Any such proposal for a joint distribution $\Pi_{\mathbb{S}}(\vb s,\tau;\Psi)$ for $\vb s\in \mathbb{S}$ provides a solution to the arrival position problem via the arrival position marginal $\overline{\Pi}_{\mathbb{S}}(\vb s;\Psi) = \int_0^\infty d\tau\,\Pi_{\mathbb{S}}(\vb s,\tau;\Psi)$ \footnote{When a finite cutoff time $T$ for the experiment is taken into account in the model, the upper bound of this integral becomes $T$ rather than $\infty$.}. The total detection probability, which we will denote by $\mathbb{P}_{\mathbb{S}}(\mathrm{D};\Psi)$, is given by $\int_{\mathbb{S}}d\vb s\,\overline{\Pi}_\mathbb{S}(\vb s;\Psi)$.

The six proposed solutions to the screen problem we will compare are the semiclassical proposal (SC), the quantum flux proposal (QF), the ``standard'' proposal (SD), the ``complex absorbing potential rule'' (CAP), the ``absorbing boundary rule'' (ABC), and Marchewka and Schuss's path-integral with absorbing boundary proposal (MS). All six proposals considered in this paper are single-particle models: their predictions concern experiments in which a single particle is prepared and released in each trial. The probability distributions are obtained experimentally by repeating this procedure over many trials.

\subsection{Survey of proposed models}\label{sec:modelSurvey}

The semiclassical proposal uses the momentum distribution of the initial wave function to predict screen distributions. In this model, particles are assumed to follow straight-line trajectories starting at the origin with velocities governed by the initial momentum distribution $\widetilde{\Psi}$. This yields the formula

\begin{equation}
\Pi_{L}^{\mathrm{SC}}(y,z,\tau;\Psi)
= \frac{m^{3}L}{\tau^{4}}
\left|
\widetilde{\Psi}\left(\frac{L\,m}{\tau},\frac{y\,m}{\tau},\frac{z\,m}{\tau}\right)
\right|^{2}\,.
\end{equation}

The semiclassical distribution and its variants are standard tools in quantum theory. They are used to model scattering cross sections \cite{NewtonScattering} and time-resolved screen distributions in the ``far-field regime'' \cite{gliserin2016high,wolf2000ion,kurtsiefer1995time,copley1993neutron,kothe2013time}. This regime is typically understood to consist of settings where the particle is effectively free and the detectors are placed far enough from the initial wave function's support that its ``source'' can be treated as a single point. This model is well-corroborated in such settings, and it provides a reasonable model of the $y$-position marginal in our experiments as well, since the only non-trivial potential after the traps are turned off is the gravitational potential, whose gradient is parallel to the screen. It will serve as a baseline in this paper against which other proposals will be compared.

The semiclassical model is only intended to give an approximate solution to the screen problem under certain idealized conditions. It is therefore natural to wonder what it might be an approximation \textit{of}. One prominent answer to this question is based on the quantum probability current, which, for a spin-$0$ particle of mass $m$ described by a wave function $\Psi$, is given by $\vb{j}(\vb x,t;\Psi)=\frac{\hbar}{m}\mathrm{Im}\big(\Psi^*(\vb{x},t)\grad\Psi(\vb x,t)\big)$. When $\vb{n}(\vb{s})$ is the unit normal to the screen locus $\mathbb{S}$ at $\vb{s}$, pointing away from the initial support of the wave function, $\vb{j}(\vb{s},t;\Psi)\cdot\vb{n}(\vb s)$ gives the (signed) probability flux across the screen per unit area~\cite{mott1965theory}. When the quantum flux is positive at the screen locus for all $t>0$ (which will be the case of interest in this paper), the flux can be treated as the probability density of arrival at the point $\vb{s}$ at time $t$. The case where the screen locus $\mathbb{S}_L$ is given by the plane $x=L$ yields the formula  
\begin{equation}
\Pi_{L}^{\mathrm{QF}}(y,z,\tau;\Psi)
= \vb{j}_x(L,y,z,\tau;\Psi)\,,
\end{equation}
where $\vb{j}_x$ denotes the $x$-component of the probability current.

Fixing $\Psi$ and taking the $L\to\infty$ limit, the quantum flux distribution and the semiclassical distribution converge, as an adaptation of the argument in \cite{daumer1997} to flat screens indicates (Appendix~\ref{app:FarField}). For any finite $L$, however, there exist wave functions $\Psi$ and values of $y$, $z$, and $\tau$ such that $\Pi_{L}^{\mathrm{QF}}(y,z,\tau;\Psi)$ is strictly negative, in which case the ``current positivity condition'' (CPC) fails, and the quantum flux model cannot be straightforwardly interpreted as a probability distribution function. Since positivity of the current is a necessary condition for POVM compatibility \cite{vona2013does}, $\Pi_L^{\mathrm{QF}}$ is not POVM-compatible on general domains of initial states. Importantly, however, for the wave functions and values of $L$ described in our experiment, this issue does not arise (see Appendix~\ref{app:CPC}), and $\Pi_{L}^{\mathrm{QF}}$ gives a well-defined probability distribution function. The quantum flux model has been justified from a number of different perspectives \cite{halliwell2009quantum, boonchui2013arrival, leavens1993arrival, damborenea2002measurement,muga1995time,Hannstein2005}, and has been studied in relativistic settings via streamlines of the Dirac current \cite{challinor1997tunnelling} and in settings where the CPC fails using pilot wave theory \cite{PhysRevA.51.2748, ruggenthaler2005times, Das2019Arrival}.

The third proposal we will consider has arisen independently in several settings, and has come to be called the ``standard'' arrival time proposal \cite{egusquiza2008standard}. This proposal originated with Aharonov and Bohm's paper on the time-energy uncertainty relations \cite{Aharonov-Bohm}, and was developed from an axiomatic perspective by Kijowski \cite{Kijowski1974}. It was subsequently extended from the case of free particles to more general situations in recent decades \cite{baute2000time, leon2000time}, and its interpretation, including the role of interference between the positive- and negative-momentum sectors, has been further analyzed \cite{sombillo2016}. The associated screen distribution can be computed using the momentum space wave function $\widetilde{\Psi}(\vb{p},\tau)$ via the formula
\begin{align}\label{SDEquation}
\Pi_{L}^{\mathrm{SD}}(y,z,\tau;\Psi) &= \sum_{\alpha=\pm} \bigg| \int_{\mathbb{R}^2} dp_y dp_z\frac{ e^{\frac{i}{\hbar} (yp_y+zp_z)}}{\sqrt{(2\pi \hbar)^{3} m }} \\ \nonumber
&~~~\times\int_{\mathbb{R}} \!dp_x \, e^{\frac{i}{\hbar} L p_x} \Theta(\alpha p_x) |p_x|^{\frac{1}{2}} \widetilde{\Psi}(\vb{p}, \tau) \bigg|^2\,,
\end{align}
where $\Theta$ is Heaviside's step function. 

The fourth proposal we consider is based on complex absorbing potentials (CAP). This model represents absorption by the screen via a complex potential function $V(\vb{x})=-i\mu(\vb{x})$ parameterized by a nonnegative real-valued function $\mu$ supported where detections are expected to occur. While originally applied in this context by Allcock as a heuristic model ``on the grounds of mathematical expediency'' \cite{allcock1969time}, this family of proposals has since been justified within the framework of continuous quantum measurements or open quantum systems, where it describes the evolution of the system conditioned on the absence of a detection event \cite{halliwell1999arrival, mackrory2010reflection, jacobs2006straightforward} and has also been recovered as a limit of appropriately parameterized repeated measurements \cite{echanobe2008disclosing, halliwell2010relationship}. This proposal can yield detection events anywhere in the support of $\mu$, so to adapt it as a solution to the (planar) screen problem, we can simply integrate out the coordinate $x$ perpendicular to the screen. The dependence of this proposal on the screen position $x=L$ can be captured by shifting the function $\mu$ in the $x$ direction. This yields the following family of proposals:
\begin{equation}\label{eq:CAPDensity}
\Pi_{L}^{\mathrm{CAP}\left(\mu\right)}(y,z,\tau;\Psi) = \frac{2}{\hbar} \int_{\mathbb{R}}\!dx\, \mu(x-L,y,z) |\Psi(\vb{x},\tau)|^2.
\end{equation}
While this is an infinite-dimensional family of proposals parameterized by the function $\mu$, restricting to the class of sigmoid potentials provides a rich finite-dimensional subfamily (see \cite{muga2004complex,jozani2026detection}). We will therefore focus on CAP applied to functions $\mu$ of the form $\mu_w^h(\vb{x})=h\left(1+e^{-\frac{x}{w}}\right)^{-1}$ of height $h$ and width $w$, where the limiting degenerate case $w=0$ yields a step potential.

The next model we will consider is derived from the ``absorbing boundary condition'' proposal (ABC). This proposal is based on a non-unitary linear wave equation first introduced by Werner in \cite{werner1987arrival}, and has seen significant recent development by other authors \cite{Dubey2021Quantum, Tumulka2022Detection, Tumulka2022Distribution, tumulka2023absorbing, tahvildar2024detection, frolov2025detecting}. In this proposal, the initial wave function $\Psi$ evolves according to the Schrödinger equation away from the screen locus, but is modified by a boundary condition of the form $\vb{n}(\vb s)\cdot \grad\Psi(\vb s,t) = \beta \Psi(\vb s,t)$ where $\beta$ is a complex constant satisfying $\mathrm{Im}(\beta)>0$. This boundary condition ensures that the evolution is a contraction, i.e., that the norm of $\Psi$ is non-increasing with time. The associated screen distribution is given by
\begin{equation}\label{eq:ABCDensity}
\Pi_{L}^{\mathrm{ABC}(\beta)}(y,z,\tau;\Psi) = \frac{\hbar}{m}\mathrm{Im}(\beta)|\Psi(L,y,z,\tau)|^2\,.
\end{equation}
While questions about the empirical scope of this proposal remain unsettled \cite{cavendish2025absorbing, frolov2025detecting}, it nevertheless provides a useful representative of the class of non-unitary linear models of the screen problem.

The final proposal we will consider, developed by Marchewka and Schuss \cite{Marchewka2001Survival,Marchewka1998Feynman,Marchewka2000Path,Marchewka2002}, is derived by modeling the screen as an idealized boundary that absorbs Feynman trajectories. This is justified by an extension of standard quantum mechanics the authors call ``Measurable Quantum Mechanics'' \cite{Marchewka2002}. The central feature of this proposal is a ``separation principle,'' which divides the wave function into two parts $\Psi(\vb{x},t)=\Psi_{\mathrm{surv.}}(\vb{x},t)+\Psi_{\mathrm{abs.}}(\vb{x},t)$, where $\Psi_{\mathrm{surv.}}$ is the ``surviving'' part of the wave function, and $\Psi_{\mathrm{abs.}}$ accounts for the ``absorbed'' part of the wave function. The proposal entails that $\big|\vb{n}(\vb s)\cdot \grad \Psi_{\mathrm{surv.}}(\vb s,t)\big|^2$ defines the probability of absorption per unit time \textit{given} that the particle has not yet been absorbed, which results in the following expression for the unconditional screen distribution in our special case:
\begin{align}\label{eq:MSDensity}
&\Pi_{L}^{\mathrm{MS}(\lambda)}(y,z,\tau;\Psi) = \frac{\lambda\hbar}{m\pi} \big|\partial_x\Psi_{\mathrm{surv.}}(L,y,z,\tau)\big|^2 \\\nonumber
&~~~\times \exp \left( -\frac{\lambda\hbar}{m\pi} \int_0^{\tau} \!ds\!\int_{\mathbb{R}^2}\! dy\,dz\,\big| \partial_x\Psi_{\mathrm{surv.}}(L,y,z,s) \big|^2  \right),
\end{align}
where $\lambda$ is a parameter characterizing the absorption strength that is to be determined experimentally. 

As the authors point out, this expression is ``not a quantum-mechanical quantity'' in the sense of the Born rule, as it cannot be expressed as a quadratic form evaluated on the initial wave function  \cite{Marchewka2001Survival}. In particular, this proposal is not POVM-compatible on general domains of initial states.


\subsection{The angular arrival position distribution}\label{sec:angArrival}

To contrast the predictions of these proposals, we focus on the angular density
\begin{align}
\mathcal{A}_L(\theta;\Psi)&=
L\sec^2(\theta)\int_0^\infty\! d\tau\!\int_{\mathbb{R}}\!dz\,\Pi_{L}(L\tan(\theta),z,\tau;\Psi)\,,
\end{align}
which describes the probability distribution function for the angular coordinate $\theta = \arctan(\tfrac{y}{L})$ of the arrival position as viewed from the origin. As we will see, these angular distributions reveal qualitative differences between the different proposals for $\Pi_{L}$ described in the previous section.

These proposals all have useful symmetry-respecting properties that permit us to reduce the analysis to two dimensions in our setting of interest. In our proposed experiments, the gradient of the gravitational potential is oriented parallel to the screen, and the initial wave function $\Psi$ can be written as a separable product state $\Psi(x,y,z)=\psi(x,y)\eta(z)$. Since the only non-trivial potential gradient points along the $z$-direction, the time evolution equations for these proposals maintain their separability. These models also have the property that the resulting detection probabilities inherit the $z$-translational symmetry of the experimental setup once the traps are turned off. In particular, the marginal distribution for the $y$-coordinate of the arrival position depends solely on $\psi(x,y)$, reducing the analysis of $\mathcal{A}_L$ to a two-dimensional problem.

Before computing the quantitative predictions of these proposals, it is important to discuss the choice of units and the role of the mass $m$. For the parameter-free proposals SC, QF, and SD, as well as for ABC and MS, whose parameters have units of inverse length and length, respectively, the predicted angular distributions $\mathcal{A}_L$ are independent of $m$ in the long accumulation-time limit. This can be verified by rewriting the time integral after an appropriate change of variables. These proposals also lead to a function $\mathcal{A}_L$ with no functional dependence on $\hbar$, so we can safely set $\hbar=m=1$ without affecting the empirical predictions. This mass independence is a significant experimental advantage, since it allows these models to be compared using different choices of particle or atom, provided that the same initial wave function can be prepared for particles of different species. This is a flexibility not shared by the arrival time marginal, which is mass-dependent for each of these proposals.

For CAP, however, this mass-invariance property generally fails. In the case of a detector described by a fixed, mass-independent potential with profile $\mu_w^h$, for instance, varying the mass of the incident particle while holding the initial spatial wave function and the potential fixed will alter the predicted angular distribution. Specifically, if the distribution for a particle of mass $m_1$ is described by $\mathcal{A}_L^{\mathrm{CAP}\left(\mu^h_w\right)}(\theta;\psi)$, the corresponding distribution for a particle of mass $m_2$ interacting with the same physical potential is given by $\mathcal{A}_L^{\mathrm{CAP}\left(\mu_w^{h'}\right)}(\theta;\psi)$, where $h'=\frac{m_2}{m_1}h$, with each height measured in units of $\hbar^2/(m\sigma^2)$. Under this assumption, the mass dependence provides a distinctive empirical signature of CAP relative to the other proposals. The assumption of a mass-independent potential is consistent with the detector models of Refs.~\cite{halliwell1999arrival, mackrory2010reflection}, in which the absorption strength does not depend on the particle's mass.


\section{Empirical predictions}\label{comp}

In this section, we compare the predictions of the proposals considered above. We start with numerical computations of $\mathcal{A}_L$ for each proposal, for both the single- and double-well setups. We then show that many of the features that can be seen in the resulting plots are reflections of structural features of these proposals that generalize to other experimental settings.

\subsection{\texorpdfstring{$\mathcal{A}_L$}{A\_L} for the single and double well}\label{sec:singleDouble}

In the case of a single well that is rotationally symmetric about the $z$-axis, the wave function $\Psi$ of the state after the cooling procedure has been carried out can be modeled as a separable three-dimensional Gaussian wave packet of the form \cite{brown2023time,wieman1999atom,leibfried2003quantum},
\begin{equation}
\Psi (x,y,z)\propto g_{\sigma}(x)g_{\sigma}(y)g_{\sigma'}(z)\,,    
\end{equation}
where
${g_{\ell} (x)=\exp\left(-x^2/4 \ell^2\right)}$, $\sigma$ is the initial dispersion in the $x$ and $y$ directions, and $\sigma'$ is the dispersion in the $z$-direction. As discussed above, the angular distribution $\mathcal{A}_L$ only depends on the $x$- and $y$-coordinate dependent factor $\psi_{\sigma}^{\mathrm{s.w.}}(x,y)$ of $\Psi$, which is defined up to normalization by $\psi_{\sigma}^{\mathrm{s.w.}}(x,y)\propto g_{\sigma}(x)g_\sigma(y)$.  In the case of the double well, $\mathcal{A}_L$ depends only on $\psi_{\sigma,d}^{\mathrm{d.w.}}(x,y)$, which is defined up to normalization by
\begin{equation}
\psi_{\sigma,d}^{\mathrm{d.w.}}(x,y)\propto g_{\sigma}(x)\big(g_{\sigma}\left(y-d/2\right)+g_{\sigma}\left(y+d/2\right)\big)\,,
\end{equation}
where $d$ is the separation between the packets.

The predictions of CAP, ABC, and MS depend on phenomenological model parameters. In the absence of data that would fix them, we choose them so as to minimize differences in far-field total detection probability for the single-well state. For MS, we take $\lambda\approx 2.73\,\sigma$, for which the far-field detection probability equals $50\%$, the value predicted by SC, QF, and SD. No choice of $\beta$ allows ABC to reach $50\%$ (see Appendix~\ref{app:ABCbound}), so we take $\beta\approx 0.333\,\tfrac{i}{\sigma}$, which maximizes its far-field detection probability at approximately $41.5\%$. Since the sigmoid family of CAP potentials has two parameters, matching a single detection probability leaves a one-parameter family of potentials. We fix the width at the natural length scale of the initial state, $w=\sigma$, and choose the height $h\approx 0.125$ (in units of $\hbar^2/(m\sigma^2)$) so that the far-field detection probability of CAP equals that of ABC. The same parameters are used for the double-well state, for each value of $L$, and for the rescaled states of Section~\ref{rescalingLimits}, which thus provide further preparations with the parameters held fixed. Because the calibration is to far-field totals, the detection probabilities at $L=10\sigma$ differ slightly from these values. The qualitative features emphasized below, such as the vanishing of the angular density at grazing incidence, hold for every choice of parameters, while Appendix~\ref{app:ParamDep} shows how the quantitative predictions change as the parameters vary.

\begin{figure*}[!htbp]
    \centering
    \begin{tikzpicture}
        \node[anchor=south west, inner sep=0] (image) at (0,0) {\includegraphics[width=\textwidth]{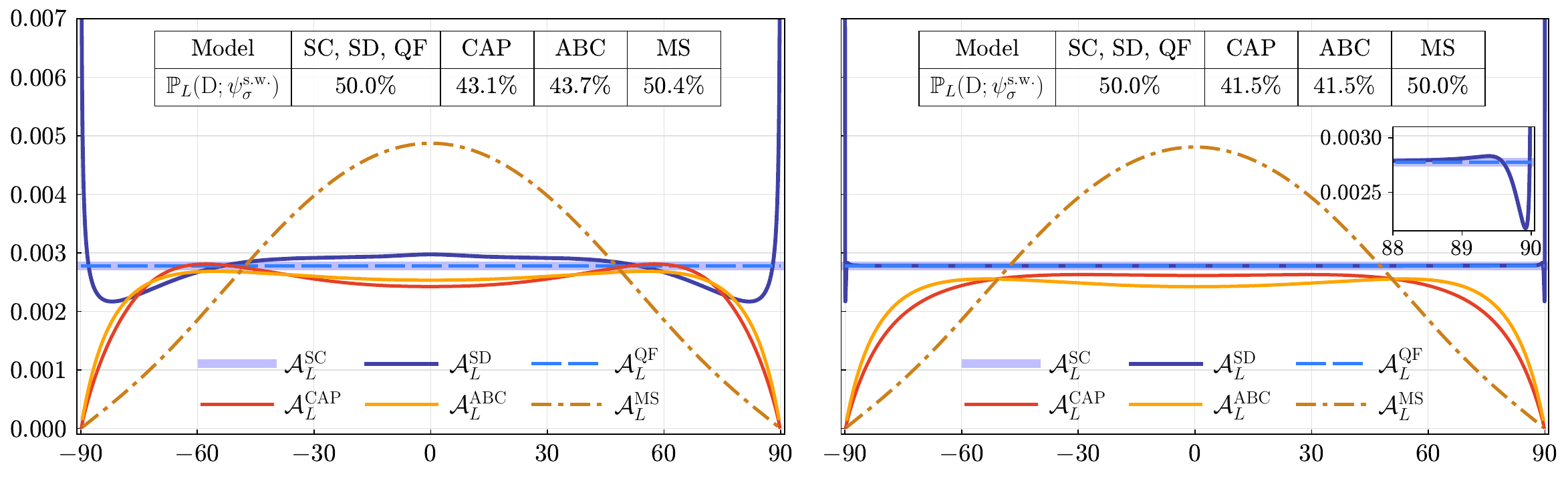}};
    \end{tikzpicture}
    \caption{Comparison of the arrival position distributions $\mathcal{A}_L\!\left(\theta;\psi^{\mathrm{s.w.}}_\sigma\right)$ for the single-well initial state with $L=10\sigma$ (left) and $L=10^3\sigma$ (right). The total detection probabilities $\mathbb{P}_L\!\left(\mathrm{D};\psi^{\mathrm{s.w.}}_\sigma\right)$ are shown in the table. The inset shows the QF and SD distributions over the angular interval from 88 to 90 degrees for $L=10^3\sigma$. For CAP, ABC, and MS, the parameters are those chosen in Section~\ref{sec:singleDouble}: a CAP potential $\mu^{h}_{\sigma}$ with $h\approx0.125$ in units of $\hbar^2/(m\sigma^2)$, $\beta\approx0.333\,\tfrac{i}{\sigma}$, and $\lambda\approx2.73\,\sigma$.}
    \label{OneWell}
\end{figure*}

\begin{figure*}[!htbp]
    \centering
    \begin{tikzpicture}
        \node[anchor=south west, inner sep=0] (image) at (0,0) {\includegraphics[width=\textwidth]{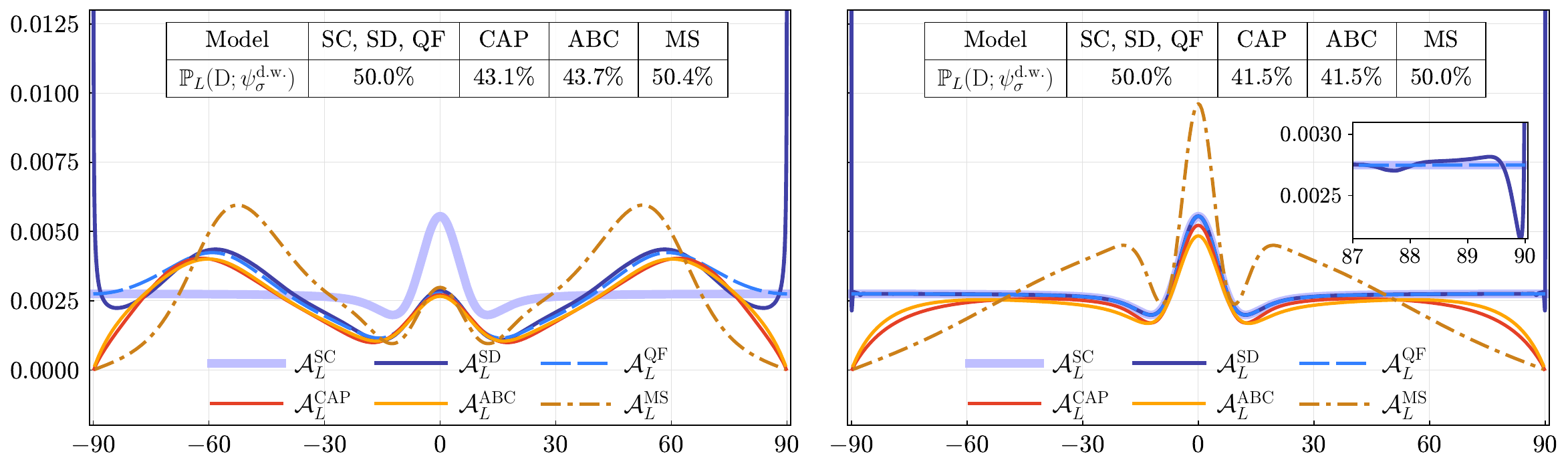}};
    \end{tikzpicture}
    \caption{Comparison of the arrival position distributions $\mathcal{A}_L(\theta;\psi^{\mathrm{d.w.}}_{\sigma,d})$ for the double-well initial state with well separation $d=20\sigma$. As in Fig.~\ref{OneWell}, the panels show results for $L=10\sigma$ (left) and $L=10^3\sigma$ (right), with total detection probabilities $\mathbb{P}_L(\mathrm{D};\psi^{\mathrm{d.w.}}_{\sigma,d})$ listed in the inset tables. All model parameters are identical to those of the single-well case. The inset shows the QF and SD distributions over the angular interval from 87 to 90 degrees for $L=10^3\sigma$.}
    \label{TwoWell}
\end{figure*}

Figs.~\ref{OneWell} and  \ref{TwoWell} show the probability distributions for the $y$-position marginals as a function of the cylindrical coordinate $\theta = \arctan(\tfrac{y}{L})$ for two different values of the ratio $\tfrac{L}{\sigma}$ between the distance to the screen and the initial dispersion. The total detection probability is shown in the table on each plot.

The total detection probabilities $\mathbb{P}_L^{\mathrm{SC}}$, $\mathbb{P}_L^{\mathrm{QF}}$, and $\mathbb{P}_L^{\mathrm{SD}}$ are exactly $50\%$, whereas those for CAP and ABC are lower. For SC, QF, and SD, this follows from the symmetry of the squared amplitude of these wave functions across the plane $x=0$. Semiclassically speaking, there is a $50\%$ chance that the particle's initial momentum is directed away from the screen, and a $50\%$ chance that it is directed toward the screen. For SC, QF, and SD, this is sufficient to determine the total detection probability, as can be derived directly from their definitions. The maps $\theta\mapsto\mathcal{A}_L^{\mathrm{SC}}\left(\theta;\psi_\sigma^{\mathrm{s.w.}}\right)$ and $\theta\mapsto\mathcal{A}_L^{\mathrm{QF}}\left(\theta;\psi_\sigma^{\mathrm{s.w.}}\right)$ are constant. This follows from the definitions and the rotational symmetry of $\psi_\sigma^{\mathrm{s.w.}}$ about the $z$-axis. This flat profile distinguishes the SC and QF predictions from those of the other four proposals.

As discussed in the following sections, several of the features of these plots have robust mathematical explanations, some of which do not depend on the initial wave function at all.


\subsection{Far-field limits}\label{sec:farfield}

In this section, we study the far-field limits $\lim_{L\to\infty}\mathcal{A}_L(\theta;\psi)$, which we denote by $\mathcal{A}_\infty(\theta;\psi)$, and present formulas for $\mathcal{A}_\infty$ for the six proposals surveyed. Since $\mathcal{A}_L^{\mathrm{SC}}$ is independent of $L$, heuristic derivations are needed only for the other five. These arguments, given in Appendix~\ref{app:FarField}, use stationary-phase methods and standard scattering-theoretic arguments, and identify the steps that would require further justification for a fully rigorous treatment. The derivations concern pointwise limits at fixed $\theta$. Convergence of the detection probability for every angular interval, which would follow from convergence in $L^1\bigl((-\tfrac{\pi}{2},\tfrac{\pi}{2}),d\theta\bigr)$, is left to future work. The formulas can also be corroborated by computing $\mathcal{A}_L$ for large $L$, as in the $L=10^3\sigma$ panels of Figs.~\ref{OneWell} and~\ref{TwoWell}, and in Fig.~\ref{fig:FarFieldConvergence} (Appendix~\ref{app:FarField}).

The simplest far-field limit is that for $\mathcal{A}_L^{\mathrm{SC}}$, since $\mathcal{A}_L^{\mathrm{SC}}$ does not depend on $L$ at all. By a change-of-variables argument from the definitions given above, $\mathcal{A}_L^{\textrm{SC}}(\theta;\psi)$ satisfies the identity
\begin{equation}\label{eq:SCFarFieldFormula}
\mathcal{A}_L^{\textrm{SC}}(\theta;\psi)=\sec^2(\theta) \int_0^\infty \!du\,u \big|\widetilde{\psi}(u, u\tan(\theta))\big|^2\,.
\end{equation}
We will therefore suppress the $L$ from our notation and write $\mathcal{A}^{\mathrm{SC}}(\theta;\psi)$ for both the far-field limit and the finite $L$ instances of this function.

The far-field limits for QF and SD exactly recover the semiclassical distribution, i.e., for all $\theta \in \left(-\frac{\pi}{2},\frac{\pi}{2}\right)$,
\begin{equation}\label{standardFF}
\mathcal{A}_\infty^{\mathrm{QF}}(\theta;\psi)= \mathcal{A}_\infty^{\mathrm{SD}}(\theta;\psi)=\mathcal{A}^{\mathrm{SC}}(\theta;\psi)\,.
\end{equation}
This limiting behavior can be seen in the $L=10^3\sigma$ panels of Figs.~\ref{OneWell} and \ref{TwoWell}, where the QF and SD curves become indistinguishable from the SC curve for central angles $\theta$. 

Note that the convergence of $\mathcal{A}_L^{\mathrm{SD}}$ to $\mathcal{A}^{\mathrm{SC}}$ is not uniform. As we will see in the next section, for any sufficiently regular $\psi$, any $L>0$, and any $M>0$, there exists an angle $\theta\in \left(-\frac{\pi}{2},\frac{\pi}{2}\right)$ such that
\begin{equation}
\left|\mathcal{A}_L^{\mathrm{SD}}(\theta;\psi)-\mathcal{A}^{\mathrm{SC}}(\theta;\psi)\right|>M\,.
\end{equation}
This behavior can be seen in the insets to the $L=10^3\sigma$ panels of Figs.~\ref{OneWell} and \ref{TwoWell}, which show that $\mathcal{A}_L^{\mathrm{SD}}$ blows up as the angular coordinate $\theta$ approaches grazing incidence.

As the $L=10^3\sigma$ panels of Figs.~\ref{OneWell} and \ref{TwoWell} suggest, the other three proposals, CAP, ABC, and MS, do not recover the standard semiclassical prediction in the far-field limit. For CAP with a sigmoid step potential, the stationary scattering problem transforms into the hypergeometric differential equation, which yields the limit
\begin{align}\label{CAPFarField}
&\mathcal{A}_\infty^{\mathrm{CAP}\left(\mu_w^h\right)}(\theta;\psi)\\\nonumber
&\qquad=\sec^2(\theta)\int_0^\infty\! du\,u\,\mathcal{T}\!\left(u;\mu_w^h\right)
\big|\widetilde{\psi}\big(u, u\tan(\theta)\big)\big|^2\,,
\end{align}
where 
\begin{equation}\label{eq:CAPTransmission}
\mathcal{T}\!\left(u;\mu_w^h\right)= 1 - \left|\frac{u+k_+}{u-k_+}\right|^2\left| \frac{\Gamma(-iw(u+k_+))}{\Gamma(iw(u-k_+))} \right|^4
\end{equation}
with $k_+^2=u^2+2ih$ and $\mathrm{Im}(k_+)>0$. The case of an absorbing step potential is given by taking the appropriate limit $w\to 0^+$, which yields
\begin{align}\label{StepFarField}
\mathcal{A}_\infty^{\mathrm{CAP}\left(\mu_0^h\right)}&(\theta;\psi)= \sec^2(\theta)\\\nonumber
&\times
\!\int_0^\infty\! du\,u\! \left(1-\left| \frac{u - k_+}{u + k_+}\right|^2\right)\big|\widetilde{\psi}\big(u, u\tan(\theta)\big)\big|^2\,.
\end{align}

Up to analytical difficulties regarding the legitimacy of exchanging limits and integrals, one can see from the above expressions that for all $w\ge 0$,
\begin{equation}\label{eq:CAPweaklim}
\lim_{h\to 0}\mathcal{A}_\infty^{\mathrm{CAP}\left(\mu_w^h\right)}(\theta;\psi)= \mathcal{A}^{\mathrm{SC}}(\theta;\psi)\,.
\end{equation}
This is closely related to a recent result of Kaimal and Tumulka~\cite{kaimal2026scattering}, who show that an imaginary step potential $-i\mu_0$ outside a sphere of radius $R$, the spherical analog of the step case in Eq.~\eqref{StepFarField}, recovers the standard scattering cross section in a joint limit where $R\to\infty$, $\mu_0\to 0$, and $R\mu_0\to\infty$, or when $R\to\infty$ is followed by $\mu_0\to0$. This does not contradict Eq.~\eqref{CAPFarField}, which describes the far-field behavior of a detector with its absorbing potential $\mu$ held fixed. For a step of strength $\mu_0$, a plane wave with normal wave number $k$ is reflected with amplitude $(k-K)/(k+K)$, where $K^2=k^2+2im\mu_0/\hbar^2$ and $\mathrm{Im}(K)>0$ (cf. Section~4 of Ref.~\cite{kaimal2026scattering}). For fixed $k>0$, this vanishes as $\mu_0\to 0$, but for fixed $\mu_0>0$ it approaches $-1$ as $k\to 0$. Weak absorbing potentials can therefore approximately recover standard scattering predictions, but how weak they must be depends on the momentum distribution of the prepared state. For a planar screen with fixed absorption strength, moreover, reflection becomes increasingly important as the angle of incidence approaches grazing, where the normal momentum vanishes \footnote{For the spherical detectors considered in Ref.~\cite{kaimal2026scattering}, outgoing waves meet the detector at normal incidence in the far-field regime, so this effect does not arise.}. Indeed, as we will see below, for fixed detector parameters the angular detection densities predicted by CAP, ABC, and MS vanish in the grazing-angle limit, providing a distinctive signature that separates them from the standard semiclassical distribution \footnote{In keeping with how these proposals are typically formulated, we treat the parameters of each absorbing proposal as properties of a given detector or detection scheme, so that they remain fixed when the same screen is moved relative to the source. On this view, a joint limit in which the absorption strength decreases as the distance increases describes a family of detector models, rather than a fixed detector being moved farther from the source. One could instead allow the effective detector parameters to depend holistically on the experimental setup, including the source--screen separation, but the proposals considered here do not specify such a dependence.}.

In the case of ABC, the far-field limit is given by the expression
\begin{align}\label{eq:ABCFarField}
\mathcal{A}_\infty^{\mathrm{ABC}(\beta)}(\theta;\psi)=&4\,\mathrm{Im}(\beta)\sec^2(\theta)\\\nonumber
&\times
\!\int_0^\infty\! du\, \frac{u^2}{|iu+\beta|^2}\big|\widetilde{\psi}\big(u, u\tan(\theta)\big)\big|^2\,.
\end{align}
From this expression, it follows that if $a$ and $b$ are real numbers with $b>0$, then
\begin{align}
\lim_{b\to 0^+} \mathcal{A}_\infty^{\mathrm{ABC}(a+ib)}(\theta;\psi)
=
\lim_{b\to \infty} \mathcal{A}_\infty^{\mathrm{ABC}(a+ib)}(\theta;\psi)
=
0\,.
\end{align}
In fact, no choice of $\beta$ even reaches the semiclassical total detection probability: for every $\beta$ with ${\mathrm{Im}(\beta)>0}$ and every $\psi$ whose momentum distribution has a right-moving component, ${\mathbb{P}^{\mathrm{ABC}(\beta)}_\infty(\mathrm{D};\psi)<\mathbb{P}^{\mathrm{SC}}_\infty(\mathrm{D};\psi)}$, as shown for one-dimensional scattering in Ref.~\cite{cavendish2025absorbing} and in the present setting in Appendix~\ref{app:ABCbound}.

For MS, the far-field limit is given by
\begin{align}\label{MSFarField}
\mathcal{A}_\infty^{\mathrm{MS}(\lambda)}(\theta&;\psi)=\!\frac{4\lambda}{\pi}\sec^2(\theta)\!\int_0^\infty\! du\,u^2 \\\nonumber
&~~\times \big|\widetilde{\psi}\big(u, u\tan(\theta)\big)\big|^2  e^{-\frac{4\lambda}{\pi}\int_{u}^\infty dv\,v\int_{\mathbb{R}}dq\,\left|\widetilde{\psi}(v,q)\right|^2}\,.
\end{align}

These results divide the proposals into two distinct classes. In the far-field limit (with model parameters held fixed), SC, QF, and SD recover the standard predictions of scattering theory, whereas CAP, ABC, and MS do not.


\subsection{Grazing-angle limits}\label{sec:grazing}
In this section, we consider the behavior of $\mathcal{A}_L$ for fixed $L$ in the grazing-angle limits $\theta \to \pm \pi/2$. This turns out to distinguish all proposals other than QF from the semiclassical prediction.

For SC, the grazing-angle limit is given by
\begin{align}
\lim_{\theta\to \pm\frac{\pi}{2}}&\mathcal{A}_L^{\mathrm{SC}}(\theta;\psi)=\int_0^\infty dp\,p\, \big|\widetilde{\psi}\big(0, \pm p\big)\big|^2\,,
\end{align}
where the limit is understood to be from below for $\frac{\pi}{2}$ and from above for $-\frac{\pi}{2}$. For QF, on the other hand, we have the identity
\begin{align}
\lim_{\theta\to \pm\frac{\pi}{2}}\mathcal{A}_L^{\mathrm{QF}}(\theta;\psi)=\!\int_0^\infty \!dp\,p\left( \big|\widetilde{\psi}\big(0, \pm p\big)\big|^2 \!+\frac{\widetilde{J}_x(0,\pm p)}{L}\right)\,,
\end{align}
where 
\begin{equation}
\widetilde{J}_x(u,v)=\hbar\,\mathrm{Im}\big(\widetilde{\psi}(u,v)^\ast\partial_u\widetilde{\psi}(u,v)\big)\,.
\end{equation}
In the numerical examples above, the wave functions satisfy $\widetilde{\psi}(u,v)=\widetilde{\psi}(-u,v)$, from which it follows that $\widetilde{J}_x(0,\pm p)=0$ for all $p$. This explains the exact agreement of the grazing-angle limit for QF and SC even in the relatively near-field case $L=10\sigma$.

As we saw in the last section, for each fixed angle, the far-field limit of the standard proposal equals the semiclassical distribution. However, at each fixed $L$, we have
\begin{equation}
\lim_{\theta\to \pm\frac{\pi}{2}} \mathcal{A}_L^{\mathrm{SD}}(\theta;\psi)=\infty\,.
\end{equation}

The absorbing proposals also differ from the semiclassical prediction at near-grazing angles, since for all $L>0$ and all choices of parameters $w,h,\beta$, or $\lambda$,
\begin{align}
\lim_{\theta\to \pm\frac{\pi}{2}}& \mathcal{A}_L^{\mathrm{CAP}\left(\mu_w^h\right)}(\theta;\psi)
\\
&=
\lim_{\theta\to \pm\frac{\pi}{2}} \mathcal{A}_L^{\mathrm{ABC}(\beta)}(\theta;\psi)
=
\lim_{\theta\to \pm\frac{\pi}{2}} \mathcal{A}_L^{\mathrm{MS}(\lambda)}(\theta;\psi)=0\,.\nonumber
\end{align}

According to each of these proposals, recording arrivals at near-grazing angles requires sufficiently large cutoff times $T$. The semiclassical heuristic for this is that reaching the screen at a near-grazing angle entails traveling a relatively large distance, which in turn requires a long time of flight. This holds even for radially symmetric initial states. For these states, SC and QF predict that the probability of detection in an interval $[\theta_1,\theta_2]$ depends only on $\theta_2-\theta_1$, so no angle is favored, yet the arrivals near grazing angles are the late ones. Experiments with short cutoff times are therefore limited in their ability to detect this grazing-angle behavior (see Appendix~\ref{ExperimentalConsiderations}).

The grazing-angle limits demonstrate three qualitatively distinct behaviors that can be seen in each panel from Figs.~\ref{OneWell} and \ref{TwoWell}: SC and QF typically predict a finite nonzero probability density, SD diverges to infinity, and the absorbing models (CAP, ABC, MS) converge to zero. This grazing-angle behavior is directly related to the long-time tail behavior of the arrival \emph{time} marginal probability distribution $\rho_L$ defined by 
\begin{equation}
\rho_L(\tau;\Psi)=\int_{\mathbb{R}^2}\!dy\,dz\,\Pi_L(y,z,\tau;\Psi)\,.
\end{equation}
For product states whose arrival-time marginal has a power-law tail $\rho_L(\tau)\sim C_L\tau^{-\gamma}$, the grazing-angle limit vanishes if $\gamma>2$, diverges if $\gamma<2$, and can only be finite and nonzero if $\gamma=2$ (Lemma~\ref{lem:grazingtail} in Appendix~\ref{app:larget}). For the proposals surveyed, the tail exponent
\begin{equation}
\gamma_{\mathrm{tail}}(L;\Psi)=-\lim_{\tau\to \infty}\frac{\log(\rho_L(\tau;\Psi))}{\log(\tau)}
\end{equation}
takes the same value for any generic wave function $\Psi$ and any $L>0$. We can therefore drop the arguments from this function and denote its generic value for proposal X by $\gamma_{\mathrm{tail}}^X$. The grazing-angle behavior seen above is then explained by the values of these tail exponents, which are $\gamma_{\mathrm{tail}}^{\mathrm{SC}}=\gamma_{\mathrm{tail}}^{\mathrm{QF}}=2$, $\gamma_{\mathrm{tail}}^{\mathrm{SD}}=\tfrac{3}{2}$, and
$\gamma_{\mathrm{tail}}^{\mathrm{CAP}\left(\mu_w^h\right)}=\gamma_{\mathrm{tail}}^{\mathrm{ABC}(\beta)}=\gamma_{\mathrm{tail}}^{\mathrm{MS}(\lambda)}=3$
(obtained directly from the explicit formula for SC and heuristically for the other proposals in Appendix~\ref{app:larget}).

The above discussion shows that a proposed solution to the screen problem can only recover the arrival position predictions of standard scattering theory (see, e.g., \cite{NewtonScattering}) at angles sufficiently close to grazing incidence if the tail behavior of the arrival time marginal follows a precise decay law. This provides an answer to the question raised earlier regarding what the semiclassical distribution might be an approximation \emph{of}: if one demands a model that is well-approximated by standard scattering predictions across all angles $\theta$ at finite screen distances $L$, QF is the sole viable candidate. All other proposals depart from standard scattering theory for some angles $\theta$.


\subsection{Behavior under spatial rescaling}\label{rescalingLimits}

This section explores the behavior of each proposal under spatial rescaling of the initial state $\Psi$, revealing structural differences that are amenable to empirical investigation.

To describe the effect of rescaling, given a function $f$ defined on $\mathbb{R}^N\times \mathbb{R}$, where the two factors are understood as the space and time coordinates, and given $\alpha>0$, let $f\cdot\alpha$ be defined by $(f\cdot\alpha)(\vb x,t)=f(\alpha^{-1} \vb{x},\alpha^{-2} t)$. If $f$ is a function of spatial coordinates alone, let $(f\cdot \alpha)(\vb x) = f(\alpha^{-1} \vb x)$. Note that if $\Psi(\vb{x},t)$ is a normalized solution to the Schrödinger equation with (potentially complex, potentially time-dependent) potential $V$, then $\alpha^{-\frac{N}{2}}\Psi\cdot \alpha$ is a normalized solution to the Schrödinger equation with potential $\alpha^{-2}\,V\cdot \alpha$. One consequence of this scaling behavior is that given a single trapped particle in $\mathbb{R}^3$ confined to a trap with ground state $\Psi$, one can construct a trap with a dilated ground state $\alpha^{-\frac{3}{2}}\Psi\cdot\alpha$ by appropriately scaling both the trap geometry and the potential strength. For a harmonic trap, $\alpha^{-2}V\cdot \alpha = \alpha^{-4}V$, so simply weakening or strengthening the potential produces a well with a dilated or contracted version of the original ground state.

Recall that the angular distribution $\mathcal{A}_L$ depends only on the normalized two-dimensional factor $\psi(x,y)$ of a separable state $\Psi(x,y,z)=\psi(x,y)\eta(z)$. Under the three-dimensional dilation $\Psi\mapsto\alpha^{-3/2}\Psi\cdot\alpha$, the two-dimensional factor transforms as $\psi\mapsto\alpha^{-1}{\psi\cdot\alpha}$. Thus, the scaling relations for $\mathcal{A}_L$ below involve the normalized two-dimensional dilation $\alpha^{-1}\psi\cdot\alpha$.

The scaling behavior of the angular distribution $\mathcal{A}_L$ provides another way to distinguish SC, QF, and SD from the other proposals. For these models, the angular arrival position distribution function satisfies
\begin{equation}
\mathcal{A}_{\alpha L}^{\mathrm{SC,QF,SD}}\left(\theta ;\alpha^{-1}\, \psi\cdot \alpha\right)=\mathcal{A}_{L}^{\mathrm{SC,QF,SD}}(\theta ; \psi)\,.
\end{equation}
Physically, this implies that the angular distributions for these models are determined by the ratio between the screen distance $L$ and the spatial scale of the initial state. In the far-field limit $L\to \infty$, we find that these angular distributions are strictly invariant under dilation.

The absorbing models (CAP, ABC, and MS), on the other hand, satisfy non-trivial scaling laws, even in the far-field limit. Specifically, for CAP we have
\begin{equation}\label{CAPScaling}
\mathcal{A}_{\infty}^{\mathrm{CAP}\left(\mu\right)}\left(\theta ;\alpha^{-1}\, \psi\cdot \alpha\right)=\mathcal{A}_{\infty}^{\mathrm{CAP}\left(\alpha^2\mu\cdot \alpha^{-1}\right)}(\theta ; \psi)\,,
\end{equation}
i.e., rescaling the initial state by $\alpha$ for a fixed CAP profile $\mu$ yields the same far-field distribution as fixing the initial state and using the CAP profile $\alpha^2\mu\cdot\alpha^{-1}$. For ABC, dilating $\psi$ by $\alpha$ and using ABC parameter $\beta$ yields the same far-field distribution as fixing $\psi$ and using ABC parameter $\alpha\beta$, and the analogous result for MS entails changing $\lambda$ to $\alpha^{-1}\lambda$. It follows that rescaling the wave function produces far-field angular distributions that appear as though they had been gathered by detector screens with different parameters.

\begin{figure}[ht]
    \centering
    \begin{tikzpicture}

        \node[anchor=south west, inner sep=0] (image) at (0,0) {\includegraphics[trim=-10 25 0 10, width=\columnwidth]{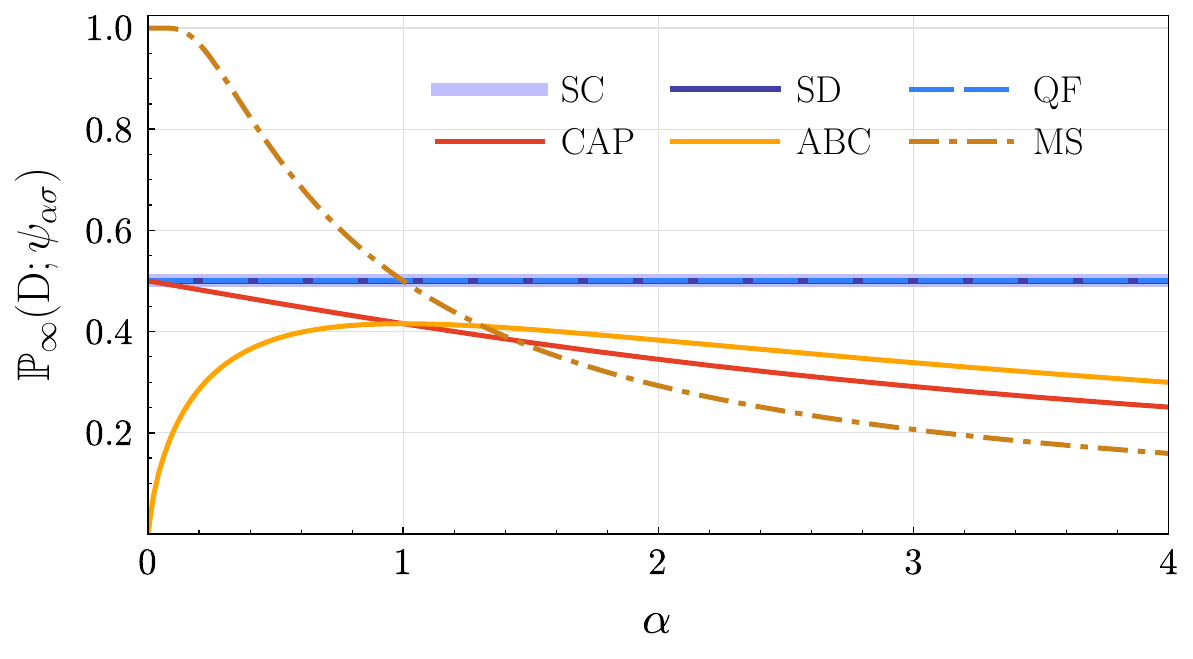}};
            
    \end{tikzpicture}
    \caption{The total detection probabilities for a Gaussian initial state with initial width $\alpha\sigma$ as a function of the rescaling parameter $\alpha$. The detector parameters are fixed across rescalings, at the values calibrated for $\alpha=1$ (see Section~\ref{sec:singleDouble}): $\lambda\approx2.73\,\sigma$, $\beta\approx0.333\,\tfrac{i}{\sigma}$, and a CAP potential $\mu^{h}_{\sigma}$ with $h\approx0.125$ in units of $\hbar^2/(m\sigma^2)$.}
    \label{ScalingBehavior}
\end{figure}

To illustrate the empirical consequences of these different scaling laws, consider a single harmonic well and a detector screen in the far field. As previously established, the ground state of this system is a Gaussian wave packet, and tuning the strength $V_0$ of the trap potential controls the width parameter $\sigma$. This yields a one-parameter family of initial Gaussian states $\psi_{\alpha\sigma}$. The total detection probability for these initial states according to each proposal is given in Fig.~\ref{ScalingBehavior}. For each of the absorbing proposals (CAP, ABC, and MS), $\mathbb{P}_\infty(\mathrm{D};\psi_{\alpha\sigma})$ decays like $\alpha^{-1}$ for large $\alpha$. By the scaling relations above, dilating the state is equivalent to changing the effective detector parameters, and for large $\alpha$ the resulting detection probability at normal momentum $u$ becomes proportional to $\tfrac{u}{\alpha}$. For ABC with $\beta=ib$, the integrand of Eq.~\eqref{eq:ABCFarField} carries the factor $4bu/(u+b)^2$, which approaches $\tfrac{4u}{b}$ with $b\propto\alpha$. For MS, the effective parameter is $\tfrac{\lambda}{\alpha}$, so the exponential in Eq.~\eqref{MSFarField} tends to $1$ and the prefactor decays like $\alpha^{-1}$. For CAP, expanding Eq.~\eqref{eq:CAPTransmission} with $w\to\tfrac{w}{\alpha}$ and $h\to\alpha^2h$ gives $\mathcal T\propto\tfrac{u}{\alpha}$.

For the examples considered, Fig.~\ref{ScalingBehavior} suggests that this asymptotic behavior sets in at relatively small values of $\alpha$. It follows that doubling the spatial scale of the initial state cuts the detection probability in half as soon as this asymptotic behavior provides a good approximation. This is in contrast to SC, QF, and SD, for which the predicted total detection probability remains invariant with respect to the potential strength $V_0$. Since rescaling the initial state is equivalent to rescaling the ABC parameter, the maximum of the ABC curve in Fig.~\ref{ScalingBehavior}, attained at $\alpha=1$, is the upper bound over all $\beta$ derived in Appendix~\ref{app:ABCbound}.

The scaling behavior as $\alpha$ goes to $0$ or $\infty$ is also revealing. For the family CAP$(\mu_w^h)$, Equation~\eqref{CAPScaling} implies that for all $\alpha>0$,
\begin{align}
\mathcal{A}_\infty^\mathrm{CAP\left(\mu_w^h\right)}&\left(\theta; \alpha^{-1}\psi\cdot\alpha\right) =\mathcal{A}_{\infty}^{\mathrm{CAP}\left(\mu_{w/\alpha}^{\alpha^{2}h}\right)}(\theta;\psi)\,.
\end{align}
When $\alpha\to \infty$, the effective potential function $\mu$ becomes increasingly sharp and tall, so the ``quantum Zeno effect'' is expected to result in strong reflections. This is corroborated by Equation~\eqref{CAPFarField} above, which shows that
\begin{equation}
\lim_{\alpha \to \infty} \mathcal{A}_\infty^\mathrm{CAP\left(\mu_w^h\right)}(\theta; \alpha^{-1}\psi\cdot\alpha)= 0\,.
\end{equation}
This shows that sufficiently dilated wave functions are unlikely to be detected by any detector modeled by CAP$(\mu_w^h)$. On the other hand, since the transmission coefficient $\mathcal{T}\!\left(u;\mu_{w/\alpha}^{\alpha^2 h}\right)$ given in Equation~\eqref{eq:CAPTransmission} converges to $1$ as $\alpha$ goes to $0$, the semiclassical proposal is recovered in the tight-confinement limit, i.e.,
\begin{equation}
\lim_{\alpha \to 0} \mathcal{A}_\infty^\mathrm{CAP\left(\mu_w^h\right)}(\theta; \alpha^{-1}\psi\cdot\alpha)=\mathcal{A}^{\mathrm{SC}}(\theta;\psi)\,.
\end{equation}

Applying the ABC scaling relation ($\beta\to\alpha\beta$) and using the fact that the probability of detection goes to zero as $\mathrm{Im}(\beta)$ goes to $0$ or $\infty$, we have
\begin{align}
\lim_{\alpha \to 0^+}& \mathcal{A}_\infty^{\mathrm{ABC}(\beta)}\left(\theta; \alpha^{-1}\psi\cdot\alpha\right)\\
&=\lim_{\alpha\to \infty} \mathcal{A}_\infty^{\mathrm{ABC}(\beta)}\left(\theta;\alpha^{-1}\psi\cdot\alpha\right)=0\,.\nonumber
\end{align}
This implies that, according to ABC, a sufficiently tightly or loosely confined rescaling of $\psi$ is unlikely to be detected.

For MS, a similar argument shows the effective absorption strength scales as $\alpha^{-1}\lambda$, which implies
\begin{equation}
\lim_{\alpha \to \infty} \mathcal{A}_\infty^{\mathrm{MS}(\lambda)}(\theta; \alpha^{-1}\psi\cdot\alpha)=0\,.
\end{equation}
To understand the tight-confinement limit $\alpha \to 0$, note that integrating the far-field distribution $\mathcal{A}_\infty^{\mathrm{MS}(\lambda)}$ from Equation~\eqref{MSFarField} over $\theta$ yields the total detection probability
\begin{align}
\mathbb{P}^{\mathrm{MS}(\lambda)}_\infty(\mathrm{D};\psi)=1-e^{-\frac{4\lambda}{\pi}\int_0^\infty\! du\,u\int_{\mathbb{R}}\!dq\,\left|\widetilde{\psi}(u,q)\right|^2}\,.
\end{align}
Since the effective parameter scales as $\alpha^{-1}\lambda$ in the far field, taking $\alpha \to 0$ is equivalent to taking the limit of this expression as the parameter $\lambda$ goes to infinity. Provided the support of $\widetilde{\psi}$ intersects the half-space $p_x>0$, we get
\begin{equation}
\lim_{\alpha \to 0} \mathbb{P}_\infty^{\mathrm{MS}(\lambda)}(\mathrm{D}; \alpha^{-1}\psi\cdot\alpha)=1\,.
\end{equation}
This implies that for any fixed parameter $\lambda$, a tightly confined rescaling of $\psi$ is almost certain to be detected. In particular, this prediction holds for radially symmetric initial states, where the semiclassical prediction gives a total detection probability of $\frac{1}{2}$. As discussed above, the semiclassical reasoning behind this value is that half of the momentum distribution of the initial wave function corresponds to momentum vectors that point away from the screen, so there is a $50\%$ chance that the particle ``travels away from the screen.'' The MS proposal therefore predicts that even the part of the wave packet $\psi$ that heads in the wrong direction will be detected when the initial confinement of the particle is sufficiently tight.


\section{Discussion}

The results presented above show that the screen problem is not merely an interpretational issue related to the  ``observability'' of time, but a concrete empirical question with signatures that can be resolved using tabletop laboratory equipment. The strong contrast between the proposals surveyed for both single-well and double-well potential geometries demonstrates that these models can be distinguished in a range of experimental settings.

Appendix~\ref{ExperimentalConsiderations} examines the proposed experiment under practical constraints: a finite cutoff time $T$, a detector of finite extent, and ${}^4\mathrm{He}$ atoms in traps with $\sigma=1\,\mu\mathrm{m}$. With a light-sheet detector at $L=100\sigma$ and $T=100\,\mathrm{ms}$, a test on the total detection probability distinguishes CAP, ABC, and MS from SC in about $1600$ runs.

Since the distributions $\mathcal{A}_L(\theta)$ for SC, QF, and SD all converge pointwise to the same distribution in the far-field limit, the distinctions between these proposals are most visible for small values of $\tfrac{L}{\sigma}$. At $L=10\sigma$, with a spatial resolution of about $1\,\mu\mathrm{m}$, of order $1000$ runs suffice to distinguish QF or SD from SC, while distinguishing SD from QF requires more than $10^4$. Larger traps relax the resolution requirement at the cost of longer cutoff times. Non-radially symmetric traps like the double well may also help, since interference features can create additional intervals of contrast (e.g., between QF and SD at around $87.5^\circ$ in the $L=10^3\sigma$ panel of Fig.~\ref{TwoWell}). These estimates indicate the scale of the required experiments. A detailed feasibility study is left to future work.

Shifting the focus from arrival time to arrival position has several advantages. Whereas the arrival time problem is studied through the temporal statistics of detection, the arrival position marginal requires no temporal data at all. The mass independence of the arrival position distribution for most proposals is also a practical advantage, granting experimentalists the flexibility to select whichever particle species best suits the capabilities of their laboratory without affecting the angular distribution $\mathcal{A}_L(\theta)$. As mentioned above, this also provides a useful experimental strategy for distinguishing CAP, which among the proposals surveyed is the only one that is mass-dependent, from the others \footnote{This assumes that the parameters $\mu$, $\beta$, and $\lambda$ are fixed properties of the detector, which may not be the case, since the proposals themselves do not specify the physical origin of these phenomenological parameters and researchers may have different approaches to deriving them. In some implementations, such as the laser-based detection schemes discussed in Ref.~\cite{mackrory2010reflection}, the effective detector parameters may depend on the internal structure of the particle, in which case they would need to be determined separately for each species. Other detector models may also yield parameters that depend on the particle's mass.}.

As Figs.~\ref{OneWell} and \ref{TwoWell} show, the angular distributions for ABC, CAP, and MS exhibit distinctive qualitative features that persist even in the far-field limit, making their theoretical predictions relatively straightforward to distinguish from those of standard semiclassical scattering theory. The vanishing of $\mathcal{A}_L(\theta;\Psi)$ as $\theta$ approaches $90^\circ$, for instance, holds for generic $\Psi$ and for all parameters of CAP, ABC, and MS (cf.\ Section~\ref{sec:grazing} and Appendix~\ref{app:larget}), and thus serves as a robust theoretical signature of the non-semiclassical behavior of these proposals. Suppression of detections for angles near $90^\circ$ is already visible in Figs.~\ref{OneWell} and~\ref{TwoWell} at around $80^\circ$, and can be explored experimentally even in settings where precise control over the initial state of the particle cannot be achieved. Finally, the invariance of the far-field distributions for SC, QF, and SD under rescaling distinguishes standard semiclassical behavior from the behavior of CAP, ABC, and MS. Taken together, these results suggest that the predicted deviations from semiclassical behavior present in CAP, ABC, and MS should be empirically detectable in a number of ways.

The proposals considered here do not exhaust the proposed solutions to the screen problem in the literature (see, e.g., \cite{naidon2024inequivalence,ayatollah2023can}). It will be of interest to study the predictions of these other models as well, provided that they yield unambiguous predictions for the experiment in question. As we have seen, a useful tool for determining whether a given proposal agrees with the predictions of standard scattering theory is to compute the decay rate $\gamma_{\mathrm{tail}}$ of the arrival-time marginal. For product states, and for proposals whose joint arrival density factorizes appropriately (see Lemma~\ref{lem:grazingtail} in Appendix~\ref{app:larget}), if $\gamma_{\mathrm{tail}}$ is well-defined and takes any value other than $2$, agreement with standard scattering theory at angles sufficiently close to grazing incidence is impossible.

It should be noted that none of the proposals surveyed has been derived from a detailed model of a particular physical detection screen (though some, including CAP, have been justified by analyses of idealized detector models \cite{halliwell1999arrival, mackrory2010reflection}). In all six proposals, the physical composition of the screen does not enter explicitly: the screen is represented by its location in space ($x=L$) in the case of SC, QF, and SD, and by its location together with phenomenological parameters like $\mu$, $\beta$, and $\lambda$ in the case of CAP, ABC, and MS, respectively. It is likely that the precise quantitative details of the arrival position distribution often depend on the physical structure of the screen, e.g., whether it is implemented as a laser sheet, a microchannel plate, or a scintillating screen. While SC has been validated in a wide range of far-field settings, empirical data are scarce outside of this regime, and strongly detector-dependent cross-sections may well be the norm. However, this hypothesis is itself testable. As pointed out by Maudlin~\cite{Maudlin2025ActualPhysics}, if detector details are essential to making these empirical predictions, the recorded data will change significantly between two instances of the experiment that use identically placed screens with different physical compositions. As the results above show, if CAP, ABC, or MS are to yield correct predictions, this dependence on the screen's physical composition must persist even in the far field.

As the above discussion shows, the arrival position problem remains unsolved, despite its apparent simplicity. Although further theoretical work is certainly needed, there is no substitute for empirical data spanning a broad spectrum of experimental implementations. Even in the far-field regime, where semiclassical reasoning has been successful, several of these proposals predict measurable departures from standard scattering theory. A variety of feasible experiments can thus distinguish these proposals and begin to close this surprising gap in our current understanding of quantum-mechanical phenomena. Until such data is available and we have the theoretical resources to accurately describe it, the empirical adequacy of our best quantum theories should not be taken for granted.

\smallskip
\noindent\textbf{Acknowledgements.} We thank Herman Batelaan, Siddhant Das, MohammadJavad Kazemi, and Tim Maudlin for helpful comments and discussions.

\smallskip
\noindent\textbf{Code availability.}
Python code reproducing the figures and corroborating the far-field formulas is
available at \url{https://github.com/wpcavendish/arrival-position-problem}.

\appendix

\section*{Appendix}

The mathematical results in the appendices are of two different kinds, which we distinguish typographically. Statements labeled ``Lemma'' are followed by a proof, marked at its end by $\square$. Statements labeled ``Claim'' are followed by a heuristic derivation, marked at its end by $\lozenge$. A heuristic derivation identifies the mechanism responsible for the stated result, but leaves one or more steps unjustified, typically the interchange of a limit with an integral or the use of an asymptotic form outside the regime in which it has been established. Each heuristic derivation is followed by a short discussion of what would be needed to make it rigorous. Some lemmas take a formula obtained heuristically as their starting point. In such cases, the proof establishes the stated result conditional on that formula, and we indicate this explicitly.

The claims should be read as conjectures. Making them rigorous and determining their precise scope (e.g., the regularity conditions required and the strongest form of convergence that holds) remain open. We hope the derivations given here will serve as a starting point for future work. 

\section{Proof of current positivity}\label{app:CPC}

In this appendix, we show that the current positivity condition (CPC) holds for the wave functions discussed in Section~\ref{comp}. Note that the CPC only requires nonnegativity of the current, not strict positivity. As in Section~\ref{comp}, let $g_\sigma$ denote a one-dimensional Gaussian wave packet, here normalized (this differs from the convention in Section~\ref{comp} only by an overall constant). Its free evolution is given by
\begin{equation}\label{eq:gaussEq}
g_\sigma(x,t)
=
\frac{\exp\left(
-\frac{x^2}{4\sigma^2(1+i\hbar t/(2m\sigma^2))}
\right)}{(2\pi\sigma^2)^{1/4}\sqrt{1+i\hbar t/(2m\sigma^2)}}\,.
\end{equation}
Let $H_0$ denote the free Hamiltonian in three dimensions. For a real-valued potential $V(y,z)$, define the transverse Hamiltonian
\begin{equation*}
H_\perp=-\frac{\hbar^2}{2m}(\partial_y^2+\partial_z^2)
+V(y,z).
\end{equation*}
We assume that $H_\perp$ is self-adjoint and that the associated unitary time-evolution operators $e^{-\frac{it}{\hbar}H_\perp}$ preserve $\mathcal{S}(\mathbb{R}^2)$. For $\Phi\in\mathcal{S}(\mathbb{R}^2)$, let
\begin{equation*}
\Phi(y,z,t)
=\left(e^{-\frac{it}{\hbar}H_\perp}\Phi\right)(y,z).
\end{equation*}
Since $H=H_0+V(y,z)$ separates into longitudinal and transverse Hamiltonians, an initial wave function $\Psi(x,y,z)=g_\sigma(x)\Phi(y,z)$ stays separable under time evolution, yielding
\begin{equation*}
\Psi(x,y,z,t)=g_\sigma(x,t)\,\Phi(y,z,t)
\end{equation*}
for all $t$.

\begin{lemma}\label{lem:curpos} Let $\sigma>0$, $\Phi\in\mathcal S(\mathbb R^2)$, and $\Psi(x,y,z)=g_\sigma(x)\Phi(y,z)$, and let $H=H_0+V(y,z)$ with $V$ as above. Then $\mathbf{j}_x(L,y,z,\tau;\Psi)\ge0$ for all $L>0$, $\tau>0$, and $(y,z)\in\mathbb R^2$.
\end{lemma}

\begin{proof} Using Equation~\eqref{eq:gaussEq}, we have 
\begin{align*}
\mathbf{j}_x(L,y,z,\tau;\Psi)&=\mathrm{Im}\left(\frac{\hbar}{m}\Psi^\ast(L,y,z,\tau)\,\partial_x\Psi(L,y,z,\tau)\right)\\
&=\frac{\hbar}{m}\mathrm{Im}\left(g_\sigma(L,\tau)^\ast \partial_xg_\sigma(L,\tau)\right)\left|\Phi(y,z,\tau)\right|^2\\
&= \frac{\hbar}{m} \mathrm{Im}\left(\frac{-2L\left|g_\sigma(L,\tau)\right|^2}{4\sigma^2\left(1+\frac{i\hbar \tau}{2m\sigma^2}\right)}\right)\left|\Phi(y,z,\tau)\right|^2\\
&=\frac{L\hbar\,|\Psi(L,y,z,\tau)|^2}{2m\sigma^2\left(1+\frac{\hbar^2 \tau^2}{4m^2\sigma^4}\right)}\mathrm{Im}\left(\frac{i\hbar \tau}{2m\sigma^2}-1\right)\\
&= L\tau\hbar^2 \,\frac{|\Psi(L,y,z,\tau)|^2}{4m^2\sigma^4+\hbar^2 \tau^2}\,.
\end{align*}
Thus, when $L>0$ and $\tau>0$, the current is nonnegative.
\end{proof}
For the single- and double-well initial states considered in Section~\ref{comp}, the three-dimensional wave functions are proportional to $g_\sigma(x)g_\sigma(y)g_{\sigma'}(z)$
and 
\begin{equation*}
g_\sigma(x)[g_\sigma(y-d/2)+g_\sigma(y+d/2)] g_{\sigma'}(z)\,,
\end{equation*}
respectively. In both cases, the transverse factor belongs to $\mathcal{S}(\mathbb{R}^2)$. Since the gravitational potential is of the form $V(y,z)=cz$, the transverse evolution consists of free evolution in $y$ and evolution under a constant force in $z$. The latter is equivalent to free evolution followed by a spatial translation and multiplication by a phase linear in $z$ (up to a constant phase), all of which preserve the Schwartz class. Thus, the assumptions of Lemma~\ref{lem:curpos} are satisfied, and the CPC holds for both initial states.

\section{Heuristic derivations of far-field formulas}\label{app:FarField}

This appendix gives heuristic derivations of the far-field formulas stated in Section~\ref{sec:farfield}. Throughout this appendix, all initial wave functions are assumed
to be normalized in their respective $L^2$ spaces. For each proposal $X$, each $\theta\in (-\pi/2,\pi/2)$, and each initial state $\Psi$ we write
\begin{equation*}
\mathcal{A}_\infty^X(\theta;\Psi)
=\lim_{L\to\infty}\mathcal{A}_L^X(\theta;\Psi)
\end{equation*}
whenever the limit exists. The semiclassical angular density of Eq.~\eqref{eq:SCFarFieldFormula} is independent of $L$, and we denote it by $\mathcal{A}^{\mathrm{SC}}(\theta;\Psi)$.

In the examples studied here, the three-dimensional initial state $\Psi$ factorizes as
$\Psi(x,y,z)=\psi(x,y)\,\eta(z)$, with both factors normalized. The dynamics preserve this factorization, so integrating over $z$ leaves an angular density that depends only on $\psi$, which we will denote $\mathcal{A}_L^X(\theta;\psi)$. In this appendix, we therefore work in two dimensions. We will set $\hbar=m=1$, and will write $\psi_t=e^{-iH_0t}\psi$, with $H_0=-\Delta/2$, for the free evolution in the $(x,y)$-plane. Our Fourier transform convention is $\widetilde\psi(\mathbf{p})=\frac{1}{2\pi} \int_{\mathbb{R}^2}d^2\mathbf{x}\,e^{-i\mathbf{p}\cdot\mathbf{x}}\psi(\mathbf{x})$.

At the screen point $(L,L\tan(\theta))$ and time $t=\tfrac{L}{u}$, the incident propagation phase has its stationary point at $\mathbf{p}=(u,u\tan(\theta))$. We write $\simeq$ for equality of leading terms as $t\to\infty$ with $\tfrac{\mathbf{x}}{t}$ fixed. At the screen point with $t=\tfrac{L}{u}$, this is the limit $L\to\infty$ at fixed $u$ and $\theta$. The
arguments below use the resulting large-$L$ asymptotics at fixed $u>0$ and then integrate over $u$ to obtain the angular density. In each case we indicate what estimates are needed to justify this passage and any approximations specific to the proposal.

\subsection{Far-field heuristics for QF}

\begin{claim}
For $\psi\in\mathcal S(\mathbb{R}^2)$ and fixed $\theta\in (-\tfrac{\pi}{2},\tfrac{\pi}{2})$,
\begin{equation*}
\mathcal{A}_\infty^{\mathrm{QF}}(\theta;\psi)=\mathcal{A}^{\mathrm{SC}}(\theta;\psi).
\end{equation*}
\end{claim}

\begin{heuristicderivation} For large $t$ with $\mathbf{v}=\tfrac{\mathbf{x}}{t}$ fixed, expanding the phase in the free propagator gives
\begin{equation*}
\psi_t(t\mathbf{v})=\frac{e^{it|\mathbf{v}|^2/2}}{it}\left[\widetilde\psi(\mathbf{v})+O(t^{-1})\right].
\end{equation*}
It follows that $\nabla\psi_t(\mathbf{x})\simeq i\frac{\mathbf{x}}{t}\,\psi_t(\mathbf{x})$, and so
\begin{equation*}
\mathbf{j}(\mathbf{x},t;\psi)=\mathrm{Im}\big(\psi_t^\ast\,\nabla\psi_t\big)
\simeq \frac{\mathbf{x}}{t^3}\,\big|\widetilde\psi(\tfrac{\mathbf{x}}{t})\big|^2 .
\end{equation*}
The QF angular density is
\begin{equation*}
\mathcal{A}_L^{\mathrm{QF}}(\theta;\psi)=L\sec^2(\theta)\int_0^\infty\!dt\, \mathbf{j}_x(L,L\tan(\theta),t;\psi)\,.
\end{equation*}
Substituting $u=\tfrac{L}{t}$ gives the exact expression
\begin{equation*}
\mathcal{A}_L^{\mathrm{QF}}(\theta;\psi)=\sec^2(\theta)\int_0^\infty du\,\frac{L^2}{u^2}\,\mathbf{j}_x\!\left(L,L\tan(\theta),\tfrac{L}{u};\psi\right).
\end{equation*}
At each fixed $u>0$, the limit $L\to\infty$ corresponds to $t=\tfrac{L}{u}\to\infty$ with $\tfrac{\mathbf{x}}{t}=(u,u\tan(\theta))$ fixed, so the leading-order term of the current shows that the integrand converges to $u\,|\widetilde\psi(u,u\tan(\theta))|^2$. Passing the limit through the integral yields
\begin{align*}
\lim_{L\to\infty}\mathcal{A}_L^{\mathrm{QF}}(\theta;\psi)
&=\sec^2(\theta)\int_0^\infty\!du\, u\,\big|\widetilde\psi(u,u\tan(\theta))\big|^2\\
&=\mathcal{A}^{\mathrm{SC}}(\theta;\psi).\qedhere
\end{align*}
\end{heuristicderivation}

To make this heuristic argument rigorous, one must justify interchanging the limit and the integral. While pointwise convergence holds for any fixed $u>0$, one still needs to bound the asymptotic regions $u\to 0$ and $u\to \infty$ (corresponding to very late and very early times, respectively). The analogous integrated-flux result for spherical detection surfaces, the free flux-across-surfaces theorem, is proven for Schwartz-class initial states in Ref.~\cite{daumer1996flux}.

\subsection{Far-field heuristics for SD}

\begin{claim}
For $\psi\in\mathcal S(\mathbb{R}^2)$ and fixed $\theta\in (-\tfrac{\pi}{2},\tfrac{\pi}{2})$,
$\mathcal{A}_\infty^{\mathrm{SD}}(\theta;\psi)=\mathcal{A}^{\mathrm{SC}}(\theta;\psi)$.
\end{claim}

\begin{heuristicderivation}
For $\epsilon=\pm1$, define
\begin{equation*}
F_\epsilon(\mathbf{x},t)
=\frac{1}{2\pi}\int_{\epsilon p_x>0} d^2\mathbf{p}\,
\sqrt{|p_x|}\,
e^{i\mathbf{p}\cdot\mathbf{x}-it|\mathbf{p}|^2/2}\,
\widetilde\psi(\mathbf{p}).
\end{equation*}
Writing $\mathbf{x}_\theta=(L,L\tan(\theta))$, the SD angular density is
\begin{equation*}
\mathcal{A}_L^{\mathrm{SD}}(\theta;\psi)
=L\sec^2(\theta)\int_0^\infty\!dt\!
\sum_{\epsilon=\pm1}|F_\epsilon(\mathbf{x}_\theta,t)|^2\, .
\end{equation*}
Substituting $u=\tfrac{L}{t}$ gives the exact expression
\begin{equation*}
\mathcal{A}_L^{\mathrm{SD}}(\theta;\psi)
=\sec^2(\theta)\int_0^\infty du\,\frac{L^2}{u^2}
\sum_{\epsilon=\pm1}\big|F_\epsilon\big(\mathbf{x}_\theta,\tfrac{L}{u}\big)\big|^2 .
\end{equation*}
At fixed $u>0$, the phase $\mathbf{p}\cdot\mathbf{x}_\theta-t|\mathbf{p}|^2/2$ with $t=\tfrac{L}{u}$ has a single stationary point, $\mathbf{p}_0=\tfrac{\mathbf{x}_\theta}{t}=(u,u\tan(\theta))$, which lies in the positive-momentum sector. Stationary phase gives
\begin{equation*}
F_+\big(\mathbf{x}_\theta,\tfrac{L}{u}\big)
\simeq \frac{u^{3/2}}{iL}\,
e^{i\frac{Lu}{2}\sec^2(\theta)}\,
\widetilde\psi(u,u\tan(\theta)).
\end{equation*}
The negative-momentum sector contains no stationary point and contributes only through the endpoint $p_x=0$ (where the weight $\sqrt{|p_x|}$ also vanishes), so its contribution is of lower order. Hence, at each fixed $u>0$, the integrand converges to $u\,|\widetilde\psi(u,u\tan(\theta))|^2$, and passing the limit through the integral yields
\begin{align*}
\mathcal{A}_\infty^{\mathrm{SD}}(\theta;\psi)
&=\sec^2(\theta)\int_0^\infty\!du\, u\,\big|\widetilde\psi(u,u\tan(\theta))\big|^2\\
&=\mathcal{A}^{\mathrm{SC}}(\theta;\psi). \qedhere
\end{align*}
\end{heuristicderivation}

The step requiring justification is this interchange of limit and integral, which requires control of the regions $u\to0$ and $u\to\infty$ and of the endpoint contributions at $p_x=0$ in both momentum sectors.

\subsection{Far-field heuristics for ABC}

\begin{claim}
For a fixed value of $\beta$ with ${b=\mathrm{Im}\beta>0}$, fixed ${\theta\in (-\tfrac{\pi}{2},\tfrac{\pi}{2})}$, and $\psi \in \mathcal{S}(\mathbb{R}^2)$,
\begin{equation*}
\mathcal{A}_\infty^{\mathrm{ABC}(\beta)}(\theta;\psi)
=4b\sec^2(\theta)\!\int_0^\infty\!
\frac{du\,u^2}{|iu+\beta|^2}
\big|\widetilde\psi(u,u\tan(\theta))\big|^2.
\end{equation*}
\end{claim}

\begin{heuristicderivation}

Let $\psi_t^{(L)}$ denote the solution of the Schr\"odinger equation on the half-plane $x<L$ with initial condition $\psi$ and boundary condition $\partial_x\psi_t^{(L)}=\beta\psi_t^{(L)}$ at $x=L$. By Eq.~\eqref{eq:ABCDensity}, the ABC detection density on the screen is $b\,|\psi_t^{(L)}|^2$, so that, after substituting $u=\tfrac{L}{t}$,
\begin{align*}
\mathcal{A}_L^{\mathrm{ABC}(\beta)}(\theta;\psi)
&=Lb\sec^2(\theta)\int_0^\infty\!dt\,\big|\psi_t^{(L)}(L,L\tan(\theta))\big|^2\\
&=b\sec^2(\theta)\int_0^\infty \!du\,\frac{L^2}{u^2}\,
\big|\psi_{L/u}^{(L)}(L,L\tan(\theta))\big|^2 .
\end{align*}

For any momentum $\mathbf{p}=(p,q)$, combining a plane wave $e^{i(px+qy)}$ with a reflected wave gives a solution ${\big(e^{ipx}+r_\beta(p)\,e^{ip(2L-x)}\big)e^{iqy}}$ satisfying the boundary condition at $x=L$, where $ip(1-r_\beta)=\beta(1+r_\beta)$ fixes $r_\beta(p)=\frac{ip-\beta}{ip+\beta}$ and $1+r_\beta(p)=\frac{2ip}{ip+\beta}$. We approximate $\psi_t^{(L)}$ on $x<L$ by superposing these waves with amplitude $\widetilde\psi(p,q)$ (for compactly supported initial data and $\mathrm{Re}\beta<0$, this matches the exact integral representation obtained via the method of images in Ref.~\cite{cavendish2025absorbing}, Eq.~(A8), together with free evolution along $y$). 

At $t=0$, the factor $e^{ip(2L-x)}$ places the reflected term behind the screen near $x=2L$ (for $\psi$ localized near the origin), so its norm on $x<L$ vanishes as $L\to\infty$ and the sum reproduces the initial state $\psi$. At $x=L$, the incident and reflected waves add, multiplying the Fourier amplitude by $1+r_\beta(p)$, and stationary phase selects $(p,q)=(u,u\tan(\theta))$ to give
\begin{equation*}
\psi_{L/u}^{(L)}(L,\!L\tan(\theta))
\simeq \!\frac{u}{iL}e^{\tfrac{iLu}{2}\!\sec^2(\theta)}
\big[1+r_\beta(u)\big]\widetilde\psi(u,\!u\tan(\theta)).
\end{equation*}
Hence, at each fixed $u>0$, the integrand of $\mathcal{A}_L^{\mathrm{ABC}(\beta)}(\theta;\psi)$ converges to $|1+r_\beta(u)|^2\,|\widetilde\psi(u,u\tan(\theta))|^2$, and passing the limit through the integral shows that $\mathcal{A}_\infty^{\mathrm{ABC}(\beta)}(\theta;\psi)$ equals
\begin{equation*}
b\sec^2(\theta)\!\int_0^\infty\!du\,
\big|1+r_\beta(u)\big|^2 \big|\widetilde\psi(u,u\tan(\theta))\big|^2\,.
\end{equation*}
Since $b\,|1+r_\beta(u)|^2=u\big(1-|r_\beta(u)|^2\big)=4bu^2/|iu+\beta|^2$, it follows that $\mathcal{A}_\infty^{\mathrm{ABC}(\beta)}(\theta;\psi)$ equals
\begin{equation*}
4b\sec^2(\theta)\!\int_0^\infty\!
\frac{du\,u^2}{|iu+\beta|^2}\,
\big|\widetilde\psi(u,\!u\tan(\theta))\big|^2 \,.\qedhere
\end{equation*}
\end{heuristicderivation}

The steps requiring justification are the control of the error in this packet approximation and the passage of the limit through the $u$-integral, in the regions $u\to0$ and $u\to\infty$. For Gaussian initial states, including the single-well state considered here, the limit follows from the exact solution given in Ref.~\cite{cavendish2025absorbing} (cf.\ Eq.~(30) therein, with the screen perpendicular to the $x$-axis and zero mean momentum).

\subsection{Far-field heuristics for CAP}

The far-field formula for CAP is expressed in terms of the scattering solutions of a one-dimensional ODE, which we first describe. Let $\mu\ge0$ be a continuous profile function for an absorbing potential that tends to $0$ as $s\to-\infty$ and to a constant $h>0$ as $s\to+\infty$, with $\mu$ integrable on $(-\infty,0]$ and $\mu-h$ integrable on $[0,\infty)$. The sigmoid potentials $\mu^h_w$ of Section~\ref{sec:modelSurvey} satisfy these conditions, with exponential approach to both limits. For $p>0$, consider the ODE
\begin{equation*}
-\tfrac{1}{2} f''(s)-i\mu(s)f(s)=\tfrac{p^2}{2}f(s).
\end{equation*}
By Levinson's asymptotic theorem~\cite[Sec.~1.3]{eastham1989asymptotic}, this equation has solutions asymptotic to $e^{\pm ik_+s}$ as $s\to+\infty$, where $k_+^2=p^2+2ih$ and $\mathrm{Im}(k_+)>0$, and to $e^{\pm ips}$ as $s\to-\infty$. Up to a constant factor, there is exactly one solution $\chi_p$ that decays as $s\to+\infty$, and as $s\to-\infty$, $\chi_p$ has the form $Ae^{ips}+Be^{-ips}+o(1)$. Multiplying the equation by $\chi_p^*$ and taking imaginary parts gives the flux identity $J_p'=-2\mu|\chi_p|^2$ for $J_p=\mathrm{Im}(\chi_p^*\chi_p')$, so $J_p$ is non-increasing and vanishes as $s\to+\infty$. If $A$ were zero, this would give $-p|B|^2=J_p(-\infty)\ge0$, forcing $\chi_p=0$. Hence $A\neq0$, and we may normalize the decaying solution so that $\chi_p(s)=e^{ips}+r_\mu(p)\,e^{-ips}+o(1)$ as $s\to-\infty$, which defines the reflection coefficient $r_\mu(p)$.

\begin{claim} Let $\psi\in\mathcal S(\mathbb{R}^2)$, let $\mu$ be an absorbing profile function as above, and let $\mathcal{T}(p;\mu)=1-|r_\mu(p)|^2$. Then, for fixed $\theta \in (-\tfrac{\pi}{2},\tfrac{\pi}{2})$,
\begin{equation*}
\mathcal{A}_\infty^{\mathrm{CAP}(\mu)}(\theta;\psi)=\sec^2(\theta)\!\int_0^\infty\!du\, u\,\mathcal{T}(u;\mu)\,\big|\widetilde\psi(u,\!u\tan(\theta))\big|^2\,.
\end{equation*}
\end{claim}

\begin{heuristicderivation}
Let $\psi_t^{(L)}$ denote the solution of the Schr\"odinger equation with potential $-i\mu(x-L)$ and initial condition $\psi$. By Eq.~\eqref{eq:CAPDensity}, the CAP detection density at height $y$ and time $t$ is $\int dx\,2\mu(x-L)\,|\psi_t^{(L)}(x,y)|^2$, so that, after substituting $u=\tfrac{L}{t}$ and $s=x-L$,
\begin{align*}
\mathcal{A}_L^{\mathrm{CAP}(\mu)}(\theta;\psi)=\sec^2(\theta)\int_0^\infty& du\,\frac{L^2}{u^2}\int_{-\infty}^\infty ds\,2\mu(s)\\
&\times\bigl|\psi^{(L)}_{L/u}(L+s,L\tan(\theta))\bigr|^2 .
\end{align*}
Let ${J_p=\mathrm{Im}(\chi_p^\ast\chi_p')}$ as above. Since $J_p(-\infty)=p\,\mathcal{T}(p;\mu)$ and $J_p(+\infty)=0$, 
\begin{equation*}
\int_{-\infty}^\infty ds\,2\mu(s)\,|\chi_p(s)|^2=-\int_{-\infty}^\infty \,ds\, J'_p(s)=p\,\mathcal{T}(p;\mu)\,.
\end{equation*}
Now consider
\begin{align*}
G_L(\mathbf{x},t)=\frac{1}{2\pi}&\!\int_{p_x>0}\!d^2\mathbf{p}\,\widetilde\psi(\mathbf{p})\,e^{ip_xL}\chi_{p_x}\!(x-L)\,e^{ip_yy-\frac{it}{2}|\mathbf{p}|^2}
\\
&+\frac{1}{2\pi}\!\int_{p_x<0}\!d^2\mathbf{p}\,\widetilde\psi(\mathbf{p})\,e^{i\mathbf{p}\cdot\mathbf{x}-\frac{it}{2}|\mathbf{p}|^2}\,.
\end{align*}
The first term solves the Schr\"odinger equation with the absorbing potential exactly. At $t=0$ and fixed $\mathbf{x}$, the $p_x>0$ part of $\psi$ is recovered, plus a reflected term with phase $e^{ip_x(2L-x)}$. That term vanishes as $L\to\infty$ by the Riemann--Lebesgue lemma, and so does the component of $G_L$ inside the absorbing potential, which carries the phase $e^{ip_xL}$. Thus $G_L$ matches $\psi$ at $t=0$, and we approximate $\psi_t^{(L)}$ by $G_L$. The $p_x<0$ term has no stationary point on the screen and, as in the derivation for SD, contributes only at lower order. At $x=L+s$, $y=L\tan(\theta)$, $t=\tfrac{L}{u}$, the phase of the first term is stationary at $\mathbf{p}=(u,u\tan(\theta))$, giving
\begin{equation*}
\psi^{(L)}_{L/u}(L+s,L\tan(\theta))\simeq\frac{u}{iL}\,e^{iLu\frac{\sec^2(\theta)}{2}}\,\chi_u(s)\,\widetilde\psi(u,u\tan(\theta)).
\end{equation*}
Substituting this into the expression for $\mathcal{A}_L^{\mathrm{CAP}(\mu)}$ cancels the factor $(\tfrac{L}{u})^2$, and the flux identity then gives
\begin{equation*}
\mathcal{A}_\infty^{\mathrm{CAP}(\mu)}(\theta;\psi)=\sec^2(\theta)\!\int_0^\infty\!du\, u\,\mathcal{T}(u;\mu)\,|\widetilde\psi(u,\!u\tan(\theta))|^2\,.\qedhere
\end{equation*}
\end{heuristicderivation}

The steps not justified here are the approximation of $\psi_t^{(L)}$ by $G_L$, the interchange of the limit with the $s$- and $u$-integrals (including the regions $u\to0$ and $u\to\infty$), and the contribution near $p_x=0$.

For the sigmoid $\mu^h_w(s)=h/(1+e^{-s/w})$, the reflection coefficient has a closed form: $r_\mu(p)=C_2/C_1$, where $C_1$ and $C_2$ are the amplitudes obtained in Ref.~\cite{landau1965quantum}, \S25, Problem~3, with $\alpha=\tfrac{1}{w}$, $k_1=p$, and $k_2=k_+=\sqrt{p^2+2ih}$, $\mathrm{Im}(k_+)>0$. Explicitly,
\begin{equation*}
r_\mu(p)=\frac{\Gamma(2iwp)\,\Gamma(a)\,\Gamma(1+a)}{\Gamma(-2iwp)\,\Gamma(b)\,\Gamma(1+b)}\,,
\end{equation*}
where $a=-iw(p+k_+)$ and $b=iw(p-k_+)$. Using ${\Gamma(1+z)=z\Gamma(z)}$ and $|\Gamma(2iwp)|=|\Gamma(-2iwp)|$, $\mathcal{T}(p;\mu)=1-|r_\mu(p)|^2$ reduces to Eq.~\eqref{eq:CAPTransmission}.

\subsection{Far-field heuristics for MS}

\begin{claim}
Let $\psi\in\mathcal S(\mathbb{R}^2)$ and $\lambda>0$, and write ${c_\lambda=\tfrac{4\lambda}{\pi}}$ and $F_\psi(v)=\int_{\mathbb{R}}|\widetilde\psi(v,q)|^2\,dq$. Then, for fixed $\theta\in (-\tfrac{\pi}{2},\tfrac{\pi}{2})$, $\mathcal{A}_\infty^{\mathrm{MS}(\lambda)}(\theta;\psi)$ is given by
\begin{equation*}
c_\lambda\sec^2(\theta)\!\int_0^\infty\!du\, u^2|\widetilde\psi(u, u\tan(\theta))|^2\,S_\infty(u;\psi),
\end{equation*}
where $S_\infty(u;\psi)=\exp\!\left[-c_\lambda\int_u^\infty dv\, vF_\psi(v)\right]$.
\end{claim}

\begin{heuristicderivation}
In the MS proposal, $\psi_{\mathrm{surv.}}$ evolves by the free Schr\"odinger equation on $x<L$ with a Dirichlet condition at $x=L$. To obtain a Dirichlet-compatible initial state, we define the antisymmetrized state $D_{L,0}=N_L\bigl[\psi(x,y)-\psi(2L-x,y)\bigr]$, which then gives rise to the time-dependent solution
\begin{equation*}
D_{L,t}(x,y)=N_L\bigl[\psi_t(x,y)-\psi_t(2L-x,y)\bigr]
\end{equation*}
for $x<L$, where $N_L$ normalizes $D_{L,0}$ on $x<L$. If $\psi$ is supported in $x<L$, $N_L=1$. In general, $D_{L,0}$ converges in $L^2$ to $\psi$ as $L\to \infty$ and $N_L$ converges to $1$. At the screen, $\partial_xD_{L,t}(L,y)=2N_L\,\partial_x\psi_t(L,y)$.

By Eq.~\eqref{eq:MSDensity}, the MS detection density at position $y$ and time $t$ is $h_L(y,t)\,S_L(t)$, where $h_L(y,t)=\tfrac{\lambda}{\pi}\bigl|\partial_xD_{L,t}(L,y)\bigr|^2$ and 
\begin{equation*}
S_L(t)=\exp\!\left[-\int_0^t ds\int_{\mathbb{R}}dy'\,h_L(y',s)\right],
\end{equation*}
so that, after substituting $u=\tfrac{L}{t}$, $\mathcal{A}_L^{\mathrm{MS}(\lambda)}(\theta;\psi)$ equals
\begin{equation*}
\sec^2(\theta)\!\int_0^\infty \!du\,\frac{L^2}{u^2}\,h_L\!\left( L\tan(\theta),\tfrac{L}{u}\right)\!S_L\!\left(\tfrac{L}{u}\right)\!.
\end{equation*}
For $t=\tfrac{L}{v}$ and $y=tq$ with $v>0$ and $q$ fixed, applying two-dimensional stationary phase to the Fourier integral for $\partial_x\psi_t(L,y)$ selects the stationary point $\mathbf{p}=(v,q)$, with the $x$-derivative supplying a factor $iv$ and stationary phase giving a $\tfrac{1}{t}$ amplitude factor. Using $h_L(y,t)=c_\lambda N_L^2\bigl|\partial_x\psi_t(L,y)\bigr|^2$ and $N_L\to 1$, we obtain $h_L(tq,t)\simeq c_\lambda\,\tfrac{v^2}{t^2}\,|\widetilde\psi(v,q)|^2$.
In the survival exponent, we set $y'=sq$, so that $dy'=s\,dq$. Since the exponent integrates over all values of $y'$, all tangential momenta $q$ contribute, and
\begin{align*}
-\log S_L(\tfrac{L}{u})&\simeq c_\lambda\int_0^{\tfrac{L}{u}}ds\,s^{-1}\left(\tfrac{L}{s}\right)^2F_\psi\left(\tfrac{L}{s}\right)\\
&=c_\lambda\int_u^\infty \!dv\,v\,F_\psi(v)\,,
\end{align*}
where the second line follows from the change of variables $v=\tfrac{L}{s}$, so $S_L(\tfrac{L}{u})\to S_\infty(u;\psi)$. At $y=L\tan(\theta)$ and $t=\tfrac{L}{u}$, we have $q=u\tan(\theta)$, and
\begin{equation*}
\frac{L^2}{u^2}\,h_L\left(L\tan(\theta),\tfrac{L}{u}\right)\simeq c_\lambda u^2\,\big|\widetilde\psi(u,u\tan(\theta))\big|^2 .
\end{equation*}
Hence $\mathcal{A}_\infty^{\mathrm{MS}(\lambda)}(\theta;\psi)$ equals
\begin{equation*}
c_\lambda\sec^2(\theta)\int_0^\infty\!du\, u^2\,\big|\widetilde\psi(u,u\tan(\theta))\big|^2\,S_\infty(u;\psi)\,.\qedhere
\end{equation*}
\end{heuristicderivation}
The steps not justified here are the replacement of $\psi$ by $D_{L,0}$, the stationary-phase approximation inside the survival integral (including early times $s\to 0$), and the interchange of the limit with the final $u$-integral (including $u\to 0$ and $u\to\infty$).

\subsection{Numerical plots of the behavior of \texorpdfstring{$\mathcal{A}_L$}{A\_L} in the far-field limit}

\begin{figure*}[!htbp]
    \centering
    \includegraphics[width=\textwidth]{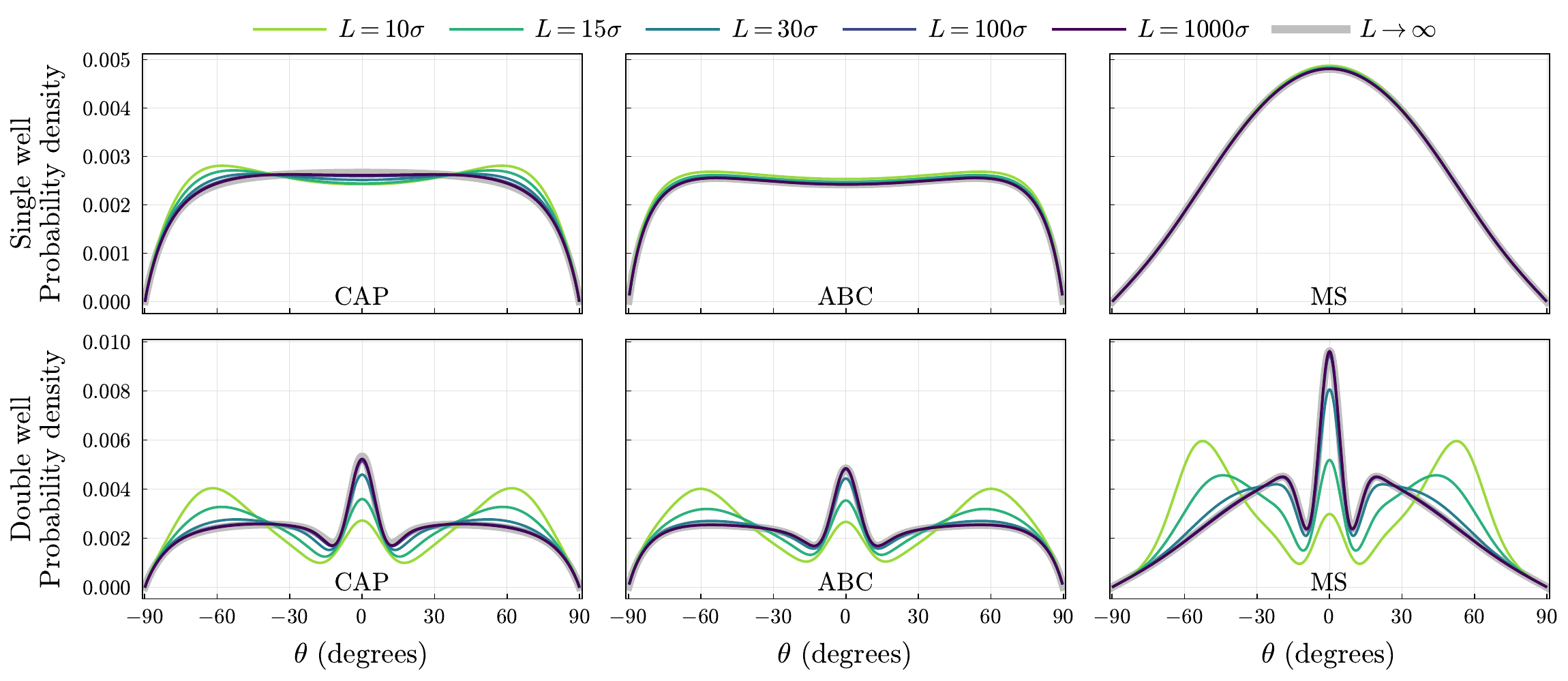}
    \caption{Angular distributions $\mathcal{A}_L$ for CAP, ABC, and MS (columns) for the single-well (top) and double-well (bottom) initial states, at $\tfrac{L}{\sigma}=10, 15, 30, 100$, and $10^3$. The thick gray curves show the far-field formulas, Eqs.~\eqref{CAPFarField}, \eqref{eq:ABCFarField}, and \eqref{MSFarField}. Model parameters are as in Section~\ref{sec:singleDouble}, and densities are shown per degree, as in Figs.~\ref{OneWell} and~\ref{TwoWell}.}
    \label{fig:FarFieldConvergence}
\end{figure*}

Figure~\ref{fig:FarFieldConvergence} compares the far-field formulas for CAP, ABC, and MS, Eqs.~\eqref{CAPFarField}, \eqref{eq:ABCFarField}, and \eqref{MSFarField}, with direct computations of $\mathcal{A}_L$ at finite $L$. The model parameters are those of Section~\ref{sec:singleDouble}. The derivations above apply to arbitrary $\psi\in\mathcal{S}(\mathbb{R}^2)$. Here we check them numerically for two states: the single-well state $\psi^{\mathrm{s.w.}}_\sigma$ and the double-well state $\psi^{\mathrm{d.w.}}_{\sigma,d}$ with $d=20\sigma$, which is neither Gaussian nor rotationally symmetric. In each case, the finite-$L$ distributions approach the far-field formula at each fixed angle as $L$ increases. The interference structure of the double-well state makes this a sharper test, since the far-field formulas must reproduce the positions and heights of its peaks.

For QF and SD, the far-field formula is the semiclassical distribution itself, Eq.~\eqref{standardFF}, and the convergence can already be seen in the $L=10^3\sigma$ panels of Figs.~\ref{OneWell} and~\ref{TwoWell}. For SD it is not uniform, as the insets of those figures show near grazing incidence.

\section{Parameter dependence for CAP, ABC, and MS}\label{app:ParamDep}

The predictions of CAP, ABC, and MS shown in Figs.~\ref{OneWell} and \ref{TwoWell} use the parameter values chosen in Section~\ref{comp}. To show how these predictions depend on that choice, we consider the far-field distributions, which are independent of $L$, for the single-well state. Since this state has the single length scale $\sigma$, its far-field distributions depend on the parameters only through the dimensionless combinations $b\sigma$ (for ABC, with $\beta=ib$), $\tfrac{\lambda}{\sigma}$ (for MS), and $\tfrac{w}{\sigma}$ and $h$ in units of $\hbar^2/(m\sigma^2)$ (for CAP), so the results below hold for every value of $\sigma$.

\begin{figure}[ht]
    \centering
    \includegraphics[width=\columnwidth]{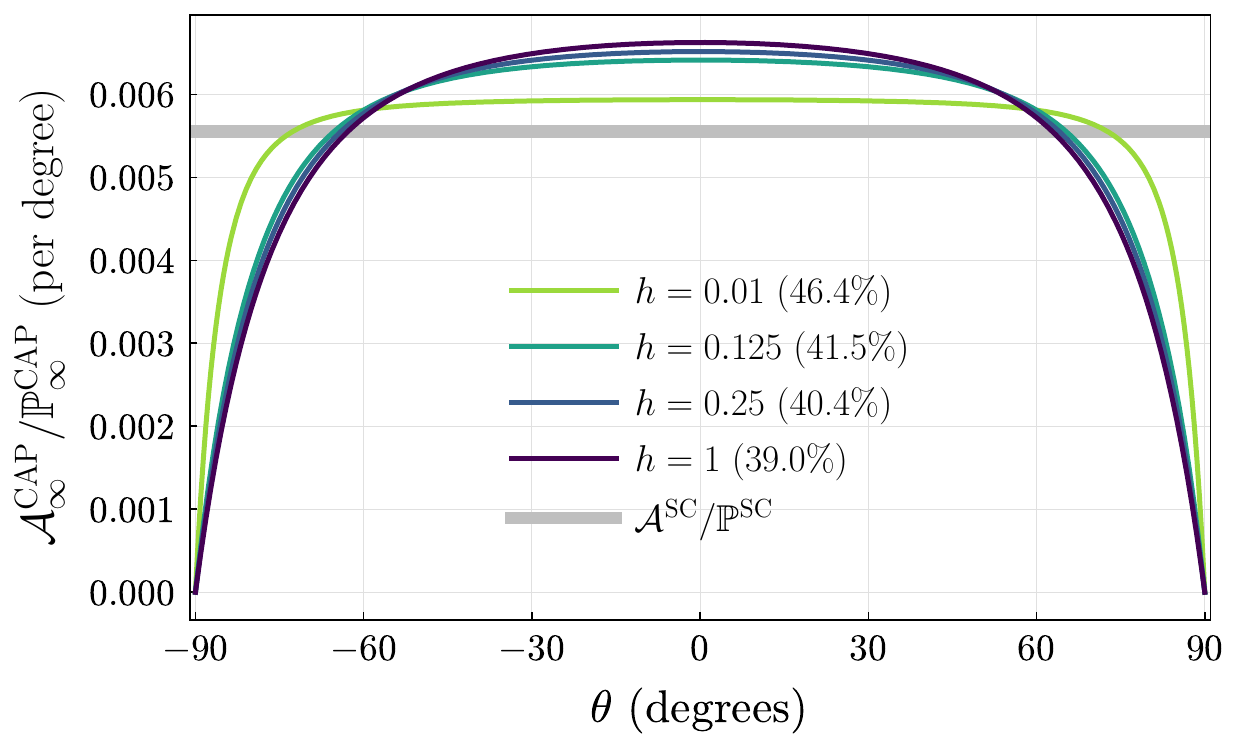}
    \caption{Conditional far-field angular distributions $\mathcal{A}^{\mathrm{CAP}}_\infty\!/\,\mathbb{P}^{\mathrm{CAP}}_\infty$ for the single-well state, with $w=\sigma$ and several values of $h$ (in units of $\hbar^2/(m\sigma^2)$). The far-field total detection probability for each value is given in parentheses. The value $h\approx0.125$ is the one used in Section~\ref{comp}. The thick gray line is the semiclassical distribution. Densities are shown per degree.}
    \label{fig:ParamCAP}
\end{figure}

\begin{figure}[ht]
    \centering
    \includegraphics[width=\columnwidth]{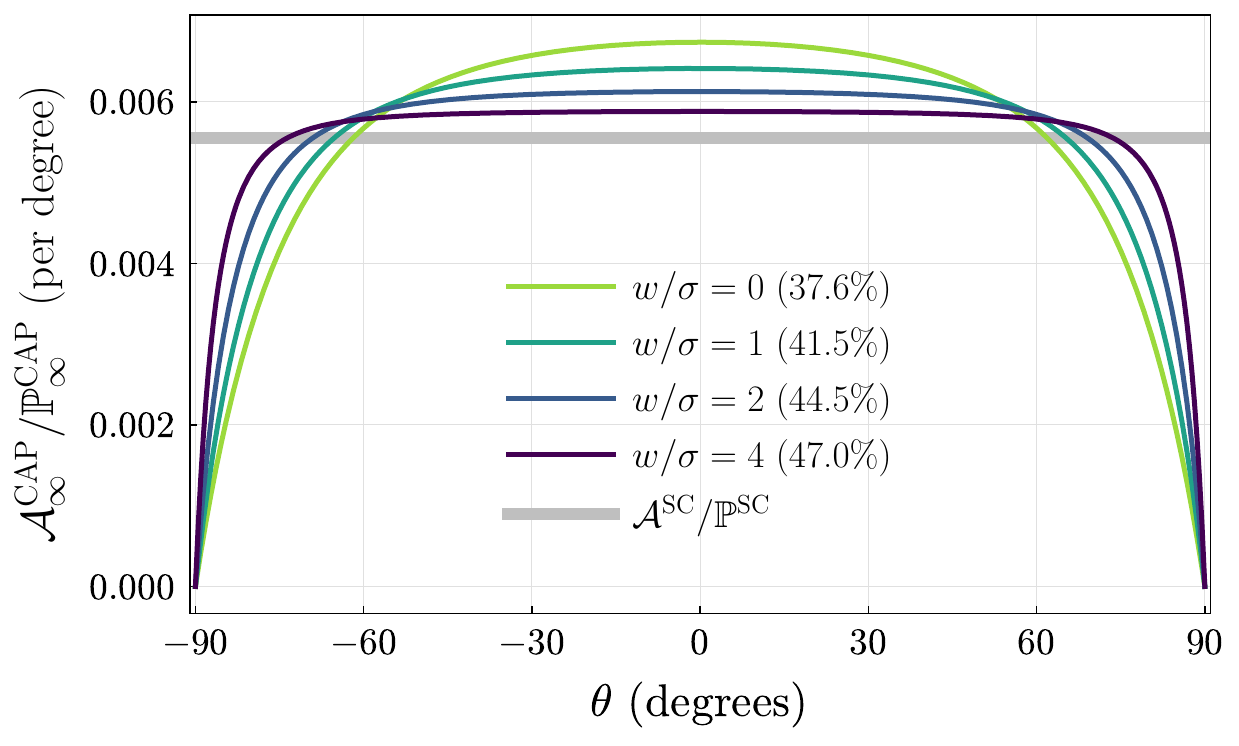}
    \caption{As in Fig.~\ref{fig:ParamCAP}, for CAP with $h\approx0.125$ (in units of $\hbar^2/(m\sigma^2)$) and several values of $\tfrac{w}{\sigma}$. The value $\tfrac{w}{\sigma}=1$ is the one used in Section~\ref{comp}, and $w=0$ corresponds to the step potential, Eq.~\eqref{StepFarField}.}
    \label{fig:ParamCAPw}
\end{figure}

For ABC and MS, the dependence of the far-field total detection probability on the parameter is already shown in Fig.~\ref{ScalingBehavior}. Since rescaling the initial state by $\alpha$ is equivalent to replacing $\beta$ by $\alpha\beta$ or $\lambda$ by $\tfrac{\lambda}{\alpha}$ (Section~\ref{rescalingLimits}), the ABC and MS curves in that figure give the totals as functions of $b\sigma\approx0.333\,\alpha$ and $\tfrac{\lambda}{\sigma}\approx \tfrac{2.73}{\alpha}$, respectively. For CAP, rescaling changes $w$ and $h$ together, so we instead vary each separately, holding the other at its value from Section~\ref{comp}. At fixed $w=\sigma$, the total decreases from the semiclassical value $\tfrac{1}{2}$ as $h$ increases (Fig.~\ref{fig:ParamCAP}), consistent with the $h\to 0$ limit in Section~\ref{sec:farfield}, and changes little for $h\gtrsim 1$, approaching about $38.3\%$ for large $h$. At fixed $h\approx0.125$, the total increases with $w$, from approximately $37.6\%$ for the step potential ($w=0$, see Eq.~\eqref{StepFarField}) to $47.0\%$ at $w=4\sigma$ (Fig.~\ref{fig:ParamCAPw}).

\begin{figure}[ht]
    \centering
    \includegraphics[width=\columnwidth]{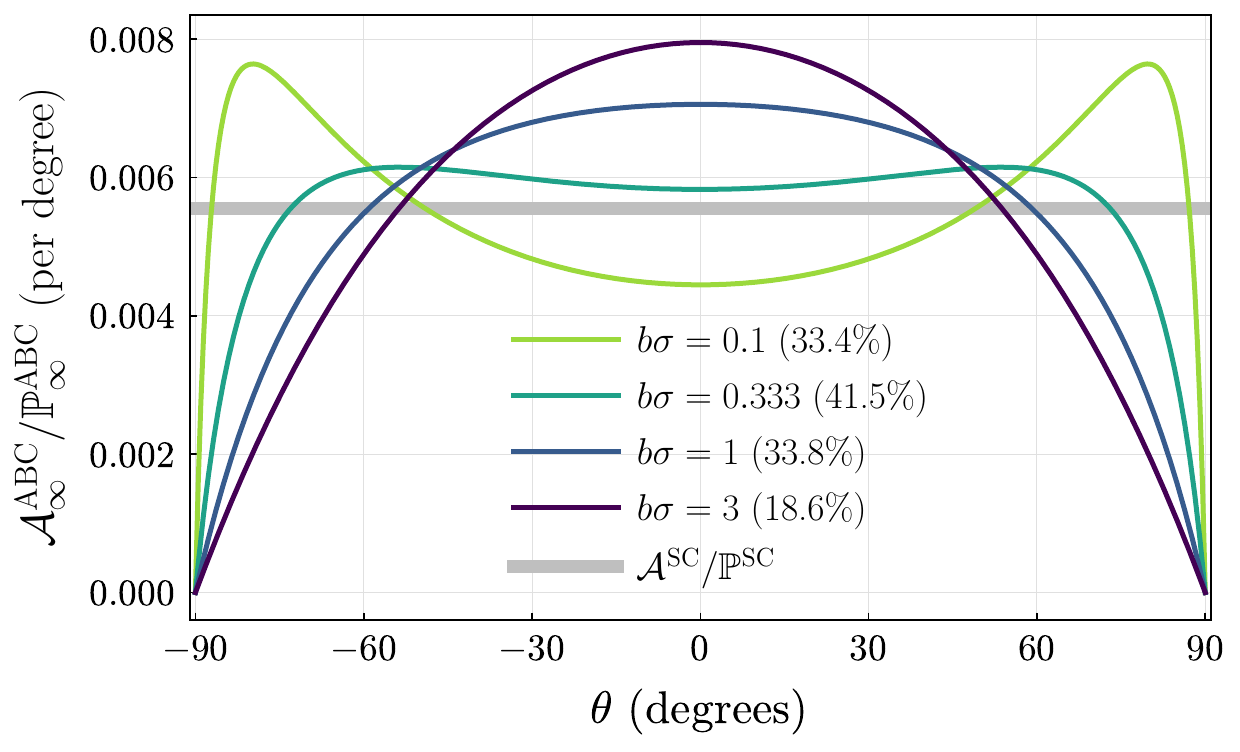}
    \caption{As in Fig.~\ref{fig:ParamCAP}, for ABC with $\beta=ib$ and several values of $b\sigma$. The value $b\sigma\approx0.333$ is the one used in Section~\ref{comp}.}
    \label{fig:ParamABC}
\end{figure}

\begin{figure}[ht]
    \centering
    \includegraphics[width=\columnwidth]{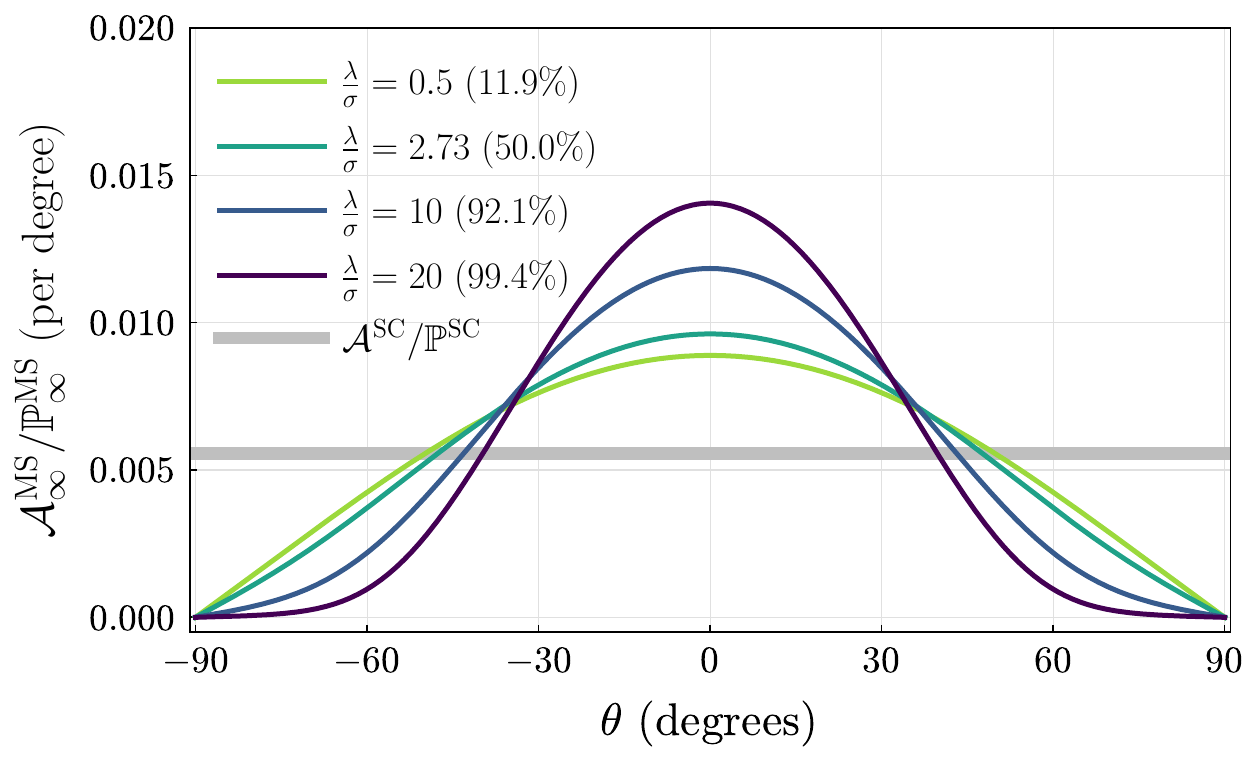}
    \caption{As in Fig.~\ref{fig:ParamCAP}, for MS with several values of $\tfrac{\lambda}{\sigma}$. The value $\tfrac{\lambda}{\sigma}\approx2.73$ is the one used in Section~\ref{comp}.}
    \label{fig:ParamMS}
\end{figure}

Figs.~\ref{fig:ParamCAP}--\ref{fig:ParamMS} also show the effect of the parameters on the shape of the far-field distributions. To separate shape from total detection probability, each curve shows the conditional distribution $\mathcal{A}_\infty(\theta;\psi^{\mathrm{s.w.}}_\sigma)/\,\mathbb{P}_\infty(\mathrm{D};\psi^{\mathrm{s.w.}}_\sigma)$, computed from Eqs.~\eqref{CAPFarField}, \eqref{StepFarField}, \eqref{eq:ABCFarField}, and \eqref{MSFarField}, and compared with the semiclassical distribution, which is uniform for this state. The shape of the CAP distribution depends only weakly on $h$, approaching the semiclassical distribution away from grazing incidence for small $h$. Increasing $w$ at fixed $h$ flattens the CAP distribution, likewise bringing it closer to the semiclassical distribution away from grazing incidence. The ABC distribution depends strongly on $b$: small values of $b\sigma$ produce local maxima near grazing incidence, while large values concentrate detections around $\theta=0$. In particular, $b\sigma=0.1$ and $b\sigma=1$ give similar total detection probabilities but clearly different angular distributions. For MS, increasing $\lambda$ concentrates detections around $\theta=0$ while the total detection probability approaches $1$. For every parameter value, the distributions vanish at grazing incidence, as follows from the far-field formulas.

\section{Large \texorpdfstring{$t$}{t} behavior and grazing angle limits}\label{app:larget}

In Section~\ref{sec:grazing}, the grazing-angle behavior of each proposal was related to the long-time tail of its arrival-time marginal. In this appendix, we make this relation precise for product states. Unlike the far-field results of Appendix~\ref{app:FarField}, which are supported by heuristic derivations, this relation admits a short complete proof under the hypotheses stated below. The tail exponent for SC follows directly from its explicit formula in Section~\ref{sec:GAL}, while heuristic arguments for the tail exponents of the other five proposals are given in Section~\ref{sec:heuristicTails}.

\subsection{Grazing-angle limits for product states}\label{sec:GAL}

Throughout this appendix, we work in units with $\hbar=m=1$ and in the two-dimensional setting of Section~\ref{comp}, with the screen at $x=L$. We use the unitary Fourier transform
\begin{equation*}
\widetilde f(k)=\mathcal F(f)(k)=\frac{1}{\sqrt{2\pi}}\int_{-\infty}^{\infty}du\,e^{-iku}f(u)\,.
\end{equation*}
For a proposal with joint arrival density $\Pi_L(y,t;\psi)$ on the screen, the arrival-time marginal is ${\rho_L(t;\psi)=\int dy\,\Pi_L(y,t;\psi)}$, and, since $y=L\tan(\theta)$, the angular density is
\begin{equation*}
\mathcal A_L(\theta;\psi)=L\sec^2(\theta)\int_0^\infty dt\,\Pi_L\big(L\tan(\theta),t;\psi\big).
\end{equation*}

\begin{lemma}\label{lem:grazingtail} Let $\psi(x,y)=g(x)\phi(y)$, where $g,\phi\in \mathcal{S}(\mathbb{R})$ and $\phi$ is normalized as an element of $L^2(\mathbb{R})$. Fix $L>0$, and suppose $\Pi_L(y,t;\psi)$ can be written as a product
\begin{equation*}
\Pi_L(y,t;\psi)=\rho_L(t;g)\,|\phi_t(y)|^2\,,
\end{equation*}
where $\phi_t(y)$ is the freely evolved transverse wave function, $t\mapsto\rho_L(t;g)$ is integrable on $[0,t_0]$ for every $t_0>0$, and there exist $C_L,\gamma >0$ such that $\displaystyle{\lim_{t\to\infty} t^\gamma\rho_L(t;g)=C_L}$. Then, with the limits taken within $\big(-\tfrac{\pi}{2},\tfrac{\pi}{2}\big)$,
\begin{align*}
\lim_{\theta\to \pm \frac{\pi}{2}} \mathcal{A}_L(\theta;\psi) &\sec^{\gamma-2}(\theta) \\
&= C_LL^{1-\gamma}\!\int_0^\infty dq\,q^{\gamma-1} \big|\widetilde{\phi}(\pm q)\big|^2\,.
\end{align*}
\end{lemma}

\begin{proof} Consider first the limit $\theta\to +\frac{\pi}{2}$, which corresponds to $y \to +\infty$, and let $F_t(k;\phi)= \mathcal{F}\left( e^{i\frac{u^2}{2t}} \phi(u)\right)(k)$, where $\mathcal{F}$ is the Fourier transform. Expressing $\phi_t$ in terms of the free propagator and expanding the exponent, we have
\begin{align}\label{eq:propRewrite}
\phi_t(y) &= \frac{1}{\sqrt{2\pi i t}} \int_{-\infty}^{\infty} du\, e^{i\frac{(y-u)^2}{2t}}\phi(u)\nonumber\\ 
&= \frac{e^{i\frac{y^2}{2t}}}{\sqrt{it}} \left[ \frac{1}{\sqrt{2\pi}} \int_{-\infty}^{\infty} du\, e^{-i\frac{y}{t}u} \left( e^{i\frac{u^2}{2t}} \phi(u) \right) \right]\nonumber\\
&= \frac{e^{i\frac{y^2}{2t}}}{{\sqrt{it}}}F_t\left(\tfrac{y}{t};\phi\right)\,.
\end{align}
Note that as $t$ goes to infinity, $e^{i\frac{u^2}{2t}}\phi(u)$ converges to $\phi(u)$ in the Schwartz topology, since the Schwartz semi-norms of $(e^{i\frac{u^2}{2t}} - 1)\phi(u)$ are $O(t^{-1})$. Because the Fourier transform is continuous on $\mathcal{S}(\mathbb{R})$, the family of functions $F_t(\cdot\,;\phi)$ converges to $\widetilde{\phi}$ in the Schwartz topology. Moreover, for any $t_0>0$, each derivative of $e^{iu^2/(2t)}\phi(u)$ is $e^{iu^2/(2t)}$ times a finite sum of terms $c\,u^j t^{-k}\partial^\ell\phi(u)$ with $t$-independent constants $c$ and integers $j,k,\ell\ge0$. Since $|e^{iu^2/(2t)}|=1$ and $t^{-k}\le t_0^{-k}$ for all $t\ge t_0$, the family ${\{e^{iu^2/(2t)}\phi : t\ge t_0\}}$ is bounded in $\mathcal S(\mathbb R)$. Since the Fourier transform is continuous on $\mathcal S(\mathbb R)$, the family ${\{F_t(\cdot\,;\phi) : t\ge t_0\}}$ is also bounded, i.e., it satisfies uniform Schwartz bounds.

Because $\lim_{t\to\infty} t^\gamma\rho_L(t;g)=C_L$, we can choose ${t_0>0}$ sufficiently large such that $|\rho_L(t;g)| \le M t^{-\gamma}$ for all $t \ge t_0$ and some constant $M>0$. By Equation~\eqref{eq:propRewrite} and the uniform boundedness of $\{F_t(\cdot\,;\phi) : t\ge t_0\}$, there is a constant $K_0$ such that $|\Pi_L(y,t;\psi)| \le M K_0 t^{-\gamma-1}$ for all $t \ge t_0$ and $y\in\mathbb{R}$, which is integrable on $[t_0,\infty)$ since $\gamma>0$. On $[0,t_0]$, free evolution is continuous in $t$ on $\mathcal S(\mathbb R)$, so the family $\{\phi_t\mid t\in[0,t_0]\}$ is compact and hence uniformly Schwartz. Together with the integrability of $\rho_L(\cdot\,;g)$ on $[0,t_0]$, this ensures that $\int_0^\infty dt\,\Pi_L(y,t;\psi)$ converges for every $y$ and allows us to apply the dominated convergence theorem on $[0,t_0]$, giving
\begin{align*}
\lim_{y\to \infty} y^\gamma&\int_{0}^{t_0}dt\,\Pi_L(y,t;\psi) \\
&=\int_{0}^{t_0}dt\,\rho_L(t;g)\lim_{y\to \infty}\left(y^\gamma|\phi_t(y)|^2\right)=0\,.
\end{align*}
Hence $\lim_{y\to \infty} y^\gamma\int_0^\infty dt\,\Pi_L(y,t;\psi)$ exists if and only if $\lim_{y\to \infty} y^\gamma\int_{t_0}^\infty dt\,\Pi_L(y,t;\psi)$ exists, and the two are equal. For $y>0$, using Equation~\eqref{eq:propRewrite} and making the substitution ${q=\tfrac{y}{t}}$, we have
\begin{equation}\label{eq:DCTIntegral}
y^\gamma\!\int_{t_0}^{\infty}dt\,\Pi_L(y,t;\psi) =
\int_{0}^{\frac{y}{t_0}}\!dq\,\frac{y^\gamma}{q}\rho_L\!\left(\tfrac{y}{q};g\right)\left|F_{\frac{y}{q}}(q;\phi)\right|^2\,.
\end{equation}
We now evaluate the $y\to\infty$ limit of the right-hand side of Equation~\eqref{eq:DCTIntegral} using the dominated convergence theorem. To justify this, we bound the integrand independently of $y$. By our choice of $t_0$, we have $|\rho_L(t;g)| \le M t^{-\gamma}$ for all $t \ge t_0$. In the domain of integration, $t = \tfrac{y}{q} \ge t_0$, which gives $\big|\rho_L(\tfrac{y}{q};g)\big| \le M (\tfrac{q}{y})^\gamma$. This yields the uniform bound
\begin{equation*}
\frac{y^\gamma}{q} \left|\rho_L\!\left(\tfrac{y}{q};g\right)\right| \le \frac{y^\gamma}{q} M \left(\tfrac{q}{y}\right)^\gamma = M q^{\gamma-1}\,.
\end{equation*}
Furthermore, because the family ${\{F_t(\cdot\,;\phi)\mid t\ge t_0\}}$ satisfies uniform Schwartz bounds, $\sup_{q} |q^k F_t(q;\phi)|$ is uniformly bounded for all $k \ge 0$. Thus, there exists a constant $K_N$ for any integer $N$ such that ${|F_t(q;\phi)|^2 \le K_N (1+q^2)^{-N}}$ for all $t \ge t_0$ and all $q \in \mathbb{R}$. Choosing an integer $N > \tfrac{\gamma}{2}$, the full integrand is dominated by $M K_N q^{\gamma-1} (1+q^2)^{-N}$. This majorant is independent of $y$ and integrable on $(0, \infty)$, which justifies applying the dominated convergence theorem.

By the assumption on $\rho_L(t;g)$ and the convergence of $F_t(\cdot\,;\phi)$ to $\widetilde{\phi}$, for each $q>0$,
\begin{align*}
\lim_{y\to \infty} 
&\frac{y^\gamma}{q}\rho_L\!\left(\tfrac{y}{q};g\right)\left|F_{\frac{y}{q}}(q;\phi)\right|^2 \\
&= q^{\gamma-1} \lim_{t\to\infty} t^\gamma \rho_L(t;g)|F_t(q;\phi)|^2 = C_L\,q^{\gamma-1} \big|\widetilde{\phi}(q)\big|^2\,,
\end{align*}
so applying the dominated convergence theorem to Equation~\eqref{eq:DCTIntegral} yields the limit of the $[t_0,\infty)$ integral and hence
\begin{align}\label{eq:pilimit} 
\lim_{y\to \infty} y^\gamma&\int_0^\infty\!dt\,\Pi_L(y,t;\psi)\nonumber \\
&= \int_{0}^{\infty}\!dq\,\lim_{y\to \infty}\left(\frac{y^\gamma}{q}\rho_L\!\left(\tfrac{y}{q};g\right)\left|F_{\frac{y}{q}}(q;\phi)\right|^2\right)\nonumber\\
&= C_L\int_0^\infty\!dq\,q^{\gamma-1}\big|\widetilde{\phi}(q)\big|^2\,.
\end{align}
It follows that
\begin{align*}
\lim_{\theta\to \frac{\pi}{2}} &\mathcal{A}_L(\theta;\psi) \sec^{\gamma-2}(\theta)\\
&=\lim_{\theta\to \frac{\pi}{2}} L \sec^\gamma(\theta) \!\int_0^\infty dt\, \Pi_L(L \tan (\theta), t; \psi)\\
&=\lim_{\theta\to \frac{\pi}{2}} \frac{L^{1-\gamma}}{\sin^\gamma(\theta)} \big(L\tan(\theta)\big)^\gamma \int_0^\infty\!dt\,\Pi_L(L\tan(\theta),t;\psi)\\
&= L^{1-\gamma}\lim_{y\to \infty} y^\gamma\int_0^\infty \!dt\,\Pi_L(y,t;\psi)\\
&\stackrel{\eqref{eq:pilimit}}{=}C_L\,L^{1-\gamma}\int_0^\infty\!dq\,q^{\gamma-1}\big|\widetilde{\phi}(q)\big|^2\,.
\end{align*}
The limit ${\theta\to -\frac{\pi}{2}}$ follows by the same argument, taking $y\to-\infty$, replacing $y^\gamma$ and $\sin^\gamma(\theta)$ by $|y|^\gamma$ and $|\sin(\theta)|^\gamma$, respectively, and making the substitution $q=\tfrac{|y|}{t}$, which evaluates $F_t$ at $-q$ and yields $\widetilde{\phi}(-q)$ in the limit.
\end{proof}

For the product states considered here, the factorization assumed in the lemma follows directly from the definitions of QF, SD, CAP, and ABC. For MS, the survival factor is defined via Dirichlet evolution on $x<L$, so the longitudinal factor must vanish at $x=L$ for $\rho_L$ to be integrable near $t=0$. Replacing $g(x)$ on $x<L$ by the antisymmetrized Dirichlet-compatible state $N_L[g(x)-g(2L-x)]$ (as in Appendix~\ref{app:FarField}) yields the same product form with a locally integrable arrival-time density. In all five cases, the arrival-time marginal $\rho_L(t;\Psi)$ defined in Section~\ref{sec:grazing} depends only on the longitudinal factor $g$, which justifies the notation $\rho_L(t;g)$. Consequently, if $\rho_L(t;g)\,t^{\gamma}$ converges to ${C_L>0}$, the lemma shows that the angular density at grazing incidence vanishes for $\gamma>2$ and, provided the relevant transverse momentum integral is nonzero, diverges for $\gamma<2$ and has a finite nonzero limit for $\gamma=2$.

The semiclassical proposal is not covered by the lemma, since its joint distribution does not factorize with the freely evolved transverse wave function. Its arrival-time marginal is, however, explicit for every $\psi \in \mathcal{S}(\mathbb{R}^2)$:
\begin{equation*}
\rho_L^{\mathrm{SC}}(t;\psi)
=\frac{L}{t^2}\int_{\mathbb{R}}dq\,\big|\widetilde{\psi}\left(\tfrac{L}{t},q\right)\big|^2,
\end{equation*}
so $\gamma_{\mathrm{tail}}^{\mathrm{SC}}=2$ whenever $\int dq\,\big|\widetilde{\psi}(0,q)\big|^2\ne 0$. For a product state $\psi(x,y)=g(x)\phi(y)$, this reduces to $\rho_L^{\mathrm{SC}}(t;g)=\frac{L}{t^2}|\widetilde g(\tfrac{L}{t})|^2$, and the exact semiclassical formula gives
\begin{equation*}
\lim_{\theta\to\pm\frac{\pi}{2}}
\mathcal A^{\mathrm{SC}}(\theta;\psi)
=\big|\widetilde{g}(0)\big|^2
\int_0^\infty dq\,q\big|\widetilde{\phi}(\pm q)\big|^2,
\end{equation*}
as follows by the substitution $q=u|\tan(\theta)|$ and dominated convergence. This coincides with the formula in Lemma~\ref{lem:grazingtail} with $\gamma=2$ and $C_L=L|\widetilde g(0)|^2$.

\subsection{Analysis of arrival-time tails}\label{sec:heuristicTails}

We now examine the late-time arrival-time tails of the six proposals as $t\to\infty$ with the screen distance $L$ and the detector parameters held fixed. Throughout, we work in units $\hbar=m=1$ and consider product states $\psi(x,y)=g(x)\phi(y)$ with $g,\phi\in\mathcal{S}(\mathbb{R})$, as in Lemma~\ref{lem:grazingtail}.

\begin{figure}[ht]
    \centering
    \includegraphics[width=\columnwidth]{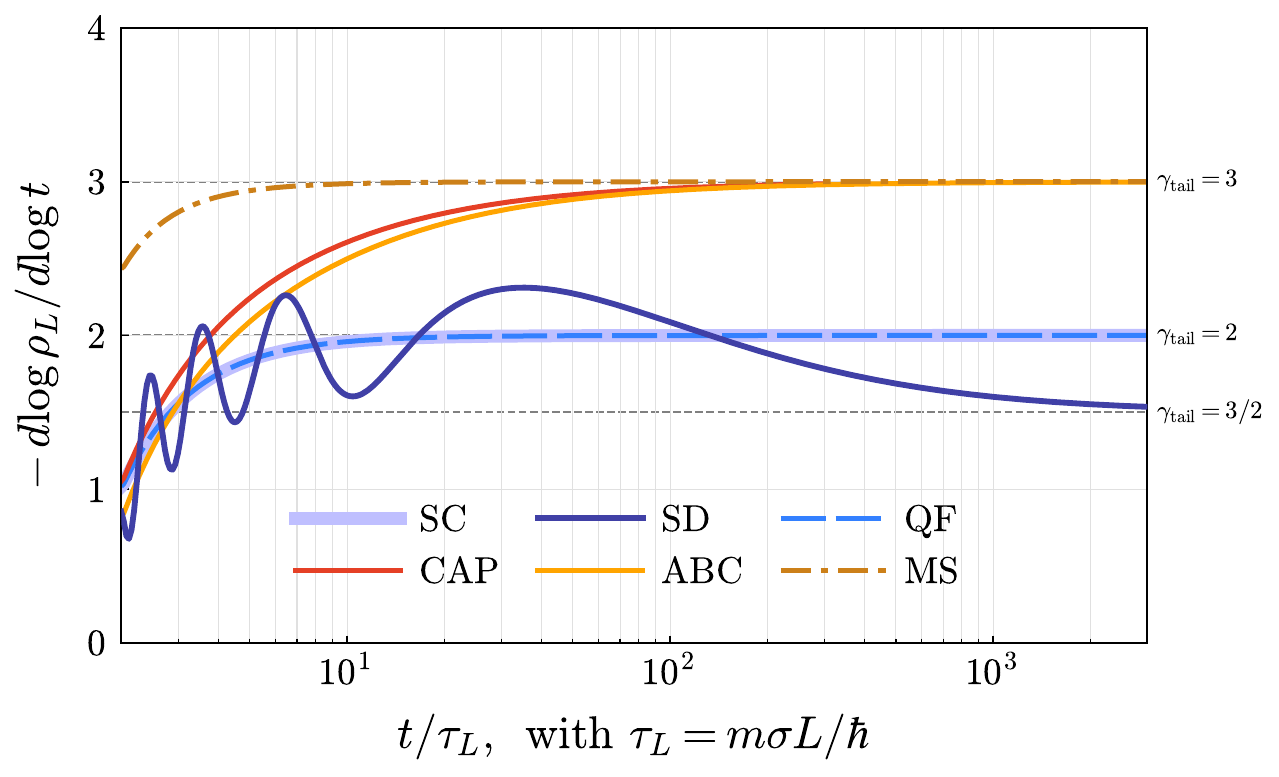}
    \caption{Local tail exponent $-d\log\rho_L/d\log t$ for the arrival-time density for the single-well state across proposals for $L=100\sigma$. Time is measured in units of $\tau_L=\tfrac{m\sigma L}{\hbar}$.}
    \label{fig:TailExponentComparisons}
\end{figure}

Figure~\ref{fig:TailExponentComparisons} illustrates these late-time scalings numerically by plotting the local exponent $-d\log\rho_L/d\log t$ for all six proposals at $L=100\sigma$ (using the centered Gaussian longitudinal factors $g_\sigma(x)$ and the sigmoid CAP potential). The curves approach three distinct asymptotes: $\gamma_{\mathrm{tail}}=\tfrac{3}{2}$ for SD, $\gamma_{\mathrm{tail}}=2$ for SC and QF, and $\gamma_{\mathrm{tail}}=3$ for MS, ABC, and CAP. The same limits hold across the other screen distances simulated ($\tfrac{L}{\sigma}=10$ to $10^3$). For the five non-SC proposals, Lemma~\ref{lem:grazingtail} then accounts for the corresponding divergent, finite, and vanishing grazing-angle behaviors. Below, we summarize the analytical and heuristic arguments behind these exponents for each proposal. Establishing the stated asymptotics rigorously for general initial states and detector parameters would require additional analysis of the relevant propagators.

\subsubsection{Semiclassical (SC)}
For SC, integrating the arrival-position density over the transverse coordinates gives the arrival-time marginal
\begin{equation*}
\rho_L^{\mathrm{SC}}(t;g)=\frac{L}{t^2}\,\big|\widetilde g\left(\tfrac{L}{t}\right)\big|^2\,.
\end{equation*}
Since $g\in\mathcal S(\mathbb R)$, its Fourier transform $\widetilde g$ is continuous, so as $t\to\infty$,
\begin{equation*}
\rho_L^{\mathrm{SC}}(t;g)\sim L\,|\widetilde g(0)|^2\,t^{-2}\,.
\end{equation*}
Whenever $\widetilde g(0)\ne 0$ (in particular, for the centered Gaussian longitudinal factors $g_\sigma(x)$ considered here), this yields $\gamma_{\mathrm{tail}}^{\mathrm{SC}}=2$, consistent with Fig.~\ref{fig:TailExponentComparisons} and the direct grazing-angle limit in Section~\ref{sec:GAL}.

\subsubsection{Quantum flux (QF)}
For a product state, the arrival-time marginal is the longitudinal probability current,
\begin{equation}\label{eq:currentExp}
\rho_L^{\mathrm{QF}}(t;g)=\mathrm{Im}\big(g_t(L)^*\,\partial_x g_t(L)\big).
\end{equation}
The free-propagator representation used in the proof of Lemma~\ref{lem:grazingtail} gives
\begin{equation*}
g_t(x)=e^{i\tfrac{x^2}{2t}}\,\frac{F_t\left(\tfrac{x}{t};g\right)}{\sqrt{it}},
\end{equation*}
where $F_t(\cdot\,;g)\to\widetilde g$ in the Schwartz topology as $t\to\infty$. In particular, $F_t(\cdot\,;g)\to\widetilde g$ and $F_t'(\cdot\,;g)\to\widetilde g\,'$ uniformly on $\mathbb{R}$, so by continuity of $\widetilde g$ and $\widetilde g\,'$ at $0$, we have $F_t(\tfrac{L}{t};g)\to\widetilde g(0)$ and $F_t'(\tfrac{L}{t};g)\to\widetilde g\,'(0)$ as $t\to\infty$. Plugging the above expression into Eq.~\eqref{eq:currentExp} and evaluating at $x=L$ yields
\begin{align*}
\rho_L^{\mathrm{QF}}(t;g)&=\frac{1}{t^2}\Big(L\,\big|F_t\left(\tfrac{L}{t};g\right)\big|^2+\mathrm{Im}\big[F_t\left(\tfrac{L}{t};g\right)^*F_t'\left(\tfrac{L}{t};g\right)\big]\Big)\\
&\sim\frac{1}{t^2}\Big(L\,|\widetilde g(0)|^2+\mathrm{Im}\big[\widetilde g(0)^*\,\widetilde g\,'(0)\big]\Big),
\end{align*}
provided the coefficient is nonzero. For the separable states with centered Gaussian longitudinal factors $g_\sigma(x)$ considered in this paper, the second term vanishes and the coefficient is positive. Thus $\gamma_{\mathrm{tail}}^{\mathrm{QF}}=2$.

\subsubsection{Standard proposal (SD)}
Integrating the SD density in Eq.~\eqref{SDEquation} over the transverse coordinates (and substituting $p=|p_x|$ in the negative-momentum sector) gives
\begin{equation*}
\rho_L^{\mathrm{SD}}(t;g)=\frac{1}{2\pi}\sum_{\alpha=\pm}\left|\int_0^\infty\!dp\,\sqrt{p}\,\widetilde{g}(\alpha p)\,e^{i\alpha pL-i\tfrac{p^2t}{2}}\right|^2\,.
\end{equation*}
For large $t$, rapid oscillation of the phase $e^{-ip^2t/2}$ away from the stationary point $p=0$ suppresses larger momenta, so the dominant contribution comes from $p\sim t^{-1/2}$. Substituting $p=s\,t^{-1/2}$ and assuming $\widetilde{g}(0)\ne 0$ gives
\begin{align*}
\int_0^\infty\!dp\,&\sqrt{p}\,\widetilde{g}(\alpha p)\,e^{i\alpha pL-i\tfrac{p^2t}{2}} \\
&= t^{-\tfrac{3}{4}}\!\int_0^\infty\!ds\,\sqrt{s}\,\widetilde{g}(\alpha s t^{-\tfrac{1}{2}})\,e^{i\alpha L s t^{-\tfrac{1}{2}}-i\tfrac{s^2}{2}} \\
&\sim t^{-\tfrac{3}{4}}\,\widetilde{g}(0)\!\int_0^\infty\!ds\,\sqrt{s}\,e^{-i\tfrac{s^2}{2}}\,.
\end{align*}
Since
\begin{equation*}
\int_0^\infty\!ds\,\sqrt{s}\,e^{-i\tfrac{s^2}{2}}=2^{-\tfrac{1}{4}}\,\Gamma(\tfrac{3}{4})\,e^{-\tfrac{3\pi i}{8}}\,,
\end{equation*}
both momentum sectors $\alpha=\pm$ have the same leading term, so summing their squared moduli cancels the factor of $2$ in the prefactor $\tfrac{1}{2\pi}$ and gives
\begin{equation*}
\rho_L^{\mathrm{SD}}(t;g)\sim\frac{\Gamma(\tfrac{3}{4})^2}{\sqrt{2}\,\pi}\,|\widetilde{g}(0)|^2\,t^{-\tfrac{3}{2}}\,.
\end{equation*}
In particular, $\widetilde{g}(0)\ne 0$ holds for the centered Gaussian longitudinal factors $g_\sigma(x)$ considered in this paper, giving $\gamma_{\mathrm{tail}}^{\mathrm{SD}}=\tfrac{3}{2}$.

\subsubsection{Absorbing boundary condition (ABC)}
For ABC, the longitudinal wave function satisfies $\partial_x g_t(L)=\beta\, g_t(L)$, where $\mathrm{Im}(\beta)>0$. The reflection amplitude for an incident plane wave of momentum $k>0$ is $r_\beta(k)=(ik-\beta)/(ik+\beta)$,
which approaches $-1$ as $k\to0$ for fixed $\beta\ne0$. The incident and reflected waves therefore cancel to leading order at low momentum, and the boundary value of the corresponding scattering state, $1+r_\beta(k)=\tfrac{2ik}{\beta}+O(k^2)$, is of order $k$ as $k\to 0$. Because this $k\to 0$ cancellation between $e^{ik(x-L)}$ and $r_\beta(k)\,e^{-ik(x-L)}$ holds pointwise on $x\le L$ (and likewise for the adjoint scattering state, with $\beta\to\beta^*$), the overlap of $g$ with the adjoint state in the scattering-state expansion also vanishes linearly in $k$. Expanding $g_t$ in scattering states of energy $\tfrac{k^2}{2}$ (which evolve with phase $e^{-ik^2t/2}$) and evaluating at $x=L$ therefore gives a low-energy contribution of the form
\begin{equation*}
g_t(L)\approx\int_0^\infty\!dk\,A(k)\,e^{-i\tfrac{k^2t}{2}},
\end{equation*}
where $A(k)$ is the product of the initial-state overlap and the scattering state's boundary value at $x=L$. Combining the two $O(k)$ factors gives $A(k)=C_{g,\beta}k^2+O(k^3)$ as $k\to 0$, while $A(k)$ decays at large $k$ due to the smoothness of $g$. Integrating by parts once using $e^{-ik^2t/2}=\tfrac{i}{kt}\tfrac{d}{dk}e^{-ik^2t/2}$, we find that the boundary terms vanish and
\begin{align*}
g_t(L)&\approx-\frac{i}{t}\int_0^\infty\!dk\,\Big(\frac{A(k)}{k}\Big)'e^{-i\tfrac{k^2t}{2}}\\
&\sim -\frac{i\,C_{g,\beta}}{t}\sqrt{\frac{\pi}{2it}}\propto t^{-\tfrac{3}{2}}.
\end{align*}
Since the ABC arrival-time marginal is $\rho_L^{\mathrm{ABC}}(t;g)=\mathrm{Im}(\beta)\,|g_t(L)|^2$, this gives $\gamma_{\mathrm{tail}}^{\mathrm{ABC}}=3$, provided the leading coefficient is nonzero.

Note that since the ABC dynamics are not generated by a self-adjoint operator when $\mathrm{Im}(\beta)>0$, the expansion in scattering states used here is itself heuristic.

The $t^{-3}$ tail, together with its coefficient, was obtained explicitly by Dubey, Bernardin, and Dhar for initial states supported on a half-line [Sec.~V A and Eq.~(69) of Ref.~\cite{Dubey2021Quantum}]. In the present notation, their result reads
\begin{equation*}
\rho_L^{\mathrm{ABC}}(t;g)\sim\frac{2\,\mathrm{Im}(\beta)}{\pi|\beta|^4}\left|\int_{-\infty}^{L}\!dx\,\big[1-\beta(L-x)\big]g(x)\right|^2\!t^{-3}.
\end{equation*}
For any nonnegative $g$ supported in $x<L$ that is not identically zero, the imaginary part of this integral, $-\mathrm{Im}(\beta)\int_{-\infty}^{L}dx\,(L-x)\,g(x)$, is nonzero. In particular, the coefficient is nonzero for the centered Gaussian factors $g_\sigma$ considered here (neglecting the exponentially small part of $g_\sigma$ beyond the screen), consistent with Fig.~\ref{fig:TailExponentComparisons}.

\subsubsection{Complex absorbing potential (CAP)}
For CAP, the arrival-time marginal is
\begin{equation*}
\rho_L^{\mathrm{CAP}}(t;g)=2\!\int_{\mathbb R}\!dx\,\mu(x-L)\,|g_t(x)|^2\,.
\end{equation*}
As with ABC, the low-energy scattering behavior provides a heuristic explanation of the CAP arrival-time tail. For the absorbing step potential, the absorption probability of an incident plane wave with normal momentum $k$ is
\begin{equation*}
\mathcal{T}(k)=1-\left|\frac{k-k_+}{k+k_+}\right|^2\,,
\end{equation*}
where $k_+^2=k^2+2ih$ and $\mathrm{Im}(k_+)>0$. For fixed $h>0$, $\mathcal{T}(k)$ vanishes linearly as $k\to 0$. The same linear vanishing holds for the sigmoid potentials $\mu^h_w$, as can be checked directly from the formula for $\mathcal{T}$ in Section~\ref{sec:farfield}. Low-energy particles are therefore almost completely reflected: for the step potential, the reflection amplitude $r(k)=(k-k_+)/(k+k_+)$ approaches $-1$ as $k\to 0$, so the transmitted amplitude into the absorbing region, $1+r(k)=2k/(k+k_+)$, vanishes linearly in $k$. As for ABC, the same $k\to 0$ cancellation suppresses the initial-state overlap linearly in $k$, giving an $O(k^2)$ low-energy integrand in the absorbing region. In the absence of exceptional threshold behavior, the low-energy contribution to the evolved wave function in this region is therefore expected to decay as $t^{-3/2}$, giving an absorption rate proportional to $t^{-3}$. This suggests $\gamma_{\mathrm{tail}}^{\mathrm{CAP}}=3$ for fixed $h>0$.

Unlike the preceding arguments, this conclusion also requires control of the spatial integral over the absorbing region, and, unlike ABC and MS, CAP has no closed-form arrival-time density against which the exponent can be checked. We therefore rely on direct simulation: as shown in Fig.~\ref{fig:TailExponentComparisons}, the local exponent for the sigmoid CAP potential approaches $3$, consistent with $\gamma_{\mathrm{tail}}^{\mathrm{CAP}}=3$ across the simulated screen distances.

\subsubsection{Marchewka--Schuss (MS)}
The surviving longitudinal wave function $g^{\mathrm{surv}}_t$ in MS evolves with a Dirichlet boundary condition at $x=L$. As in Appendix~\ref{app:FarField}, we use the antisymmetrized Dirichlet-compatible initial state $g_0^{\mathrm{surv}}(x)=N_L\bigl[g(x)-g(2L-x)\bigr]$ on $x<L$, which reduces to $g$ (with $N_L=1$) when $g$ is supported in $x<L$. Its Dirichlet evolution is $g_t^{\mathrm{surv}}(x)=N_L\bigl[g_t(x)-g_t(2L-x)\bigr]$. Differentiating at the boundary gives $\partial_x g_t^{\mathrm{surv}}(L)=2N_L\,\partial_x g_t(L)$, which, via the free propagator, takes the form
\begin{equation*}
\partial_x g_t^{\mathrm{surv}}(L)\propto t^{-\tfrac{3}{2}}\!\int_{-\infty}^{\infty}\!dx\,(L-x)\,g(x)\,e^{i\tfrac{(L-x)^2}{2t}}\,.
\end{equation*}
As $t\to\infty$, the integral approaches
\begin{equation*}
\int_{-\infty}^{\infty}\!dx\,(L-x)\,g(x)=\sqrt{2\pi}\,\big[L\,\widetilde g(0)-i\,\widetilde g\,'(0)\big],
\end{equation*}
which is generically nonzero, and in particular nonzero for the centered Gaussian longitudinal factors $g_\sigma(x)$ considered in this paper (for which $\widetilde g_\sigma(0)>0$ and $\widetilde g_\sigma\,'(0)=0$). Thus, whenever this integral is nonzero,
\begin{equation*}
|\partial_x g_t^{\mathrm{surv}}(L)|^2\sim C_g\,t^{-3}
\end{equation*}
for a constant $C_g>0$. Integrating Eq.~\eqref{eq:MSDensity} over the transverse coordinates gives the arrival-time marginal $\rho_L^{\mathrm{MS}}(t;g)=\tfrac{\lambda}{\pi}\,|\partial_x g_t^{\mathrm{surv}}(L)|^2 S(t)$, with survival factor
\begin{equation*}
S(t)=\exp\!\left[-\frac{\lambda}{\pi}\!\int_0^t\!ds\,|\partial_x g_s^{\mathrm{surv}}(L)|^2\right].
\end{equation*}
Because $g\in\mathcal S(\mathbb R)$, $|\partial_x g_s^{\mathrm{surv}}(L)|^2$ is continuous down to $s=0$ and decays as $s^{-3}$ as $s\to\infty$; hence the integral in the exponent converges on $[0,\infty)$, and the survival factor approaches a strictly positive constant. Therefore, for any $\lambda>0$, the tail exponent satisfies $\gamma_{\mathrm{tail}}^{\mathrm{MS}}=3$.

\section{Upper bounds on far-field ABC detection probabilities}\label{app:ABCbound}

This appendix establishes two upper bounds on the far-field total detection probability predicted by ABC: a general bound by the semiclassical value, valid for every initial state, and a numerical bound, uniform in $\beta$, for the Gaussian initial states of Section~\ref{comp}. The general bound was shown for one-dimensional scattering in Ref.~\cite{cavendish2025absorbing}: in the far-field limit, the total detection probability predicted by ABC is lower than the scattering-theory prediction for every $\beta$ with $\mathrm{Im}(\beta)>0$, and for fixed $\mathrm{Im}(\beta)$ the discrepancy is smallest when $\beta$ is purely imaginary (see Eqs.~(10)--(12) therein). Since the planar-screen problem separates into coordinates normal and parallel to the screen (cf.\ Eq.~(28) of Ref.~\cite{cavendish2025absorbing}), the same argument applies to the total detection probability in the experimental setting considered here.

As in Section~\ref{sec:angArrival}, we set $\hbar=m=1$ and take the far-field formula \eqref{eq:ABCFarField} as given. Write $\beta=a+ib$ with $b>0$, and let $F(u;\psi)=\int_{\mathbb{R}}dq\,\big|\widetilde{\psi}(u,q)\big|^2$ denote the marginal distribution of the longitudinal momentum, so that $\int_0^\infty du\,F(u;\psi)$ is the total probability mass of the right-moving part of the momentum distribution for $\psi$.

\begin{lemma}\label{lem:ABCbound}
If Equation~\eqref{eq:ABCFarField} holds, then for every $\beta$ with $\mathrm{Im}(\beta)>0$ and every $\psi$ with $\int_0^\infty du\,F(u;\psi)>0$, the total detection probabilities satisfy $\mathbb{P}^{\mathrm{ABC}(\beta)}_\infty(\mathrm{D};\psi)
<\mathbb{P}^{\mathrm{SC}}_\infty(\mathrm{D};\psi)$.
\end{lemma}

\begin{proof}
Integrating Equation~\eqref{eq:ABCFarField} over $\theta\in(-\pi/2,\pi/2)$ and substituting $q=u\tan(\theta)$, so that $dq=u\sec^2(\theta)\,d\theta$, gives
\begin{align}\label{eq:ABCTotalProb}
\mathbb{P}^{\mathrm{ABC}(\beta)}_\infty(\mathrm{D};\psi)
&=4b\int_0^\infty du\,\frac{u\,F(u;\psi)}{|iu+\beta|^2}\nonumber\\
&=\int_0^\infty du\,\frac{4bu\,F(u;\psi)}{a^2+(u+b)^2}\,.
\end{align}
Since $a^2+(u+b)^2-4bu=a^2+(u-b)^2\ge0$, the factor $4bu/\left(a^2+(u+b)^2\right)$ is at most 1, with equality at most at the single point $u=b$ (and only if $a=0$). Hence, by Equation~\eqref{eq:ABCTotalProb}, $\mathbb{P}^{\mathrm{ABC}(\beta)}_\infty(\mathrm{D};\psi)\le\int_0^\infty du\,F(u;\psi)$, and the inequality is strict because $F$ is positive on a set of positive measure. Finally, integrating Equation~\eqref{eq:SCFarFieldFormula} over $\theta$ shows that $\mathbb{P}^{\mathrm{SC}}_\infty(\mathrm{D};\psi)=\int_0^\infty du\,F(u;\psi)$, so it follows that $\mathbb{P}^{\mathrm{ABC}(\beta)}_\infty(\mathrm{D};\psi)<\mathbb{P}^{\mathrm{SC}}_\infty(\mathrm{D};\psi)$ as required.
\end{proof}

For both the single- and double-well states of Section~\ref{comp}, the longitudinal factor is the same Gaussian $g_\sigma$, so that $F(u;\psi)=\sqrt{2\sigma^2/\pi}\,e^{-2\sigma^2u^2}$. For fixed $b$, a nonzero real part of $\beta$ only increases the denominator, so it suffices to consider purely imaginary $\beta=ib$. After the substitution $v=\sigma u$, the detection probability depends only on the dimensionless combination $c=b\sigma$, yielding
\begin{equation*}
\mathbb P^{\mathrm{ABC}(ib)}_\infty(\mathrm D;\psi)=\sqrt{\frac{2}{\pi}}\int_0^\infty dv\,\frac{4cv}{(v+c)^2}\,e^{-2v^2}\,,
\end{equation*}
which is continuous in $c$ and tends to $0$ as $c\to 0$ and as $c\to\infty$. Numerical maximization over the parameter $c$ then gives $\sup_{\beta}\,\mathbb{P}^{\mathrm{ABC}(\beta)}_\infty(\mathrm{D};\psi)\approx 0.415$, attained at $\beta\approx 0.333\,\tfrac{i}{\sigma}$. This value of $\beta$ is the one used throughout. Since dilating the initial state by $\alpha$ is equivalent to replacing $\beta$ by $\alpha\beta$ (Section~\ref{rescalingLimits}), the ABC curve in Fig.~\ref{ScalingBehavior} gives $\mathbb{P}^{\mathrm{ABC}(ib)}_\infty(\mathrm D;\psi_\sigma)$ as a function of $c\approx0.333\,\alpha$, and its maximum, at $\alpha=1$, is the maximum above.

Granting Equation~\eqref{eq:ABCFarField}, it follows that, for these initial states, a far-field total detection probability exceeding approximately $41.5\%$ would be incompatible with every member of the ABC family.

\section{Experimental considerations}\label{ExperimentalConsiderations}

The distributions compared in the main text assume that each run of the experiment lasts indefinitely. In the lab, physical constraints, such as the height of the screen in a free-fall implementation, impose an effective cutoff time $T$ on each run. Fig.~\ref{SchematicSetup} shows the QF and MS distributions over a finite region of the screen for $T=100\,\mathrm{ms}$, including gravity. The cutoff matters because one of the most prominent differences between the predicted distributions in Figs.~\ref{OneWell} and~\ref{TwoWell} occurs at grazing angles, and in the models surveyed, grazing angles are correlated with late arrival times. This appendix asks how this contrast varies with $T$ and $L$, and how many runs are then needed to resolve it.

We use ${}^4\mathrm{He}$ with $\sigma=1\,\mu\mathrm{m}$ and, except in Section~\ref{sec:nearfield}, $L=100\sigma$, so that $\tau_L=m\sigma L/\hbar\approx6.3\,\mathrm{ms}$. During $T=100\,\mathrm{ms}$ of free fall the atom would drop about $5\,\mathrm{cm}$ along the screen. Free fall is not essential, however. The reduction to two dimensions in Section~\ref{comp} uses only that the initial state is a product $\psi(x,y)\eta(z)$ and that the potential acting on the atom depends on $z$ alone. The $z$-factor then evolves independently, and integrating over $z$ leaves the arrival distributions in $(y,t)$ unchanged. This holds for gravity, and equally for any confining potential along $z$ (cf.\ Appendix~\ref{app:CPC}). We therefore assume confinement along $z$, so that a screen of modest height suffices. To work with values known to be achievable, we take the screen's field of view ($3\,\mathrm{mm}$) and resolution (below $10\,\mu\mathrm{m}$) from the light sheet of Ref.~\cite{bucker2009single}. At $L=100\,\mu\mathrm{m}$, this field of view covers $|\theta|\le\theta_{\max}\approx86^\circ$. Except in Section~\ref{sec:nearfield}, all probabilities below refer to detections within this window. For CAP, ABC, and MS we use the parameters of Section~\ref{sec:singleDouble} ($h\approx0.125$ in units of $\hbar^2/(m\sigma^2)$ with $w=\sigma$, $\beta\approx0.333\,\tfrac{i}{\sigma}$, and $\lambda\approx2.73\,\sigma$). The numbers of runs quoted for these proposals depend on this choice.

\subsection{Total detection probability with spatiotemporal constraints}\label{sec:totals}

In this subsection, we ask whether the total detection probability can distinguish the proposals when detections are restricted to a time interval $[0,T]$ and a bounded region of the screen, namely $|y|\le1.5\,\mathrm{mm}$ at $L=100\,\mu\mathrm{m}$, corresponding to arrival angles $|\theta|\le86^\circ$. We write $\mathrm{D}_T$ for the event of a detection within this window before time $T$, so that $\mathbb{P}_L(\mathrm{D}_T;\psi^{\mathrm{s.w.}}_\sigma)$ is the corresponding detection probability. Since each run of the experiment either results in a detection before time $T$ or does not, it follows that the number of detections in $N$ runs is binomially distributed with success probability $\mathbb{P}_L(\mathrm{D}_T;\psi^{\mathrm{s.w.}}_\sigma)$. Taking the SC prediction $p_0=\mathbb{P}^{\mathrm{SC}}_L(\mathrm{D}_T;\psi^{\mathrm{s.w.}}_\sigma)$ as the null hypothesis, the number of runs needed to reject it at significance level $\alpha$ with power $1-\beta_{\mathrm{II}}$, when the true value is $p_1$, is approximately
\begin{equation}
N=\left[\frac{z_{1-\alpha/2}\sqrt{p_0(1-p_0)}+z_{1-\beta_{\mathrm{II}}}\sqrt{p_1(1-p_1)}}{p_1-p_0}\right]^2,
\label{eq:BinomialN}
\end{equation}
where $z_q$ is the $q$-quantile of the standard normal distribution \cite{NISTHandbook}. Table~\ref{tab:TotalsN} lists the total detection probabilities at $T=100\,\mathrm{ms}$ and the resulting $N$ for $\alpha=0.05$ and $1-\beta_{\mathrm{II}}=0.95$. Around $1600$ runs suffice to distinguish SC from CAP, ABC, and MS. SC, QF, and SD, on the other hand, predict nearly identical totals, so this test cannot separate them without a large number of trials. We return to them in Section~\ref{sec:nearfield}.

\begin{table}[ht]
\centering
\begin{tabular}{lccccc}
\hline\hline
 & SC, QF & SD & CAP & ABC & MS \\
\hline
$\mathbb{P}_L(\mathrm{D}_T;\psi^{\mathrm{s.w.}}_\sigma)$ & $44.9\%$ & $45.0\%$ & $40.4\%$ & $40.0\%$ & $49.7\%$ \\
$N$   &  --- & $2.5\times 10^6$  & $1600$   & $1400$  & $1400$ \\
\hline\hline
\end{tabular}
\caption{Total detection probability within the detection window $|\theta|\le86^\circ$
($|y|\le1.5\,\mathrm{mm}$) before $T=100\,\mathrm{ms}$ at $L=100\sigma$, and the number of runs $N$ needed to distinguish each proposal from SC using Eq.~\eqref{eq:BinomialN} with $\alpha=0.05$ and $1 -\beta_{\mathrm{II}}=0.95$, rounded up to the
nearest hundred. QF predicts the same total as SC, so cannot be distinguished from it by this test. SD's prediction differs from SC's by about $0.1$ percentage points, so distinguishing SD from SC requires a large number of trials. Within the window, the single- and double-well totals agree to within $0.01$ percentage points.}
\label{tab:TotalsN}
\end{table}

These values of $N$ depend on the parameters chosen for CAP, ABC, and MS. For ABC, however, a statement independent of $\beta$ is available. In the far field, the total ABC detection probability is maximized by purely imaginary $\beta$ and, for the Gaussian initial states considered here, never exceeds $41.5\%$ (Appendix~\ref{app:ABCbound}). At the present settings, maximizing numerically over purely imaginary $\beta$ gives at most $40.1\%$ within the window before $T=100\,\mathrm{ms}$. If the SC prediction is correct, a one-sided test at $\alpha=0.05$ with power $0.95$ therefore excludes ABC for every purely imaginary $\beta$ after approximately $1200$ runs.

Because this test compares the number of detections to the number of runs, it requires that every run releases exactly one atom and that the atom is not lost before reaching the screen. A run in which no atom was loaded, or in which the atom was lost, for example through background-gas collisions, is indistinguishable from one in which the atom went undetected. These requirements concern preparing exactly one atom per run and avoiding losses before detection, not the detector. In Ref.~\cite{kaufman2014two}, for example, single-atom loading is verified by imaging the trap before each experiment, and the subsequent loss probability is calibrated independently. As for the detector itself, each proposal considered here is understood to predict all recorded data, including the probability of detection. For this reason, we do not introduce a separate efficiency factor, which would be an additional hypothesis about the detector on top of what each proposal specifies.

\begin{figure}[ht]
    \centering
    \includegraphics[width=\columnwidth]{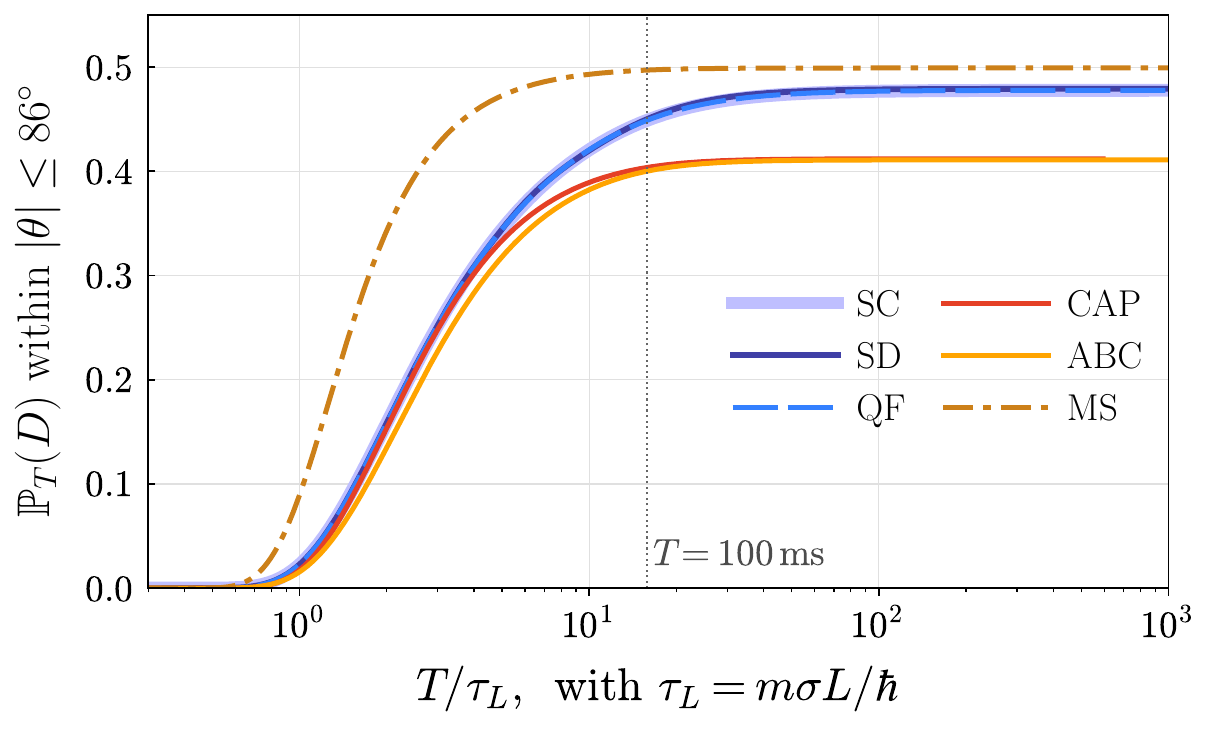}
    \caption{Total detection probability $\mathbb{P}_L(\mathrm{D}_T;\psi^{\mathrm{s.w.}}_\sigma)$ within the detection window $|\theta|\le86^\circ$ before the cutoff $T$, as a function of $T/\tau_L$ with $\tau_L=m\sigma L/\hbar$, for the single-well state at $L=100\sigma$ and the parameters of Section~\ref{sec:singleDouble}. The double-well values agree to within $0.03$ percentage points. The dotted line marks $T=100\,\mathrm{ms}$ for ${}^4\mathrm{He}$ with $\sigma=1\,\mu\mathrm{m}$ ($\tau_L\approx6.3\,\mathrm{ms}$).}
    \label{fig:CutoffDetectionProbs}
\end{figure}

Figure~\ref{fig:CutoffDetectionProbs} shows $\mathbb{P}_L(\mathrm{D}_T;\psi^{\mathrm{s.w.}}_\sigma)$ as a function of $T$. If this were computed over all angles $\theta$, it would simply give each proposal's cumulative arrival-time distribution at $x=L$. Within the $\theta$ window, however, arrivals with $|\theta|>\theta_{\max}$ are not counted. At $T=100\,\mathrm{ms}$ this lowers the totals by less than $0.1$ percentage points, since the cutoff has already suppressed most grazing arrivals, but at larger $T$ the difference grows: SC, QF, and SD approach approximately $48\%$ rather than $50\%$, while CAP and ABC lose only about half a percentage point, and MS less than $0.1$. For the parameters chosen, MS saturates earliest, by $T\approx10\tau_L$. The approach to the asymptote is governed by the arrival-time tails: over the whole screen, the probability of arrival after $T$ falls as $T^{1-\gamma}$, i.e.\ as $T^{-1}$ for SC and QF, $T^{-2}$ for CAP, ABC, and MS, and only as $T^{-1/2}$ for SD (Appendix~\ref{app:larget}).

In settings where detection times are recorded, their distribution provides a further test which, unlike the binomial test above, does not require knowing the total number of trials. We compare the observed distribution of detection times, conditional on detection within the $\theta$ window before $T$, with the SC prediction using a $\chi^2$ goodness-of-fit test. For this, the detection times are grouped into bins of $10\,\mathrm{ms}$, corresponding to an assumed timing resolution of that order. A coarser resolution would be analyzed identically with coarser bins. The curves in Fig.~\ref{fig:CutoffDetectionProbs} are the cumulative distributions of these detection times, so the probability of bin $i$ is the increment of $\mathbb{P}_L(\mathrm{D}_t;\psi^{\mathrm{s.w.}}_\sigma)$ over that bin, divided by $\mathbb{P}_L(\mathrm{D}_T;\psi^{\mathrm{s.w.}}_\sigma)$. With $p_i$ and $q_i$ the probabilities of bin $i$ under the null hypothesis (here SC) and under the alternative, define $w^2=\sum_i(q_i-p_i)^2/p_i$. Pearson's $\chi^2$ statistic for $N_{\mathrm{det}}$ detections then has approximately a noncentral $\chi^2$ distribution with noncentrality $N_{\mathrm{det}}w^2$ and one fewer degree of freedom than the number of bins~\cite[Sec.~16.3.5]{Agresti2013}, so a $\chi^2$ goodness-of-fit test reaches significance level $\alpha$ with power $1-\beta_{\mathrm{II}}$ after approximately $N_{\mathrm{det}}=\nu^*/w^2$ detections, where $\nu^*$ is the noncentrality at which that distribution exceeds the critical value with probability $1-\beta_{\mathrm{II}}$. As a common rule of thumb, the approximation is reliable when each bin has an expected count of at least five. The corresponding number of runs is $N_{\mathrm{det}}/\mathbb{P}_L(\mathrm{D}_T;\psi^{\mathrm{s.w.}}_\sigma)$, with $\mathbb{P}_L(\mathrm{D}_T;\psi^{\mathrm{s.w.}}_\sigma)$ the alternative's total detection probability. For MS and the near-field comparisons, where expected counts are small, the numbers of runs are obtained by simulating the test: the rejection threshold is set, conditional on the number of detections, to achieve significance level $\alpha$ under the relevant null hypothesis, and the number of runs is chosen to achieve power $1-\beta_{\mathrm{II}}$.

At $T=100\,\mathrm{ms}$, this test distinguishes MS from SC in about $100$ runs, since MS predicts markedly earlier detections, and CAP and ABC in about $2400$ and $4400$ runs, respectively (rounded up to the nearest hundred). It also distinguishes SD from SC in principle, through its slower tail, but requires of order $10^5$ runs. QF and SC predict the same distribution of detection times to within numerical precision at this distance.

\subsection{Shape of the angular distribution}\label{sec:shape}

Another way of discriminating proposals is to consider the shape of the angular distribution, conditional on detection
before $T$. Unlike the binomial test of Section~\ref{sec:totals}, this comparison does not require knowing the total number of trials, and any loss that does not depend on the arrival angle leaves the conditional distribution unchanged.

\begin{figure}[ht]
    \centering
    \includegraphics[width=\columnwidth]{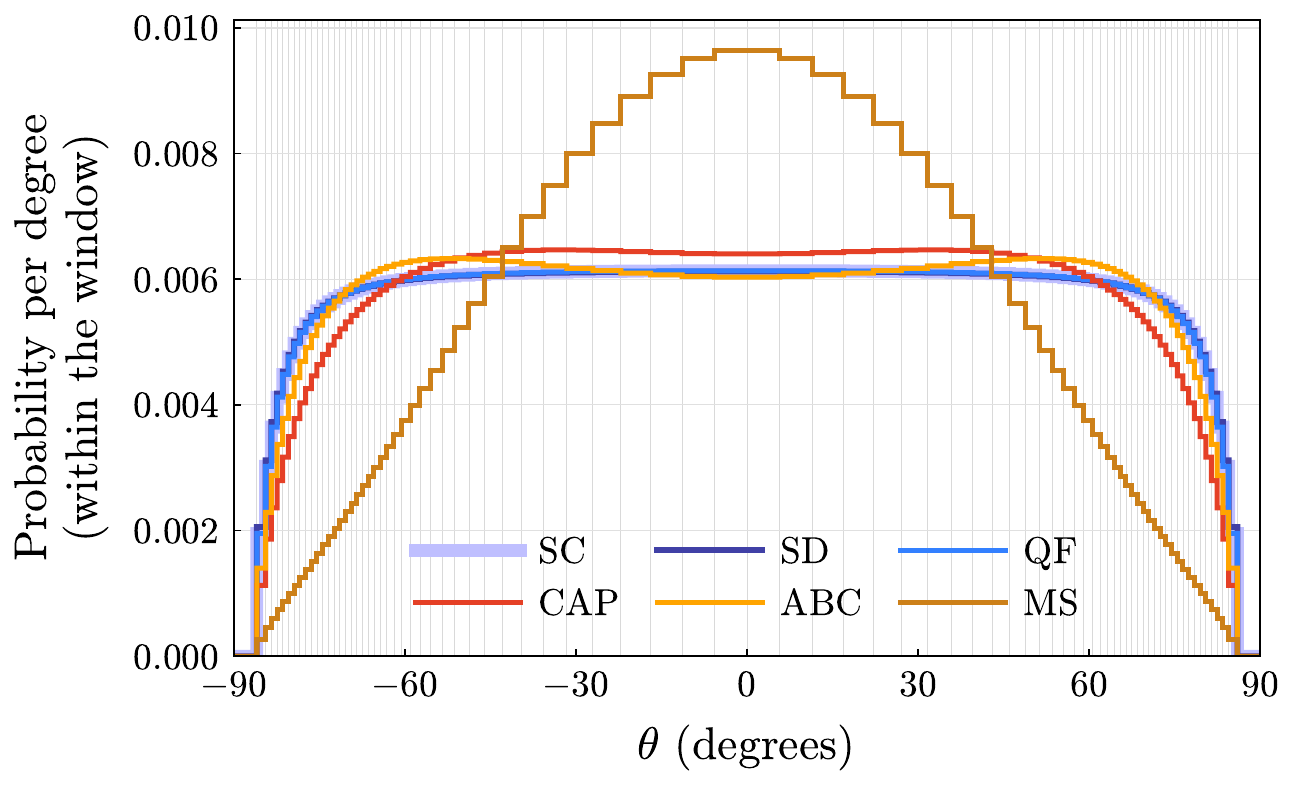}

    \caption{Binned angular distributions conditional on detection within the window $|\theta|\le86^\circ$ before $T=100\,\mathrm{ms}$, for the single-well state at $L=100\sigma$ with the parameters of Section~\ref{sec:singleDouble}, shown as probability per degree. The bins have width $\max\big(1^\circ,\cos^2(\theta)\,\tfrac{\Delta y}{L}\big)$ with $\Delta y=10\,\mu\mathrm{m}$ (faint vertical lines mark the bin edges). Nothing is recorded outside the window. The SC distribution lies above the uniform value $1/172$ per degree away from the window edges, because the arrivals lost near grazing are excluded by the conditioning. Distinguishing each proposal from SC by the $\chi^2$ test on these distributions ($\alpha=0.05$, $1-\beta_{\mathrm{II}}=0.95$) requires about $13{,}200$ runs for CAP, $46{,}800$ for ABC, and several hundred for MS. QF and SD are not distinguishable from SC.}
    \label{fig:T100Cutoff}
\end{figure}

\begin{figure}[ht]
    \centering
    \includegraphics[width=\columnwidth]{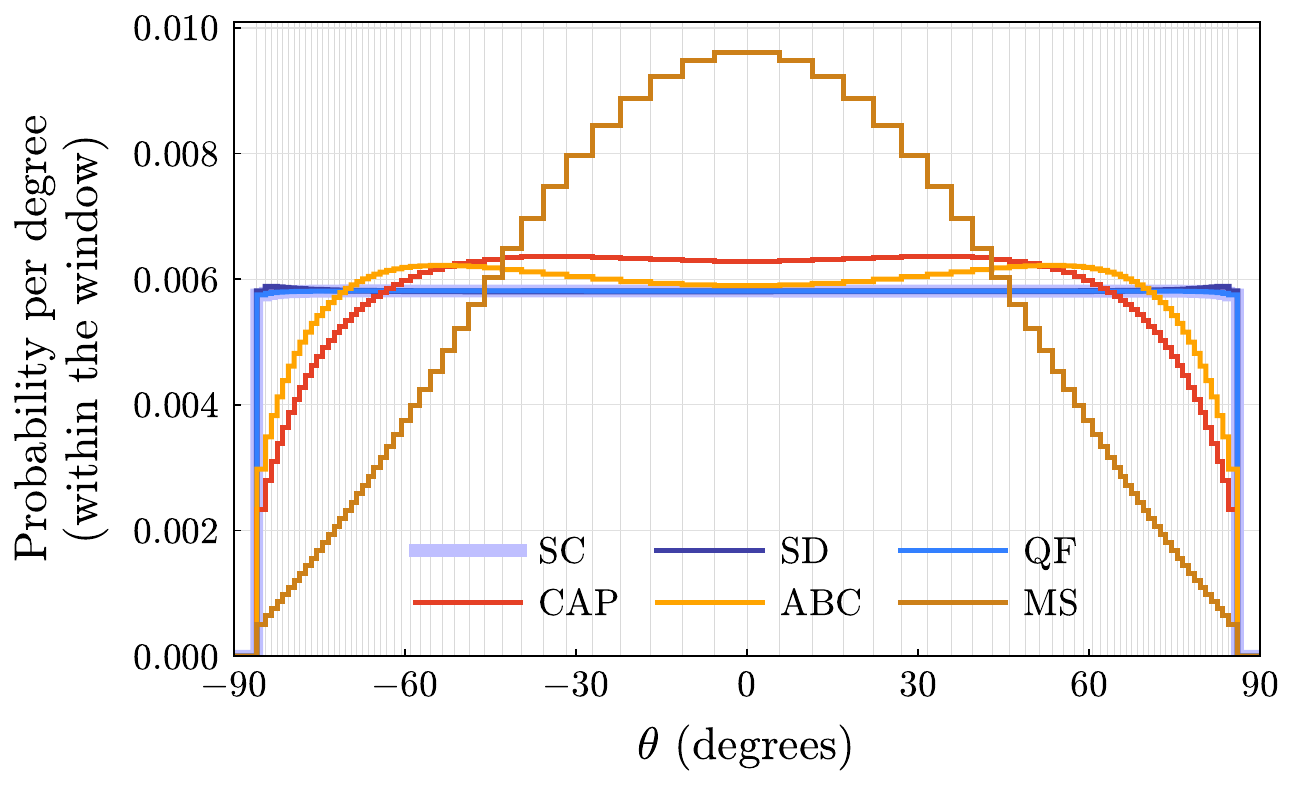}
    \caption{As in Fig.~\ref{fig:T100Cutoff}, at $T=1000\,\mathrm{ms}$. The corresponding numbers of runs are about $5000$ for CAP, $10{,}600$ for ABC, and several hundred for MS.}
    \label{fig:T1000Cutoff}
\end{figure}

A temporal cutoff affects the shape most dramatically at grazing angles, since, as noted above, arrivals there tend to be late. For SC and the single-well state, this effect can be computed exactly: an SC arrival at angle $\theta$ occurs before $T$ precisely when the momentum magnitude exceeds $mL\sec(\theta)/T$, and since that magnitude is Rayleigh-distributed with scale $\hbar/(2\sigma)$, the fraction of arrivals at angle $\theta$ that occur before $T$ is $\exp[-2(L\sec(\theta)/y_T)^2]$, with $y_T=\hbar T/(m\sigma)$. The cutoff is therefore significant only at angles where $\sec(\theta)$ is not small compared to $T/\tau_L$. Half of the arrivals at angle $\theta$ occur after $T$ when $\sec(\theta)\approx0.59\,T/\tau_L$, that is, at $\theta\approx84^\circ$ for $T=100\,\mathrm{ms}$, inside the window, and $\theta\approx89^\circ$ for $T=1000\,\mathrm{ms}$, outside it. Figures~\ref{fig:T100Cutoff} and~\ref{fig:T1000Cutoff} show the conditional distributions at these two cutoffs. At $T=100\,\mathrm{ms}$, CAP and ABC already differ visibly from SC, but the grazing region where the difference is largest is partly suppressed. At $T=1000\,\mathrm{ms}$ that region is largely restored. MS differs from SC at both cutoffs, since its distribution differs at all angles rather than only near grazing. For the double-well state at this distance (not shown), QF and SD remain nearly indistinguishable from SC, as expected from their coincidence in the far-field limit.

To quantify the number of runs needed, we apply the $\chi^2$ test of Section~\ref{sec:totals}, with SC as the null hypothesis, to the detected angles, with each distribution normalized over the window. We assume that the detection scheme resolves angles at least as finely as the bins. For a planar screen that resolves the position $y$ to within $\Delta y$, the corresponding angular resolution is $\Delta\theta\approx\cos^2(\theta)\tfrac{\Delta y}{L}$, which becomes finer toward grazing angles, where a fixed $\Delta y$ subtends a smaller angle. We therefore use bins of width $\max\big(1^\circ,\cos^2(\theta)\tfrac{\Delta y}{L}\big)$ with $\Delta y=10\,\mu\mathrm{m}$, the single-atom resolution of Ref.~\cite{bucker2009single}. This gives bins of about $6^\circ$ near the normal, narrowing to $1^\circ$ beyond $\theta\approx65^\circ$. A detection scheme with coarser resolution would be analyzed in the same way, with a larger $\Delta y$. Background events would add an approximately uniform component to the recorded distribution, and can be neglected provided that the expected number per run within the window is small compared with $\mathbb{P}_L(\mathrm{D}_T;\psi^{\mathrm{s.w.}}_\sigma)$.

The resulting numbers of runs are given in the captions of Figs.~\ref{fig:T100Cutoff} and~\ref{fig:T1000Cutoff}. Within the window, the CAP and ABC distributions are close in shape to SC's, so distinguishing them by shape requires considerably more runs than the total-probability test of Table~\ref{tab:TotalsN}. The MS distribution differs markedly in shape, and several hundred runs suffice. Extending the cutoff to $T=1000\,\mathrm{ms}$ reduces the required numbers for CAP and ABC by factors of about three and four, since more of the late arrivals near grazing incidence, where these proposals differ most from SC, are then recorded.

Neither the totals of Section~\ref{sec:totals} nor the shape analysis above separates SC, QF, and SD in practice at $L=100\sigma$. Doing so efficiently requires smaller values of $\tfrac{L}{\sigma}$, which we consider in Section~\ref{sec:nearfield}.

\begin{figure}[ht]
    \centering
    \includegraphics[width=\columnwidth]{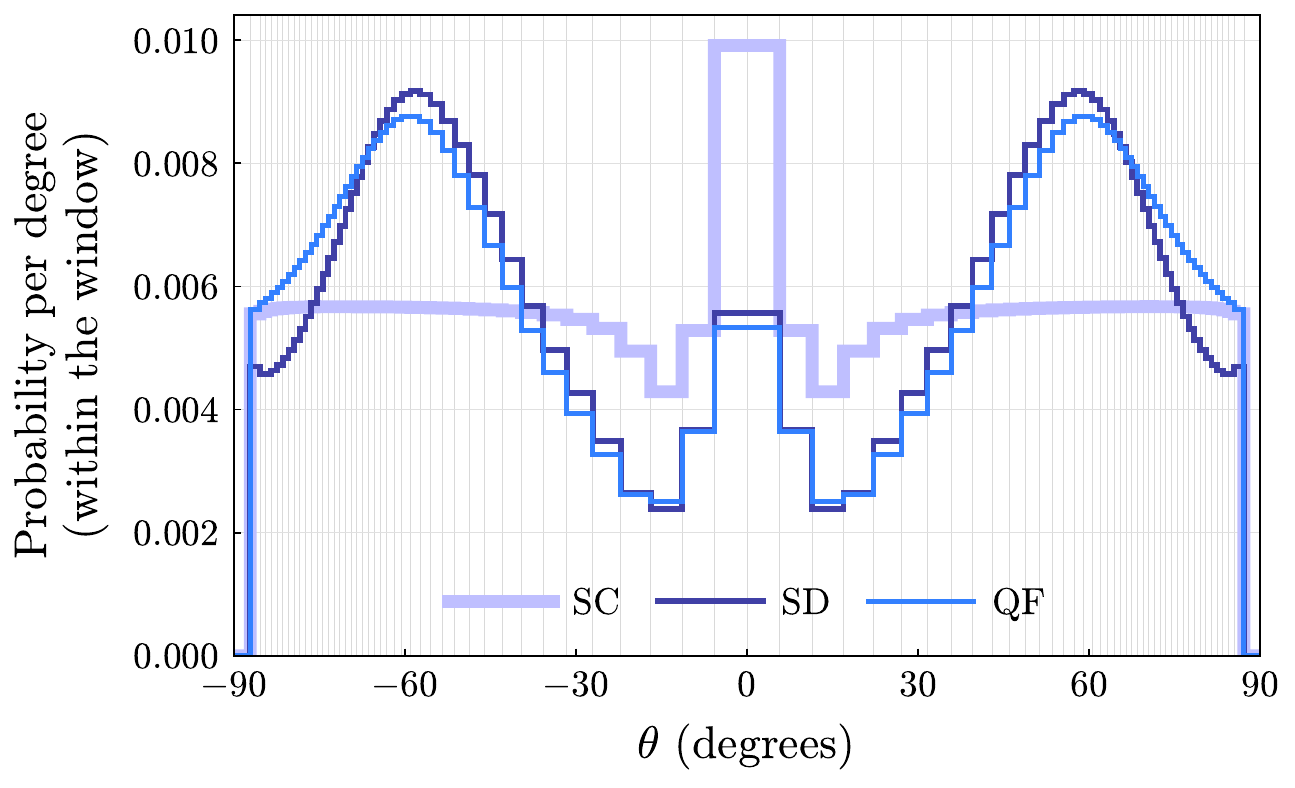}
    \caption{Binned angular distributions of the double-well state conditional on detection within a field of view of $0.4\,\mathrm{mm}$ ($|\theta|\le87^\circ$) before $T=100\,\mathrm{ms}$, at $L=10\sigma$, for SC, QF, and SD, shown as probability per degree. The bins have width $\max\big(1^\circ,\cos^2(\theta)\tfrac{\Delta y}{L}\big)$ with $\Delta y=1\,\mu\mathrm{m}$ (faint vertical lines mark the bin edges). SC retains the central peak of the momentum distribution, while QF and SD peak near $\theta\approx\pm60^\circ$.}
    \label{fig:NearFieldBinned}
\end{figure}

\subsection{Distinguishing SC, QF, and SD in the near field}\label{sec:nearfield}

\begin{figure}[hb]
    \centering
    \includegraphics[width=\columnwidth]{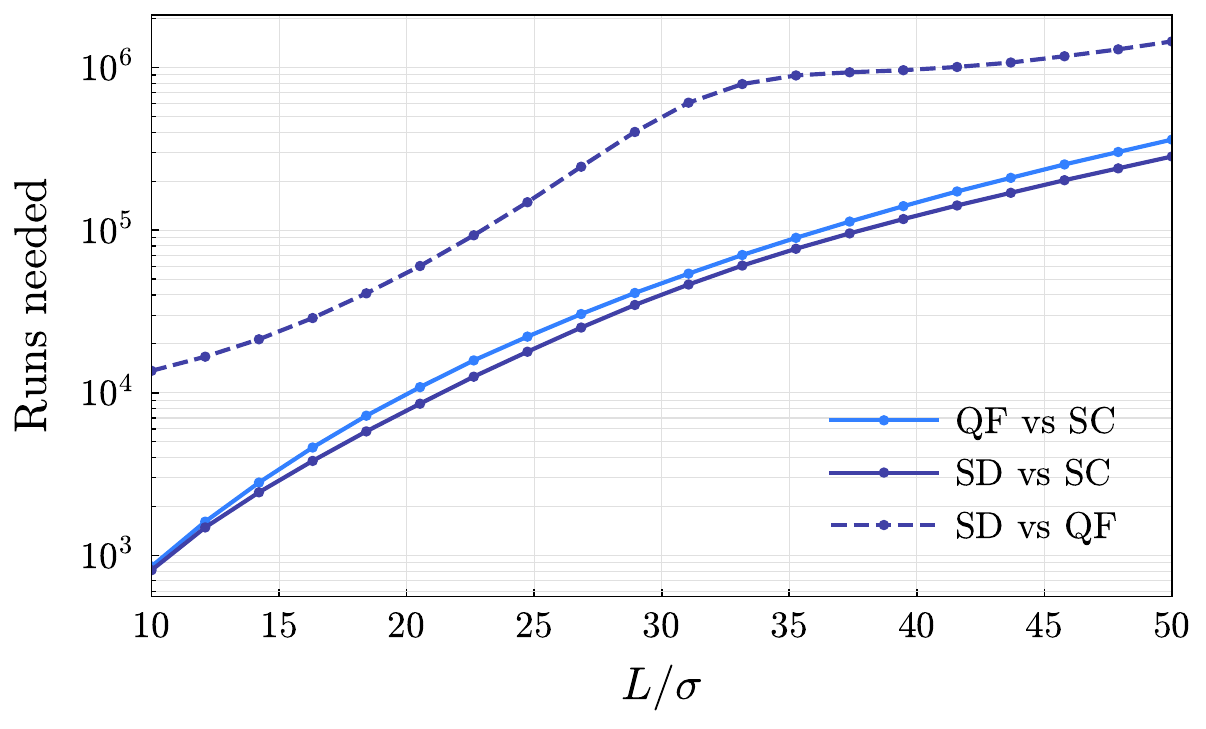}
    \caption{Number of runs needed to distinguish QF and SD from SC, and SD from QF, using the $\chi^2$ test on the binned angular distribution of the double-well state with cutoff $T=100\,\mathrm{ms}$, as a function of the screen distance $L$. The window $|\theta|\leq\arctan(0.2\,\mathrm{mm}/L)$ narrows from $87^\circ$ at $L=10\sigma$ to $76^\circ$ at $L=50\sigma$. Bins follow the rule of Section~\ref{sec:shape} with $\Delta y=1\,\mu\mathrm{m}$.}
    \label{fig:RunsVsNearness}
\end{figure}

In the far-field limit, QF and SD yield the same angular distribution as SC (Section~\ref{sec:farfield}). The three differ appreciably only when the screen distance is comparable to the separation of the wells. At $L=10\sigma$, for example, the SC distribution for the double-well state retains the central peak of the momentum distribution, while QF and SD instead peak near $\theta\approx\pm60^\circ$, reflecting the positions of the two wells (Fig.~\ref{fig:NearFieldBinned}). Figure~\ref{fig:RunsVsNearness} shows how the number of runs needed to distinguish these proposals grows with $L$, using the $\chi^2$ test of Section~\ref{sec:totals} on the binned angular distribution, as in Section~\ref{sec:shape}, at $T=100\,\mathrm{ms}$.

At $L=10\sigma$, of order $1000$ runs suffice to distinguish QF or SD from SC. The number grows rapidly with $L$: about $10^4$ at $20\sigma$, about $5\times 10^4$ at $30\sigma$, and more than $10^5$ from $40\sigma$ on. Distinguishing SD from QF (with QF as the null hypothesis) is harder, and already requires more than $10^4$ runs at $L=10\sigma$.

Smaller values of $\tfrac{L}{\sigma}$ also place stricter demands on spatial resolution. Here we take $\Delta y=1\,\mu\mathrm{m}$ and a $0.4\,\mathrm{mm}$ field of view. This is within reach of MCP-based detectors, which have demonstrated a resolution of about $1\,\mu\mathrm{m}$ over a $0.4$--$0.8\,\mathrm{mm}$ field of view~\cite{kurtsiefer19962A,kurtsiefer1997measurement}. At $L=10\sigma$, this field of view covers $|\theta|\le87^\circ$.

These requirements can be relaxed by scaling up the trap, increasing $\sigma$ and the well separation in proportion. Note that the angular distributions depend only on $\tfrac{L}{\sigma}$, the well separation in units of $\sigma$, and $\tfrac{T}{\tau_L}$. Scaling the trap up tenfold, to $\sigma=10\,\mu\mathrm{m}$, places the screen at $L=10\sigma=100\,\mu\mathrm{m}$ and enlarges the angular features tenfold in $y$, bringing them within the resolution of light-sheet detection. The tradeoff is a longer experiment: $\tau_L=m\sigma L/\hbar$ grows as $\sigma^2$ at fixed $\tfrac{L}{\sigma}$, reaching about $60\,\mathrm{ms}$ for ${}^4\mathrm{He}$. At $L=10\sigma$, the number of runs needed to distinguish QF or SD from SC is the same at $T=10\tau_L$ as at $T\approx 160\tau_L$ (i.e., $T=100\,\mathrm{ms}$ for $\sigma=1\,\mu\mathrm{m}$), while distinguishing SD from QF requires about twice as many runs at $10\tau_L$ as at $30\tau_L$. The cutoff $T$ must therefore grow to about $0.6\,\mathrm{s}$ to distinguish QF or SD from SC, and to about $2\,\mathrm{s}$ to distinguish SD from QF.

\vfill

\bibliography{bibliography}

\end{document}